\documentclass[aps,twocolumn,prb,floatfix,eqsecnum,showpacs]{revtex4-1}

\usepackage{amsmath}
\usepackage{amsfonts}
\usepackage{amssymb}
\usepackage{graphicx}
\usepackage{bm}
\usepackage{color}
\usepackage{ulem}

\usepackage[resetlabels]{multibib}
\usepackage[toc,page]{appendix}

\def\stacksymbols #1#2#3#4{\def\theguybelow{#2}
    \def\verticalposition{\lower#3pt}
    \def\spacingwithinsymbol{\baselineskip0pt\lineskip#4pt}
    \mathrel{\mathpalette\intermediary#1}}
\def\intermediary#1#2{\verticalposition\vbox{\spacingwithinsymbol
      \everycr={}\tabskip0pt
      \halign{$\mathsurround0pt#1\hfil##\hfil$\crcr#2\crcr
               \theguybelow\crcr}}}

\begin{document} 

\title{Spin-directed network model for the surface states of weak three-dimensional 
$\mathbb{Z}^{\,}_{2}$ topological insulators}

\author{Hideaki Obuse} 
\affiliation{
 Department of Applied Physics, Hokkaido University, Sapporo 060-8628, Japan
            } 

\author{Shinsei Ryu} 
\affiliation{
Department of Physics, University of Illinois, 
1110 West Green St, Urbana IL 61801, USA
            } 

\author{Akira Furusaki} 
\affiliation{
Condensed Matter Theory Laboratory,
RIKEN, Wako, Saitama 351-0198, Japan
            } 
\affiliation{
RIKEN Center for Emergent Matter Science (CEMS),
Wako, Saitama 351-0198, Japan
            } 

\author{Christopher Mudry} 
\affiliation{
Condensed matter theory group, 
Paul Scherrer Institute, CH-5232 Villigen PSI,
Switzerland
            } 

\date{April 14, 2014}

\begin{abstract}
A two-dimensional spin-directed $\mathbb{Z}^{\,}_{2}$ network model 
is constructed that describes the combined effects of 
dimerization and disorder for the surface states of 
a weak three-dimensional $\mathbb{Z}^{\,}_{2}$ topological insulator.
The network model consists of helical edge states of two-dimensional
layers of $\mathbb{Z}^{\,}_{2}$ topological insulators which are coupled by
time-reversal symmetric interlayer tunneling.
It is argued that, without dimerization of interlayer couplings,
the network model has no insulating phase for any disorder strength.
However, a sufficiently strong dimerization induces a transition from
a metallic phase to an insulating phase.
The critical exponent $\nu$ for the diverging localization length
at metal-insulator transition points
is obtained by finite-size scaling analysis of numerical data from
simulations of this network model.
It is shown that the phase transition belongs to the
\textit{two-dimensional symplectic universality class} of Anderson
transition.
\end{abstract}


\maketitle


\newpage


\section{Introduction}
\label{Sec: Introduction}

A strong $d$-dimensional topological band insulator 
is a band insulator of noninteracting electrons
which is characterized by a nontrivial topological index
under certain symmetry constraints.
Its $(d-1)$-dimensional boundary always has gapless
boundary states which are extended on the boundary
but localized in the direction normal to the boundary.
They share, because of the nontrivial topological index,
a degree of robustness to perturbations
that respect the symmetry constraints.

The simplest example of a strong two-dimensional topological band insulator
(without any symmetry constraint)
was constructed by Haldane.~\cite{Haldane88}
With periodic boundary conditions,
it has two single-particle Bloch bands separated by a gap $\Delta$.
Its nontrivial topological invariant is a non-vanishing Chern number
that is proportional to the Hall conductivity.%
~\cite{Thouless82}
This example is a representative of topological
band insulators called Chern insulators.
In open geometries, Chern insulators support gapless single-particle
boundary states, i.e., edge states.
These edge states are chiral in that they
propagate along the edge either clockwise or anti-clockwise
depending on the sign taken by the Hall conductivity of the occupied
bands. Since intra-edge backward scattering is not permitted, 
these chiral edge states are robust to perturbations.%
~\cite{Halperin82}

\begin{figure}[t]
\centering
\includegraphics[width=0.4\textwidth]{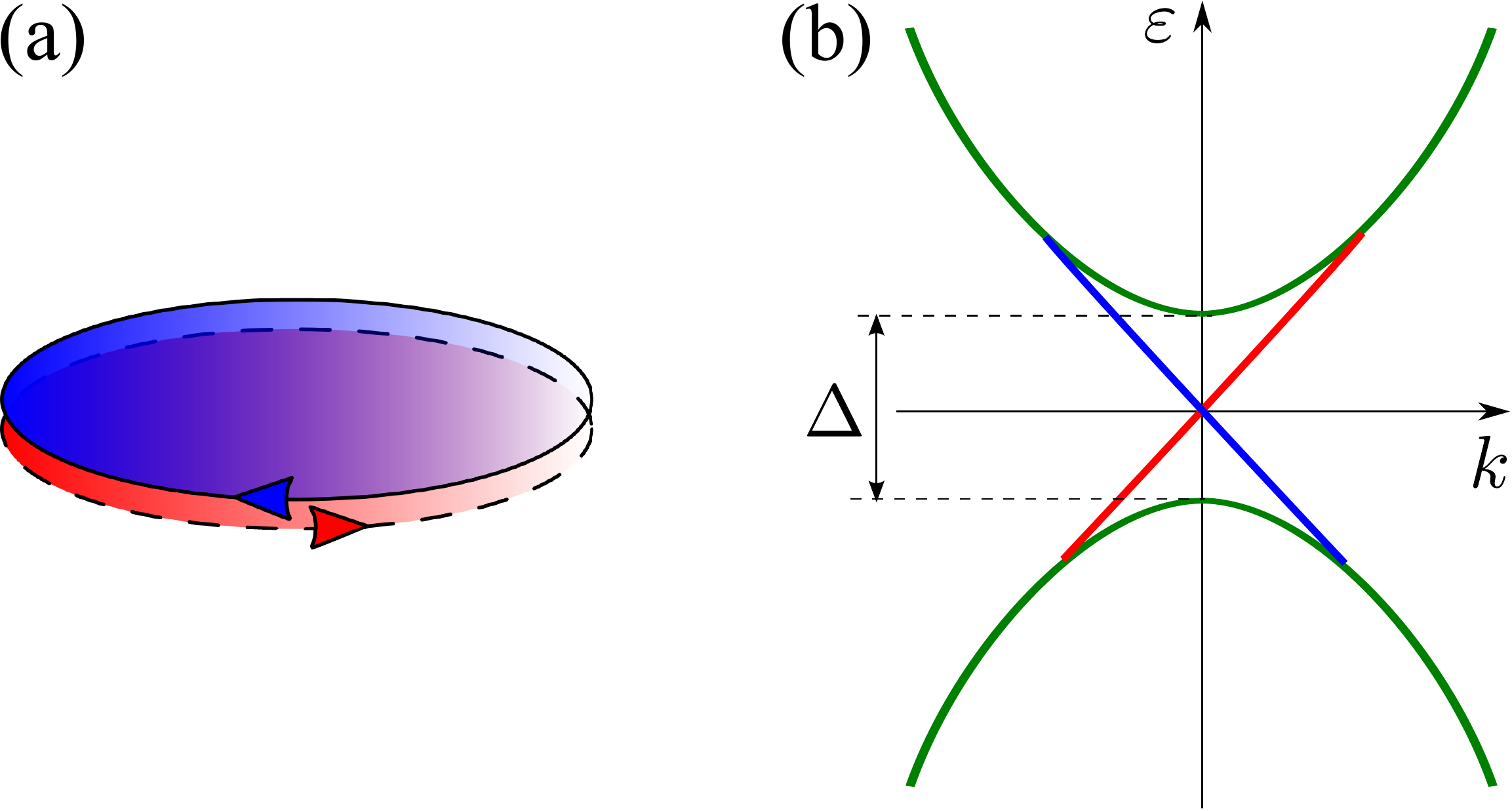}
\caption{
(Color online:)
Cartoon representation of a strong two-dimensional 
$\mathbb{Z}^{\,}_{2}$ topological band insulator with one 
connected boundary. 
(a)
Electrons are noninteracting and confined within an ellipse
in two-dimensional position space.
A single pair of counter propagating
helical edge states at zero energy
are denoted by the thick and dashed lines 
with arrows along the connected boundary of the ellipse, 
respectively. They are confined to this one-dimensional
boundary.
(b) The single-particle spectrum consists of
eigenstates separated by an energy gap $\Delta$
and eigenstates crossing the band gap $\Delta$.
The former eigenstates are the bulk eigenstates.
The latter eigenstates are the edge eigenstates.
The momentum $k$ is the momentum along the boundary.
The wave functions in position space
of bulk eigenstates are supported in the shaded region
of the ellipse.
The wave functions in position space
of edge eigenstates are extended along the edge
and labeled by the momentum quantum number $k$ along the edge,
while they decay exponentially fast away from the edge.
        }
\label{fig:z2pancake}
\end{figure}

A more intricate example of a strong two-dimensional topological band insulator
was constructed by Kane and Mele.~\cite{Kane05a,Kane05b}
With periodic boundary conditions,
it has four single-particle bands that form two Kramers' pairs of bands
as a consequence of time-reversal symmetry.
However, spin-rotation symmetry is completely broken by spin-orbit coupling.
A gap $\Delta$ separates the two pairs of bands, and
the Bloch wave functions of the occupied bands have a nontrivial topological
$\mathbb{Z}^{\,}_{2}$ index.%
~\cite{Kane05b} 
In an open geometry,
there is a Kramers' pair of edge states,
called helical edge states (Fig.~\ref{fig:z2pancake}).
This pair is robust to perturbations that respect
the time-reversal symmetry.%
~\cite{Suzuura02,Ando02,Takane04}
This example is a representative of topological
band insulators called $\mathbb{Z}^{\,}_{2}$ topological band insulators.

\begin{figure}[t]
\centering
\includegraphics[width=0.3\textwidth]{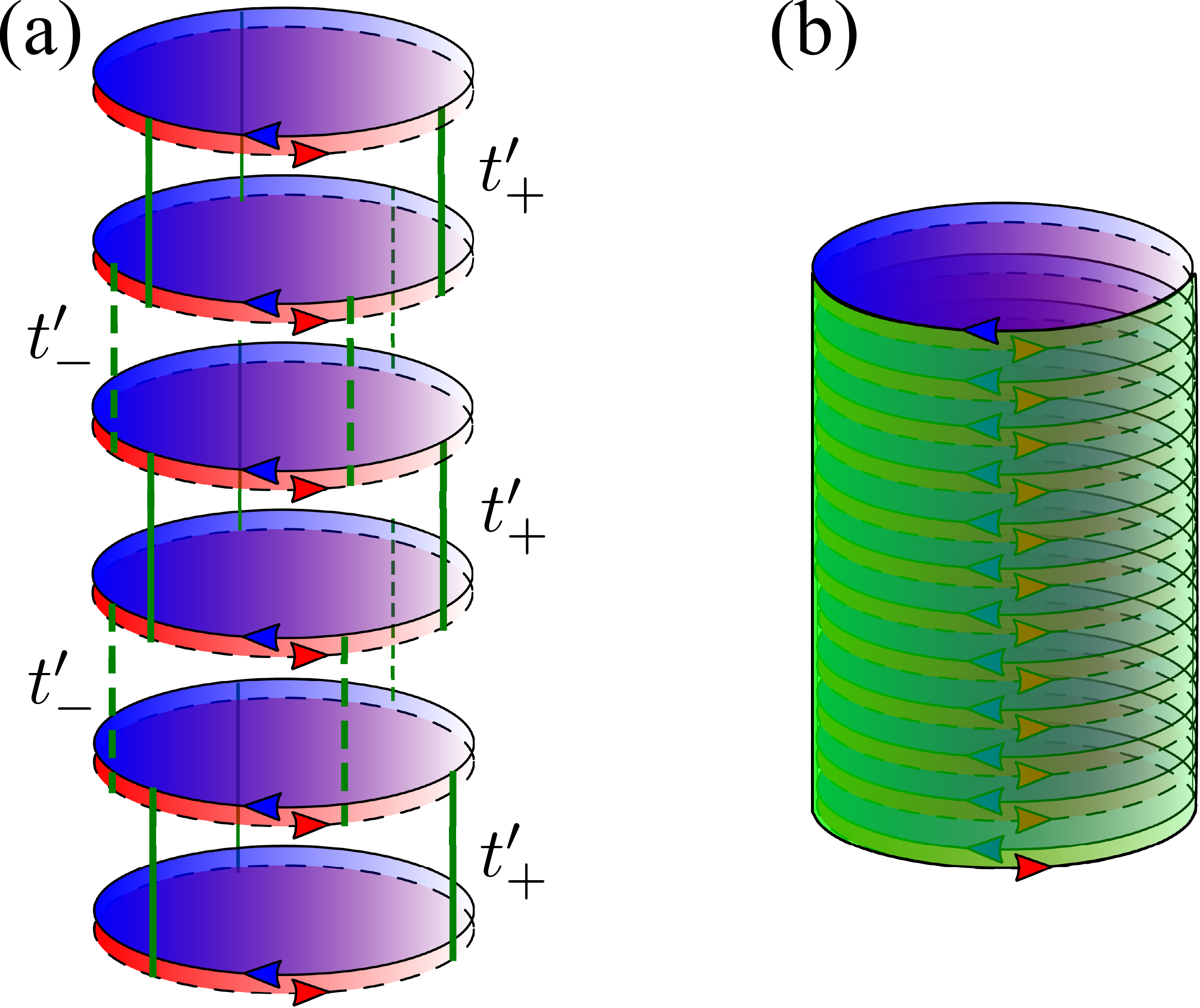}
\caption{
(Color online:)
(a) Cartoon representation of a weak layered $\mathbb{Z}^{\,}_{2}$
topological band insulator with open boundary conditions along the
layering axis.  A vertical line
represents the amplitude for the Kramers' degenerate helical
edge states of a layer consisting of a 
strong two-dimensional $\mathbb{Z}^{\,}_{2}$ topological band insulator
to hop to one of its adjacent layers. 
The vertical lines are alternatively
drawn with solid
and dotted lines in order to indicate
that the magnitude of this hopping amplitude takes two distinct values.
As a consequence of the breaking of translation invariance 
by one lattice spacing along the stacking axis, 
a dimerization gap opens for all helical edge states,
i.e., for all surface states of this weak layered $\mathbb{Z}^{\,}_{2}$
topological band insulator.
(b) The topology in (a) is that of a cylinder.
        }
\label{fig:layeredz2pancake}
\end{figure}

Weak topological band insulators are built out of strong lower-dimensional 
topological band insulators. For example, a weak three-dimensional
topological insulator can be a stack of strong two-dimensional
topological insulators as is illustrated in Fig.~\ref{fig:layeredz2pancake}.
Weak topological band insulators can support surface states
in open geometries that are inherited from the boundary states
of their lower-dimensional strong topological band building blocks.

Weak three-dimensional Chern insulators have been studied theoretically 
in the context of three-dimensional generalizations of the quantum Hall
effect.~\cite{Halperin87,Chalker95,Balents96,Kim96,Mathur97,Gruzberg97,Cho97,Balents97}
Two-dimensional transport at the surface of 
GaAs/AlGaAs multilayer structures
subjected to a large uniform magnetic field parallel to the stacking axis
was established in Ref.~\onlinecite{Druist98}.

Models for weak three-dimensional $\mathbb{Z}^{\,}_{2}$ topological 
band insulators were constructed simultaneously with models 
for strong three-dimensional 
$\mathbb{Z}^{\,}_{2}$ topological insulators.%
~\cite{FuKaneMele07,Moore07,Roy07,Fu07}
Of course, most of the excitement
generated by these works was reserved for the strong three-dimensional 
$\mathbb{Z}^{\,}_{2}$ topological insulators, especially after their
experimental discovery in $\mathrm{Bi}^{\,}_{1-x}\mathrm{Sb}^{\,}_{x}$,%
~\cite{Hsieh08}
$\mathrm{Bi}^{\,}_{2}\mathrm{Se}^{\,}_{3}$,%
~\cite{Xia09,Chen09}
and 
$\mathrm{TlBiSe}^{\,}_{2}$.%
~\cite{Lin10,Sato10,Kuroda10,Chen10}
A weak three-dimensional 
$\mathbb{Z}^{\,}_{2}$ topological insulator has been identified
experimentally in the form of 
$\mathrm{Bi}^{\,}_{14}\mathrm{Rh}^{\,}_{3}\mathrm{I}^{\,}_{9}$.%
~\cite{Rasche13}

Two-dimensional surfaces of 
strong (weak) three-dimensional $\mathbb{Z}^{\,}_{2}$ 
topological band insulators support an odd (even) number of
helical surface states. 
Each such helical surface state effectively realizes the linear dispersion 
of a massless two-component Dirac particle in two-dimensional momentum space.
A single massless two-component Dirac particle 
cannot accommodate a mass term without breaking time-reversal symmetry
in two dimensions.
In this sense, a single Dirac cone on the surface of a strong
three-dimensional $\mathbb{Z}^{\,}_{2}$ 
topological band insulator is protected by time-reversal symmetry.
This is not so for a pair of massless 
two-component Dirac particles in two dimensions.
They can accommodate a mass term that does not break time-reversal symmetry,
but breaks some of the lattice symmetries.
Thus, a pair of Dirac cones on the surface of a weak $\mathbb{Z}^{\,}_{2}$ 
topological band insulator is not protected by time-reversal
symmetry.%
~\cite{FuKaneMele07}
For this reason, weak topological band insulators have initially
attracted less interest than strong topological band insulators.

In the presence of disorder that preserves the time-reversal symmetry, 
momentum is not a good quantum number anymore. In fact, the very notion 
of a spectral gap or of gaplessness is meaningless in the presence 
of disorder. It is replaced by the notion of mobility edges 
that defines the windows of single-particle energies
for which single-particle states are localized by disorder.

\begin{figure}[t]
\centering
\includegraphics[width=0.3\textwidth]{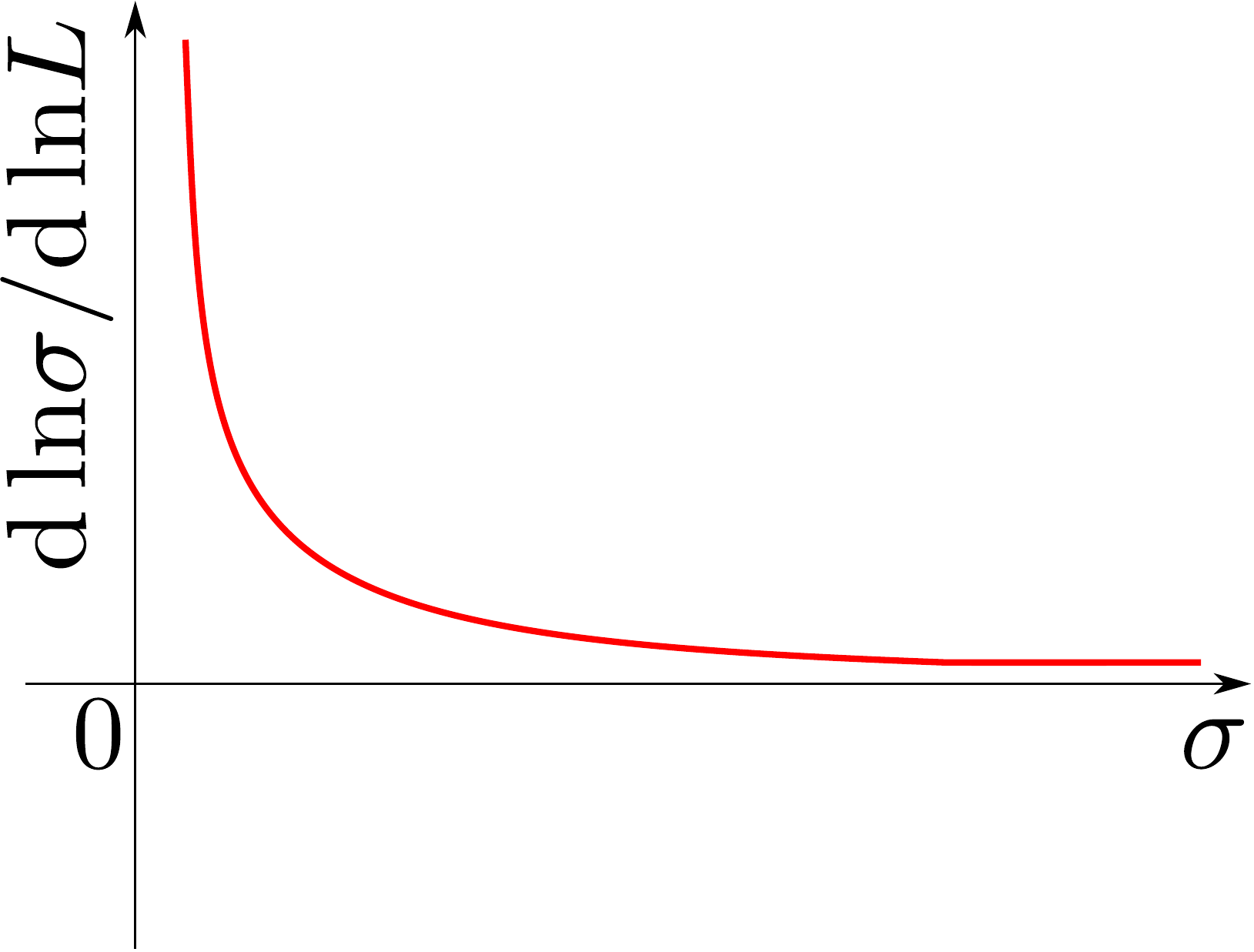}
\caption{
(Color online:)
Qualitative beta function
$d\ln\sigma/d\ln L$
for the conductivity 
$\sigma$ of a massless two-component Dirac fermion 
on the surface with the area $L^{2}$ of a strong three-dimensional 
$\mathbb{Z}^{\,}_{2}$
topological insulator, when subjected to disorder
that preserves time-reversal symmetry. The beta function
$d\ln\sigma/d\ln L$
is deduced from the numerics of 
Refs.~\onlinecite{Bardarson07}
and~\onlinecite{Nomura07}.
It is always positive and
the smaller the initial value of the conductivity is
the larger the positive value of the beta function.
         }
\label{fig:beta fct sigma at Dirac cone STI}
\end{figure}

The consequences of disorder weaker than the bulk band gap $\Delta$
for the surface states 
of a strong three-dimensional $\mathbb{Z}^{\,}_{2}$ topological 
band insulator can be accounted for by describing 
the helical surface states
as massless two-component Dirac fermions
in two-dimensional space perturbed by
local potentials that break translation invariance along the boundary 
but preserve time-reversal symmetry. 
In turn, this effective field theory can be
approximated by a nonlinear $\sigma$-model (NL$\sigma$M) 
with a two-dimensional base space and a target space
determined by the symplectic symmetry of the effective Hamiltonian, 
that is augmented by a topological term.%
~\cite{Ostrovsky07,Ryu07}
This topological term prevents the transition
from a conducting to an insulating phase as the disorder strength 
on the boundary is made arbitrarily large,%
~\cite{footnote: order limits matter here}
as is implied by the strictly monotonic
one-parameter scaling law obeyed by the conductivity 
$\sigma$ on a surface of area $L^{2}$
that is captured by the beta function 
$d\ln\sigma/d\ln L$
shown in
Fig.~\ref{fig:beta fct sigma at Dirac cone STI}.%
~\cite{Bardarson07,Nomura07}
By contrast,
no topological term augments the NL$\sigma$M description of the effect 
of time-reversal-symmetric disorder for the surface states of a 
weak three-dimensional $\mathbb{Z}^{\,}_{2}$ topological 
band insulator.~\cite{Ryu07}
One might be tempted to deduce from this fact that
strong disorder always causes localization.
This is not so however.

The even number of Dirac points on the surface 
of a weak three-dimensional $\mathbb{Z}^{\,}_{2}$ topological 
band insulator allows a time-reversal symmetric perturbation,
a dimerization,
to open a spectral gap $2|m|\ll\Delta$ 
for the surface states at the cost of breaking some lattice
symmetries.%
~\cite{FuKaneMele07,Liu12}
The effective Dirac Hamiltonian with a dimerization mass $m$
in the clean limit turns out to be the one describing
two-dimensional strong $\mathbb{Z}^{\,}_{2}$ topological insulators.%
~\cite{Liu12,Mong12,Fu12,Morimoto13}
Hence, the sign of the mass $m$ 
selects one of the two topologically distinct massive phases
of a strong two-dimensional $\mathbb{Z}^{\,}_{2}$ topological band insulator.
The surface Dirac points can thus be associated
with a quantum critical point separating two massive dimer phases
with less lattice symmetries. 
A defect of the dimerization mass $m$
along a curve on the surface of the weak three-dimensional 
$\mathbb{Z}^{\,}_{2}$ topological band insulator 
at which $m$ smoothly changes sign binds a pair of
Kramers degenerate helical edge states whose dispersion crosses the gap
$2|m|$ of the surface states.%
~\cite{Liu12,Mong12,Fu12,Morimoto13,Callan85}
Provided the disorder
is time-reversal symmetric,
this quantum critical point has the remarkable property
that, perturbed by time-reversal symmetric surface disorder, 
it turns into a metallic phase separating two
insulating dimerized phases.

\begin{figure}[t]
\includegraphics[width=0.3\textwidth]{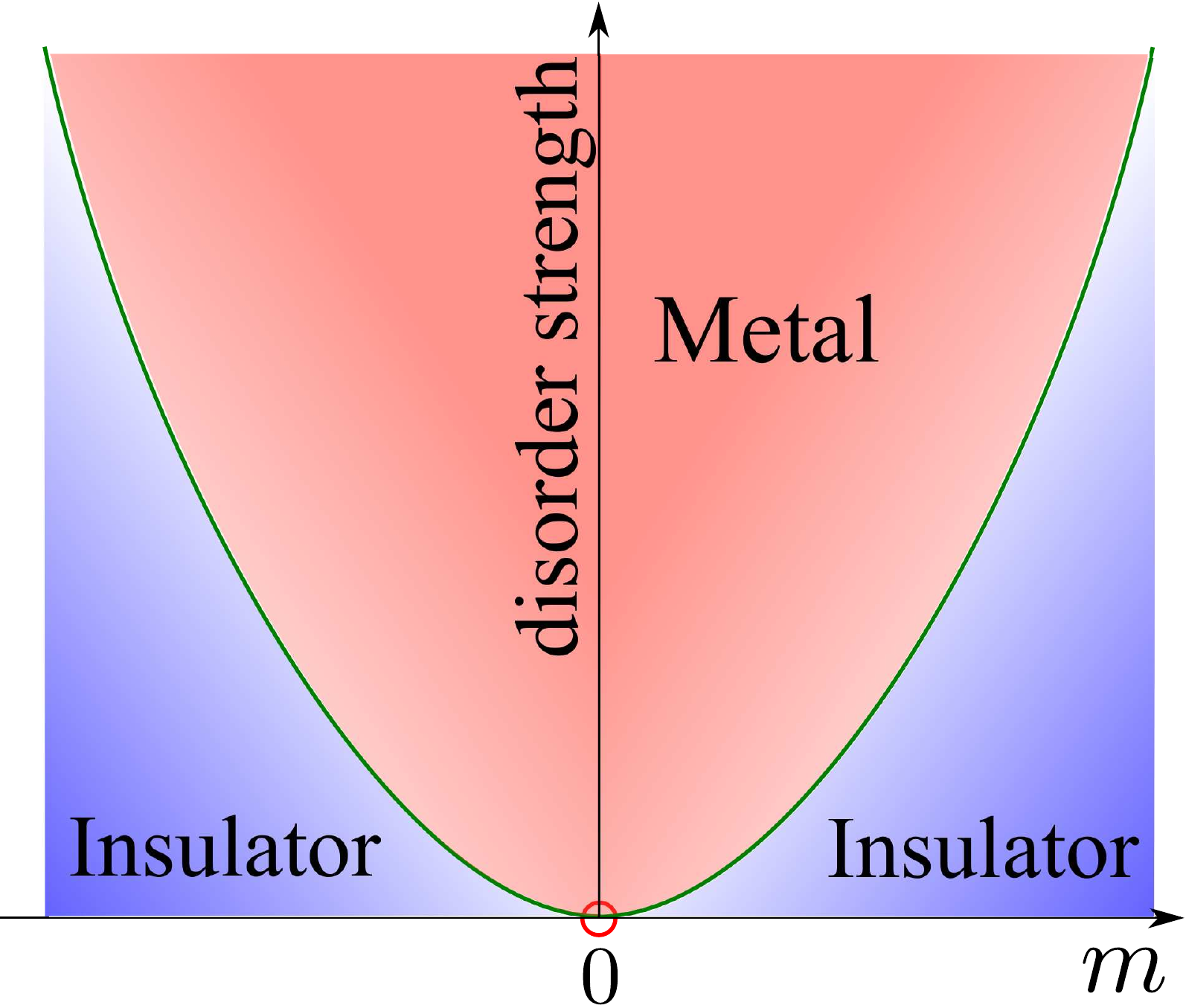}
\caption{
(Color online:)
Phase diagram from Ref.~\onlinecite{Mong12}
for the surface states of
a weak three-dimensional $\mathbb{Z}^{\,}_{2}$
topological band insulator as a function of 
the mass that opens a band gap at the Dirac points
(horizontal axis) and the disorder strength
(vertical axis).
It is assumed that the disorder does not mix
surface states localized on disconnected boundaries
of the three-dimensional $\mathbb{Z}^{\,}_{2}$
topological band insulator.%
~\cite{footnote: order limits matter here}
The origin of the phase diagram is the critical
point corresponding to an even number of
two-component massless Dirac fermions at their Dirac point.
         }
\label{fig: phase dia+RG flows WTI}
\end{figure}

This counter-intuitive conclusion was first reached by
Ringel, Kraus, and Stern for a layered model of
a weak three-dimensional $\mathbb{Z}^{\,}_{2}$ topological 
band insulator based on the sensitivity to twisted
boundary condition of spectral flows.~\cite{Ringel11}
Mong, Bardarson, and Moore did
a numerical study of the disordered Dirac equation
capturing the low-energy and long-wave-length limit
of the disordered surface states 
of a weak three-dimensional $\mathbb{Z}^{\,}_{2}$ topological 
band insulator that confirmed this prediction and, 
furthermore, showed that the clean Dirac critical point 
is smoothly connected to  a metallic phase 
as shown in Fig.~\ref{fig: phase dia+RG flows WTI}.%
~\cite{Mong12}
Fu and Kane pointed out the importance of 
$\mathbb{Z}^{\,}_{2}$ vortices in a description in terms of
a NL$\sigma$M of disordered surface states of a weak three-dimensional
$\mathbb{Z}^{\,}_{2}$ topological band insulator.%
~\cite{Fu12} 
Furthermore, they have proposed the possibility of a non-monotonous
renormalization-group flow on the line $m=0$
in Fig.~\ref{fig: phase dia+RG flows WTI}.
However, the monotonic scaling observed in 
Ref.~\onlinecite{Mong12} seems to
indicate that $d\ln\sigma/d\ln L>0$ at $m=0$, 
as is the case of surface states of a strong $\mathbb{Z}^{\,}_{2}$
topological insulator shown in Fig.~\ref{fig:beta fct sigma at Dirac cone STI}.
Finally, a numerical
study of the localization properties of a disordered
lattice model for a three-dimensional 
$\mathbb{Z}^{\,}_{2}$ topological band insulator
that interpolates from the strong to the weak regimes 
is consistent with the phase diagram of
Fig.~\ref{fig: phase dia+RG flows WTI}.%
~\cite{Kobayashi13}

A question that has been left open so far
is that of the nature of the phase transition
between the insulating and the metallic phases
in Fig.~\ref{fig: phase dia+RG flows WTI}. 
In this paper, we construct a two-dimensional 
network model for the surface states of 
a weak three-dimensional $\mathbb{Z}^{\,}_{2}$ topological band insulator. 
We call this network model 
the two-dimensional spin-directed $\mathbb{Z}^{\,}_{2}$ network model. 
It differs from a two-dimensional $\mathbb{Z}^{\,}_{2}$ network model
that we constructed in Ref.~\onlinecite{Obuse07}
to study the effects of disorder on the
phase diagram of a strong two-dimensional 
$\mathbb{Z}^{\,}_{2}$ topological band insulator.%
~\cite{Obuse08,Ryu10}
For the two-dimensional spin-directed $\mathbb{Z}^{\,}_{2}$
network model, we argue by considering several limiting cases
and by mapping it to effective Dirac Hamiltonians that,
without dimerization, there is no insulating phase even when
disorder is strong. With a finite dimerization, we establish numerically
the existence of three phases, two insulating phases
that are separated by a metallic phase.
We show that, aside from being continuous,
the quantum phase transition between the insulating
and metallic phases belongs 
to the two-dimensional universality class 
with symplectic symmetry within the theory of Anderson localization,
irrespectively of whether the insulating phases are
topologically trivial or nontrivial.

The paper is organized as follows.
The two-dimensional spin-directed $\mathbb{Z}^{\,}_{2}$ network model
that describes the combined effects of dimerization and 
disorder for the surface states of
a weak three-dimensional $\mathbb{Z}^{\,}_{2}$ band insulator
is constructed in 
Sec.%
~\ref{sec: Definitions}.
This model is studied in Sec.~\ref{sec: Numerical data with an even number of channels},
in which the main numerical results of this paper are explained.
A generalization of the spin-directed $\mathbb{Z}^{\,}_{2}$ network model to
incorporate surface states of an odd stacking number of strong 
$\mathbb{Z}^{\,}_{2}$ topological insulators is presented in 
Sec.%
~\ref{Numerical data with an odd number of channels}.
Section%
~\ref{sec: Numerical data with trimerization}
is devoted to 
the two-dimensional spin-directed $\mathbb{Z}^{\,}_{2}$ network model
that describes the combined effects of trimerization and 
disorder on the surface states of
a weak three-dimensional $\mathbb{Z}^{\,}_{2}$ band insulator.
Section~\ref{Sec: discussion}
closes the paper by summarizing our results.
We also present in Appendix  
\ref{appsec: Wire construction for the surface states of a 3D Z2 WTI}
and 
\ref{appsec: Dirac Hamiltonian from the directed Z2 network model}
the relationship between
the model from Sec.%
~\ref{subsec: Quasi-one-dimensional model for the surface states of ...}
captured by Fig.~\ref{fig:layeredz2pancake},
the two-dimensional spin-directed $\mathbb{Z}^{\,}_{2}$ network model 
from Sec.%
~\ref{sec: Two-dimensional directed Z2 network model with dimerization},
and two-dimensional Dirac fermions.\cite{suppl}
Thereby, we establish the complementarity
of our results to those from Ref.~\onlinecite{Mong12}.
The parameter sets used in our finite-size scaling analysis 
are given in Appendix \ref{appsec:finite-size-scaling}.

\section{Definitions and main results}
\label{sec: Definitions}

\subsection{Quasi-one-dimensional model for the surface states of 
a weak three-dimensional $\mathbb{Z}^{\,}_{2}$ topological insulator}
\label{subsec: Quasi-one-dimensional model for the surface states of ...}

We start from the model 
of a weak three-dimensional $\mathbb{Z}^{\,}_{2}$
topological band insulator that is depicted
in Fig.~\ref{fig:layeredz2pancake}.
It consists of a stacking of layers, each of which represents a 
strong two-dimensional $\mathbb{Z}^{\,}_{2}$
topological band insulator depicted
in Fig.~\ref{fig:z2pancake}(a).
We assume that all pancakes in 
Fig.~\ref{fig:layeredz2pancake}(a) are identical.
The single-particle spectrum corresponding to any pancake
in Fig.~\ref{fig:layeredz2pancake}(a)
is shown in Fig.~\ref{fig:z2pancake}(b).
It consists of two continua corresponding to bulk single-particle eigenstates 
separated by the band gap $\Delta$ and of a pair of Kramers' degenerate helical edge states crossing the bulk gap. We will always assume that 
$\Delta$ is much larger than the amplitude 
$\max\left\{\left|t^{\prime}_{-}\right|,\left|t^{\prime}_{+}\right|\right\}$ 
for edge states on adjacent pancakes to hop between layers.
Here, $t^{\prime}_{+}$ is depicted by
a solid line in Fig.~\ref{fig:layeredz2pancake}, while 
$t^{\prime}_{-}$ is drawn as a dotted
line in Fig.~\ref{fig:layeredz2pancake}. 
The characteristic energy
\begin{subequations}
\begin{equation}
\delta^{\prime}:=
+
\sqrt{
\frac{
\left|t^{\prime2}_{+}-t^{\prime2}_{-}\right|
     }
     {
2
     }
     }
\end{equation}
quantifies the amount by which translation symmetry 
by one stacking layer is broken, 
i.e., the amount of dimerization about
the average hopping amplitude
\begin{equation}
t^{\prime}:=
+
\sqrt{
\frac{
t^{\prime2}_{+}+t^{\prime2}_{-}
     }
     {
2
     }
     }.
\end{equation}
The hierarchy of energy scales
\begin{equation}
\Delta\gg
t^{\prime}\geq
\delta^{\prime}
\label{eq: hierarchy of energy scale for layered model 3D WTI}
\end{equation}
\end{subequations}
will be assumed. 

Assumption~(\ref{eq: hierarchy of energy scale for layered model 3D WTI})
justifies ignoring the degrees of freedom from the bulk altogether and 
keeping only the degrees of freedom living on the edges of
Fig.~\ref{fig:layeredz2pancake}
for energies below $\Delta$.
In Appendix%
~\ref{appsec: Wire construction for the surface states of a 3D Z2 WTI}, 
we present a quasi-one-dimensional Hamiltonian
that governs the dynamics of the surface states of 
Fig.~\ref{fig:layeredz2pancake}
when
\begin{equation}
\Delta\gg
t^{\prime}\gg
\delta^{\prime}.
\label{eq: hierarchy of energy scale for layered model 3D WTI bis}
\end{equation}
This approach is inspired from similar constructions 
when time-reversal symmetry is broken
(see Refs.~\onlinecite{Yakovenko91,Lee94,Kane02,Teo11}).
We show in Appendix%
~\ref{appsec: Wire construction for the surface states of a 3D Z2 WTI}, 
that there exists a continuum limit along the stacking axis that reduces
this effective Hamiltonian to the Dirac Hamiltonian
studied numerically by Mong, Bardarson, and Moore in Ref.~\onlinecite{Mong12}.

\subsection{
Two-dimensional spin-directed $\mathbb{Z}^{\,}_{2}$ network model
for the surface states of 
a weak three-dimensional $\mathbb{Z}^{\,}_{2}$ topological insulator
           }
\label{sec: Two-dimensional directed Z2 network model with dimerization}

\begin{figure}[t]
\centering
\includegraphics[width=0.23\textwidth]{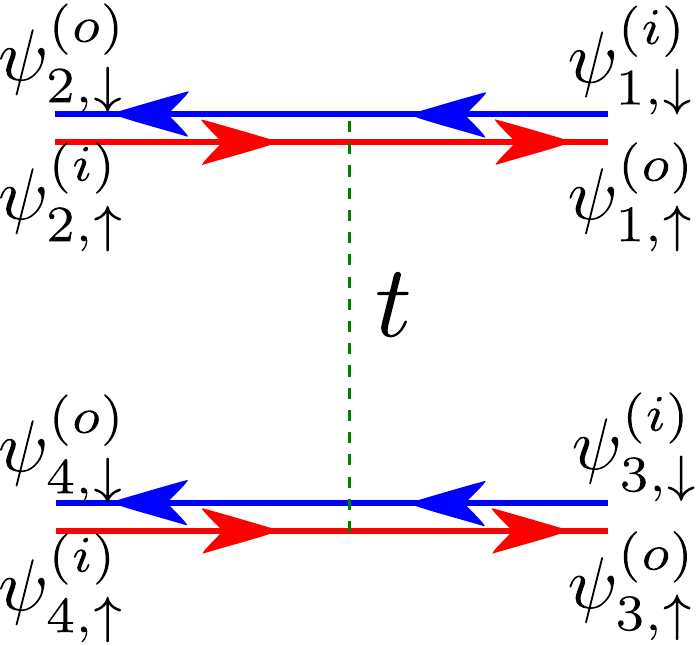}
\caption{
(Color online:)
An elementary scattering event in the
spin-directed $\mathbb{Z}^{\,}_{2}$ network model maps
four incoming plane waves into four outgoing plane waves
through a unitary $4\times4$ matrix $S$ compatible 
with the operation of time reversal.
By demanding that the operation of time reversal
is an antiunitary operation that squares to minus
the $4\times4$ identity matrix, the scattering matrix
$S$ can break spin-rotation symmetry 
(hence the spin labels on the plane waves),
but preserves time-reversal symmetry.
In other words, the scattering matrix belongs
to the symplectic class of scattering matrices.
The matrix $S$ is represented by 
Eq.~(\ref{eq:S-matrix-dimer}).
        }
\label{fig:network-dimer}
\end{figure}

Alternatively, we may encode the dynamics of the surface states of 
Fig.~\ref{fig:layeredz2pancake},
under the assumption that the hierarchy of energy scales%
~(\ref{eq: hierarchy of energy scale for layered model 3D WTI})
holds, in terms of the unitary scattering matrix of a 
two-dimensional spin-directed $\mathbb{Z}^{\,}_{2}$ network model.

The elementary building block of the 
two-dimensional spin-directed $\mathbb{Z}^{\,}_{2}$ network model
is a $4\times 4$ unitary  matrix $S$
that scatters four incoming plane waves into four outgoing plane waves.
The conventions on the labels of the scattering states
that we choose to represent $S$
are defined in Fig.~\ref{fig:network-dimer}.
If we demand that this $4\times4$ unitary matrix $S$
preserves time-reversal symmetry, whereby time reversal 
is represented by an antiunitary map on scattering states 
that squares to minus the $4\times4$ unit matrix, 
we obtain the representation
\begin{subequations}
\label{eq:S-matrix-dimer}
\begin{equation}
\begin{pmatrix}
\psi^{(o)} _{1,\uparrow}
\\
\psi^{(o)} _{2,\downarrow}
\\
\psi^{(o)} _{3,\uparrow}
\\
\psi^{(o)} _{4,\downarrow}
\end{pmatrix}
=
S\,
\begin{pmatrix}
\psi^{(i)} _{2,\uparrow}
\\
\psi^{(i)} _{1,\downarrow}
\\
\psi^{(i)} _{4,\uparrow}
\\
\psi^{(i)} _{3,\downarrow}
\end{pmatrix},
\end{equation}
where the scattering matrix that maps four incoming into four outgoing 
plane waves has the $2\times2$ block structure
\begin{equation}
S:=
e^{+{i}\phi^{\,}_{0}}
\begin{pmatrix}
r\, 
e^{+{i}\phi^{\,}_{3}}\,\sigma^{\,}_{0} 
& 
t\, 
Q 
\\
-
t\, 
Q^{\dag} 
& 
r\, 
e^{-{i}\phi^{\,}_{3}}\,
\sigma^{\,}_{0}
\end{pmatrix}.
\end{equation}
The real numbers $r$ and $t$ obey the condition
\begin{equation}
1=r^{2}+t^{2}
\Longleftrightarrow
r=\tanh x,
\quad
t=\frac{1}{\cosh x},
\qquad x\in\mathbb{R},
\end{equation}
and may be interpreted as the reflection and transmission
amplitudes, respectively, 
upon inspection of the transformation laws of the labels
1 and 2 on the one hand and 3 and 4 on the other hand from
Fig.~\ref{fig:network-dimer}.
The $2\times2$ matrix $Q$ is defined by
\begin{equation}
\begin{split}
Q=&\,
\begin{pmatrix}
e^{+{i}\phi^{\,}_{1}}\, 
\cos\theta
& 
e^{+{i}\phi^{\,}_{2}}\,
\sin\theta  
\\
e^{- {i}\phi^{\,}_{2}}\,
\sin\theta 
& 
-e^{-{i}\phi^{\,}_{1}}\,
\cos \theta
\end{pmatrix}
\\
=&\,
{i}
\sin\phi^{\,}_{1}\,
\cos\theta\,
\sigma^{\,}_{0}
+
\cos\phi^{\,}_{2}\,
\sin\theta\,
\sigma^{\,}_{1}
\\
&\,
-
\sin\phi^{\,}_{2}\,
\sin\theta\,
\sigma^{\,}_{2}
+
\cos\phi^{\,}_{1}\,
\cos\theta\,
\sigma^{\,}_{3}.
\end{split}
\end{equation}
It acts on the spin up and down labels of the incoming and outgoing
plane waves through the unit $2\times2$ matrix $\sigma^{\,}_{0}$
and the three Pauli matrices $\sigma^{\,}_{1}$, $\sigma^{\,}_{2}$,
and $\sigma^{\,}_{3}$.
The parameter $0\leq\theta\leq\pi/2$ quantifies the amount
of spin-rotation symmetry breaking. When $\theta=0$,
the $SU(2)$ spin-rotation symmetry is broken down to the
subgroup $U(1)$. Any $0<\theta\leq\pi/2$ breaks the
residual $U(1)$ symmetry by a spin-flip process.
The rate of this spin-flip tunneling is maximal for
$\theta=\pi/2$.
The remaining four phases
$\phi_j$ ($0\leq\phi^{\,}_{j}<2\pi$, $j=0,1,2,3$)
parametrize the phase arbitrariness 
of incoming and outgoing plane waves compatible with the condition 
\begin{equation}
\begin{pmatrix}
\sigma^{\,}_{2}
&
0
\\
0
&
\sigma^{\,}_{2}
\end{pmatrix}\,
S^{*}\,
\begin{pmatrix}
\sigma^{\,}_{2}
&
0
\\
0
&
\sigma^{\,}_{2}
\end{pmatrix}\,
=
S^{\dag}
\end{equation}
\end{subequations}
that implements time-reversal symmetry on the scattering matrix.
The sign of the amplitudes $t$ and $r$ can always be absorbed 
into the shifts $\phi^{\,}_{1,2}\to\phi^{\,}_{1,2}+\pi$ and
$\phi^{\,}_{3}\to\phi^{\,}_{3}+\pi$, respectively. Hence,
we may assume without loss of generality that $t$ and $r$ 
are positive numbers.

\begin{figure}[t]
\centering
\includegraphics[width=0.3\textwidth]{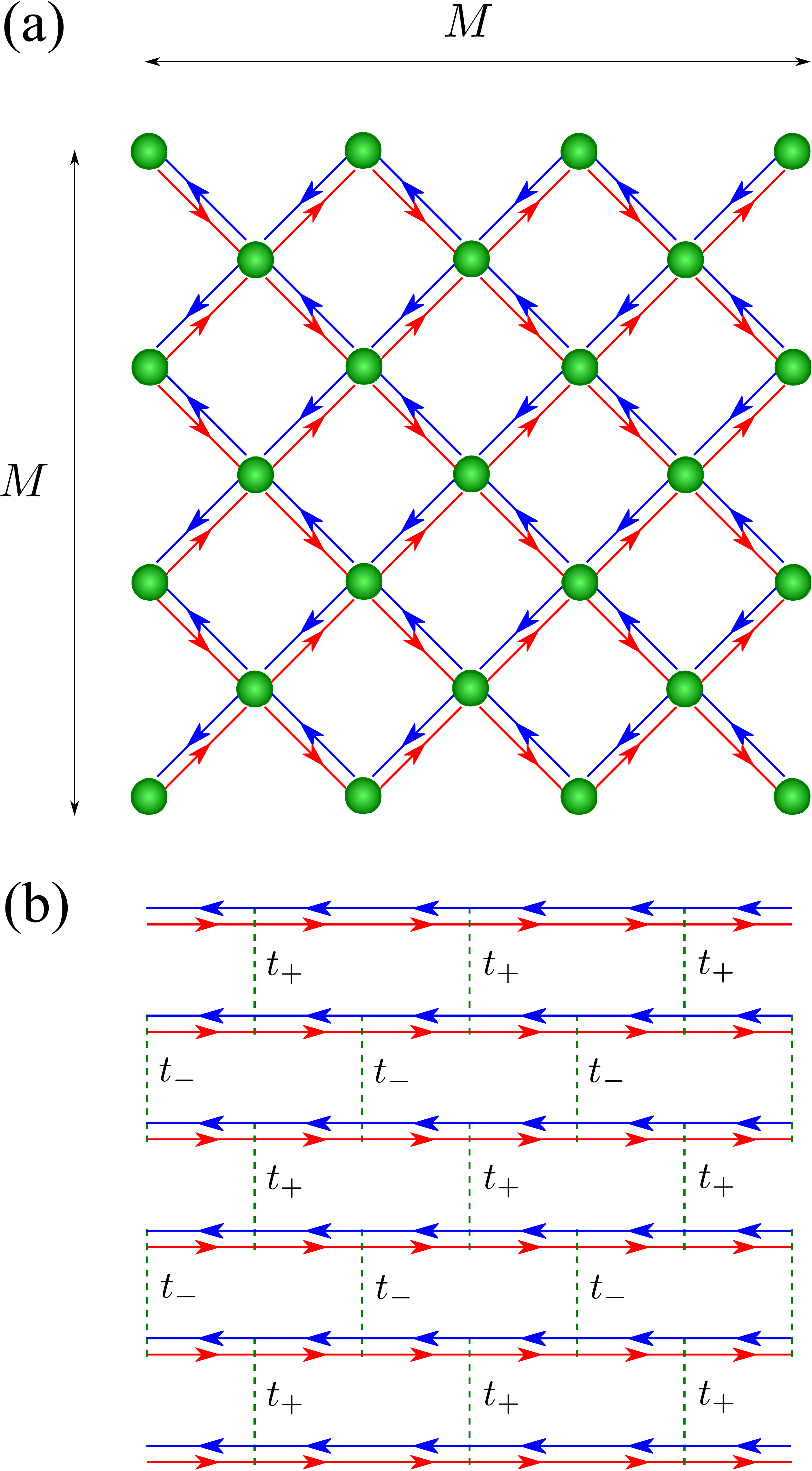}
\caption{
(Color online:)
A two-dimensional spin-directed $\mathbb{Z}^{\,}_{2}$ network model 
built out of the elementary scattering processes 
shown in Fig.~\ref{fig:network-dimer}
in the vertex (a) and the brick-wall representations (b),
respectively. The red and blue lines represent the flow of electrons 
with up and down spins, respectively.
The dimensions of both networks in (a) and (b) are
$M\times M$ with $M=6$. 
In both representations, 
one $M$ counts the number of pairs of up and down spins that move
from left to right or right to left, 
while the other $M$ counts the number of pairs of up and down spins
that move from up to down or down to up.
In the brick-wall representation, each dotted line represents 
an elementary scattering shown in Fig.~\ref{fig:network-dimer}
with the transmission amplitude 
$t^{\,}_{+}$ or $t^{\,}_{-}$.
This periodic pattern implements one of two dimerization patterns,
the other one following from interchanging $t^{\,}_{+}$ and $t^{\,}_{-}$.
        }
\label{fig:network directed Z2}
\end{figure}

A two-dimensional spin-directed $\mathbb{Z}^{\,}_{2}$ network model
is defined by arranging a collection of elementary scattering 
events, with possibly distinct values for the parameters 
$t$, $\theta$, $\phi^{\,}_{j}$ ($j=0,1,2,3$),
as is shown in Fig.~\ref{fig:network directed Z2}.
These network models are spin-directed because, if all transmission
amplitudes are chosen to vanish, there is no flipping of
the spin quantum numbers so that there are $M$ independent
pairs of Kramers' degenerate helical edge states propagating
unimpeded along $M$ one-dimensional channels.

\begin{figure}[t]
\centering
\includegraphics[width=0.3\textwidth]{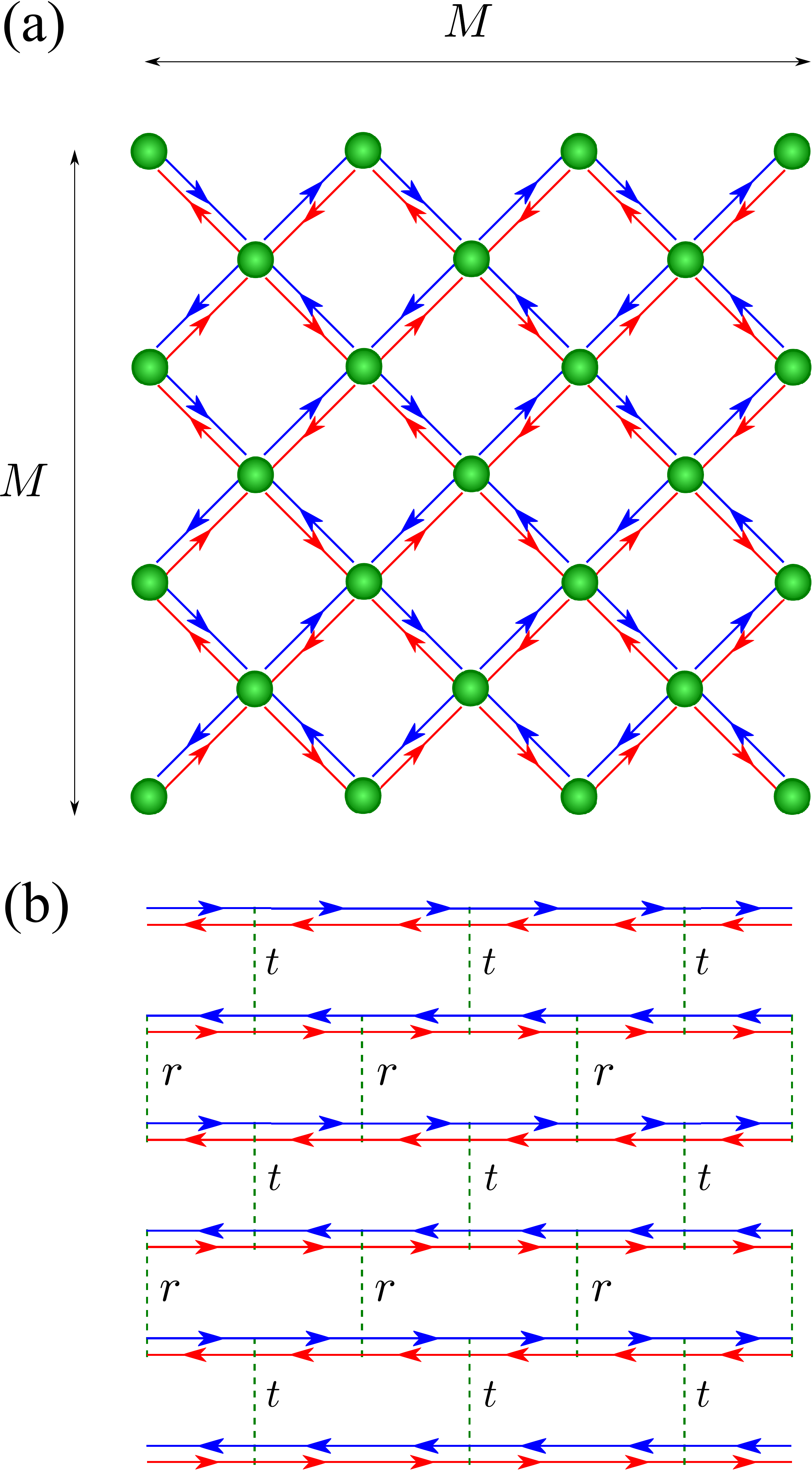}
\caption{
(Color online:)
The two-dimensional $\mathbb{Z}^{\,}_{2}$ network model studied in 
Refs.~\onlinecite{Obuse07}, \onlinecite{Obuse08}, and \onlinecite{Ryu10}
for a strong two-dimensional $\mathbb{Z}^{\,}_{2}$ topological insulator
with $M=6$ in the vertex (a) and the brick-wall representations (b), 
respectively. Note that the flow direction of the same spin component
(up or down) 
is opposite on adjacent layers in (b), in contrast to 
Fig.~\ref{fig:network directed Z2}(b).
        }
\label{fig:network undirected Z2}
\end{figure}

The two-dimensional spin-directed $\mathbb{Z}^{\,}_{2}$ network model
is thus different from 
the two-dimensional (undirected) $\mathbb{Z}^{\,}_{2}$ network model
that realizes strong two-dimensional $\mathbb{Z}^{\,}_{2}$ topological
insulators studied in Refs.~\onlinecite{Obuse07,Obuse08,Ryu10},
as can be verified by comparing
Fig.~\ref{fig:network directed Z2}
to Fig.~\ref{fig:network undirected Z2}.
In the two-dimensional (undirected) $\mathbb{Z}^{\,}_{2}$ network model,
the nodes labeled by $r$ in   
Fig.~\ref{fig:network undirected Z2}
are obtained from those labeled by $t$ through a 90 degree rotation. 
Hence, the two-dimensional (undirected) $\mathbb{Z}^{\,}_{2}$ network model
is invariant (on average) under 90 degree rotation,
whereas there is no discrete rotation symmetry for the 
two-dimensional spin-directed $\mathbb{Z}^{\,}_{2}$ network model.

Disorder is introduced in any 
two-dimensional spin-directed $\mathbb{Z}^{\,}_{2}$ network model 
by choosing the four phases
$\phi^{\,}_{j}$ ($j=0,1,2,3$)
of any elementary scattering process making up a network
model to be random numbers independently and identically 
distributed with a uniform distribution on the interval 
\begin{equation}
\left[
-\frac{\delta\phi}{2},
+\frac{\delta\phi}{2}
\right].
\end{equation}

In this paper, we study the combined effects of 
dimerization and disorder on 
the two-dimensional spin-directed $\mathbb{Z}^{\,}_{2}$ network model.
To incorporate dimerization, we assume that the squared amplitude
of the transmission (reflection) amplitude alternates 
in a periodic fashion between the two values labeled by $\pm$ of
\begin{subequations}
\label{eq: t pm}
\begin{equation}
0\leq
t^{2}_{\pm}
\leq1
\qquad
\left(r^{2}_{\pm}=
1-t^{2}_{\pm}\right),
\label{eq: t pm a}
\end{equation}
where
\begin{equation}
t^{2}_{\pm}:=
\begin{cases}
t^{2}\pm\delta^{2},
&
\hbox{if $t^{2}_{+}>t^{2}_{-}$},
\\&\\
t^{2}\mp\delta^{2},
&
\hbox{if $t^{2}_{+}<t^{2}_{-}$}.
\end{cases}
\label{eq: t pm b}
\end{equation}
\end{subequations}
Equation~(\ref{eq: t pm b}) emphasizes that 
the choice of which of the squared transmission amplitudes
$t^{2}_{+}$ and $t^{2}_{-}$ is the largest is arbitrary
for an alternating covering of the network.
When $\delta=0$, there is no dimerization. 
Parameter space $\Omega^{\,}_{4d}$
for the two-dimensional spin-directed $\mathbb{Z}^{\,}_{2}$ 
network model with disorder and dimerization is four-dimensional.
We choose the parametrization
\begin{subequations}
\label{eq: def parameter space is 4D}
\begin{equation}
\Omega^{\,}_{4d}=
\Omega^{+}_{4d}\cup\Omega^{-}_{4d} 
\label{eq: def parameter space is 4D a}
\end{equation}
with
\begin{equation}
\begin{split}
\Omega^{\pm}_{4d}:=
\Big\{&
\left(
t^{2},\theta,\delta\phi,\pm\delta^{2}\right)
\Big|
t^{2}\in[0,1],\
\theta\in[0,\pi/2],
\\
&
\delta\phi\in[0,2\pi[,\
\delta^{2}\in[0,1],\
0\leq t^{2}+\delta^{2}\leq1
\Big\}.
\end{split}
\label{eq: def parameter space is 4D b}
\end{equation}  
\end{subequations}
However, in most cases,
we will choose the disorder to be maximal in that
\begin{subequations}
\label{eq: def parameter space is 3D}
\begin{equation}
\delta\phi=2\pi,
\end{equation}
the exception being Sec.\ \ref{subsec:delta phi dependence}
where we study the dependence of the normalized localization length
on $\delta\phi$.
If so, parameter space is three-dimensional and given by
\begin{equation}
\Omega^{\,}_{3d}=
\Omega^{+}_{3d}\cup\Omega^{-}_{3d}
\label{eq: def parameter space is 3D a}
\end{equation}
with
\begin{equation}
\begin{split}
\Omega^{\pm}_{3d}:=
\Big\{&
\big(t^{2},\theta,\pm\delta^{2}\big)
\Big|
t^{2}\in[0,1],\
\theta\in[0,\pi/2],
\\
&\,
\delta^{2}\in[0,1],\qquad
0\leq t^{2}+\delta^{2}\leq1
\Big\}.
\end{split}
\label{eq: def parameter space is 3D b}
\end{equation}  
\end{subequations}
The interchange of $t^{2}_{+}$ and $t^{2}_{-}$ amounts 
to the interchange of
$\Omega^{+}_{\mathrm{x}d}$
and
$\Omega^{-}_{\mathrm{x}d}$
(here either $\mathrm{x}=4$ or $\mathrm{x}=3$).

We show in Appendix%
~\ref{appsec: Dirac Hamiltonian from the directed Z2 network model}
that there exists two continuum limits 
of the two-dimensional spin-directed $\mathbb{Z}^{\,}_{2}$ network model
in the vicinity, as measured by a small $\delta^{2}$,
of the lines $\theta=\pi/2$ with $t^{2}$ arbitrary
and $\theta=0$ with $t^{2}$ arbitrary, respectively. 
The line with $\theta=\pi/2$
delivers the Dirac Hamiltonian studied numerically by 
Mong, Bardarson, and Moore in Ref.~\onlinecite{Mong12}.

\subsection{Some limiting cases without dimerization}
\label{subsec: Some limiting cases without dimerization}

There are several lines in the two-dimensional subspace
\begin{equation}
\big(t^{2},\theta,\delta^{2}=0\big)\in
[0,1]
\times
[0,\pi/2]
\vphantom{\Bigg[}
\label{eq: def parameter space no dimerization}
\end{equation}  
of the parameter space~(\ref{eq: def parameter space is 3D})
without dimerization for which 
the two-dimensional spin-directed $\mathbb{Z}^{\,}_{2}$ network model 
simplifies.

\subsubsection{Limit $t^{2}=0$ without dimerization}
\label{subsubsec: Limit t2=0 without dimerization}

\begin{figure}[t]
\centering
\includegraphics[width=0.25\textwidth]{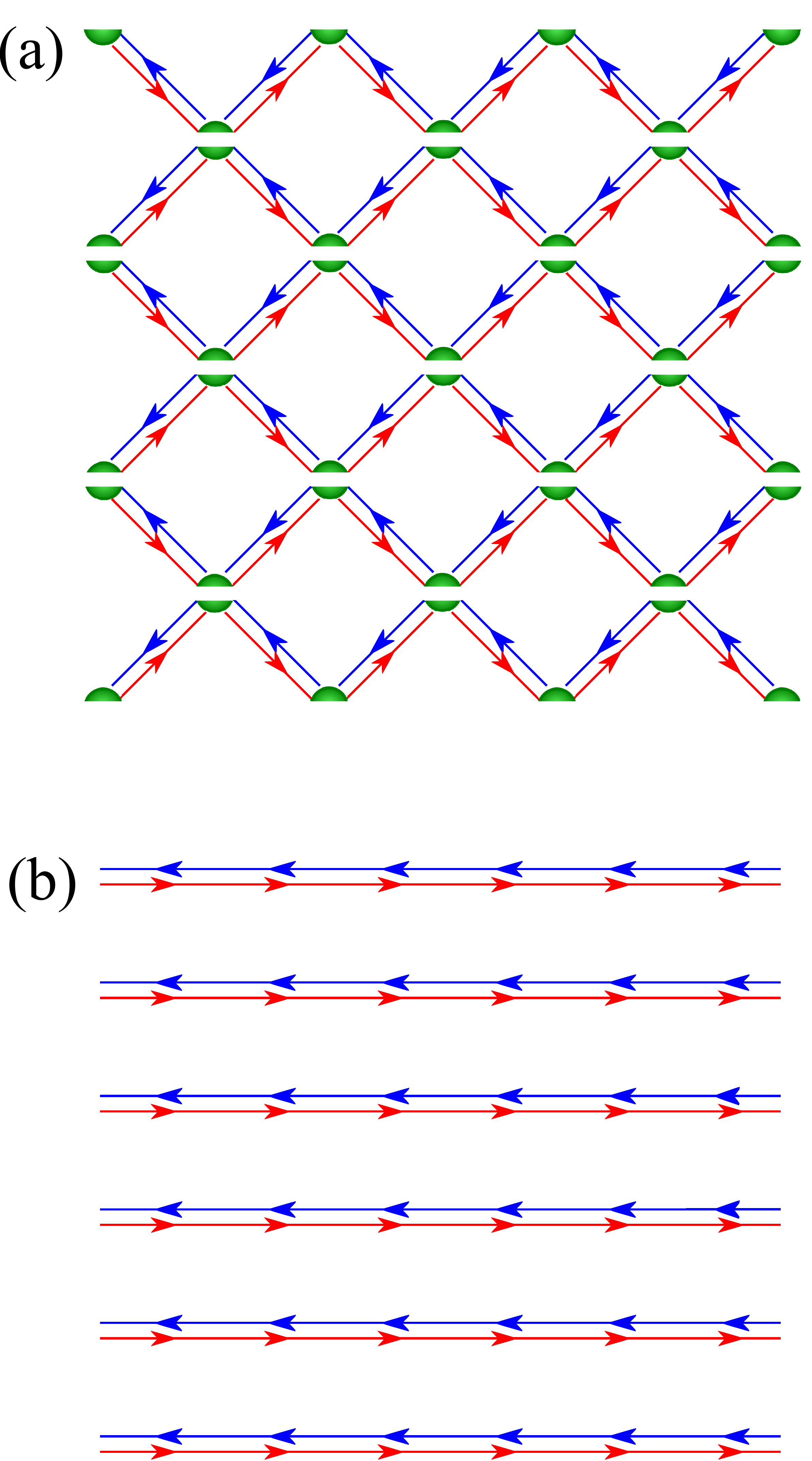}\\
\caption{
(Color online:)
When $\delta^{2}=t^{2}=0$, 
the two-dimensional spin-directed $\mathbb{Z}^{\,}_{2}$
network model from Fig.~\ref{fig:network directed Z2}
decouples into $M$ one-dimensional channels, each of which consists of
a single Kramers' degenerate pair of helical edge states that propagate
unimpeded by the disorder. Panels (a) and (b) are drawn using
the vertex and the brick-wall representations, respectively.
        }
\label{fig: network t=0}
\end{figure}

We set 
\begin{equation}
\delta^{2}=t^{2}=0.
\label{eq: delta=t=0}
\end{equation}
When $t^{2}=0$, the elementary scattering matrix from Eq.%
~(\ref{eq:S-matrix-dimer})
is block diagonal and independent of $\theta$,
\begin{equation}
S=
e^{+{i}\phi^{\,}_{0}}\,
\begin{pmatrix}
e^{+{i}\phi^{\,}_{3}}\,
\sigma^{\,}_{0} 
& 
0 
\\
0 
& 
e^{-{i}\phi^{\,}_{3}}\,
\sigma^{\,}_{0}
\end{pmatrix}.
\label{eq: S matrix t=0}
\end{equation}
The two-dimensional spin-directed $\mathbb{Z}^{\,}_{2}$ model of Fig.%
~\ref{fig:network directed Z2}
is shown in Fig.%
~\ref{fig: network t=0}
when $\delta^{2}=t^{2}=0$.
It represents $M$ decoupled one-dimensional channels, each of which consists
of a Kramers' degenerate pair of helical edge states that propagates
unimpeded by the disorder. The localization length is
infinite in the direction of propagation of these 
helical edge states, while it is vanishing in the orthogonal
direction. The fixed point~(\ref{eq: delta=t=0}) is thus
the \textit{quasi-one-dimensional symplectic metallic fixed point}.

\subsubsection{Limit $\theta=0$ without dimerization}
\label{subsubsec: Limit theta=0 without dimerization}

\begin{figure}[t]
\centering
\includegraphics[width=0.4\textwidth]{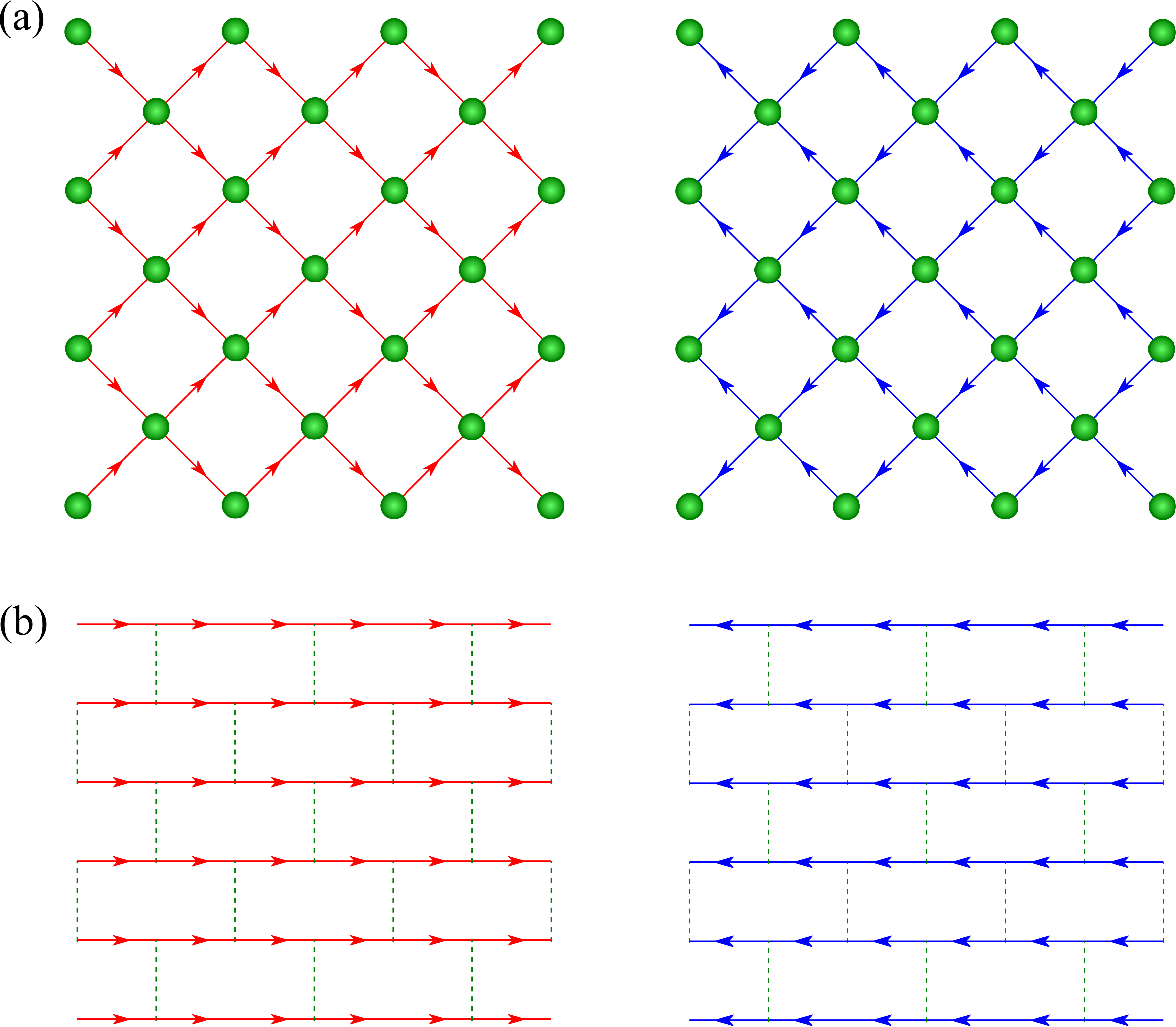}
\caption{
(Color online:)
When $\delta^{2}=\theta=0$, 
the two-dimensional spin-directed $\mathbb{Z}^{\,}_{2}$ network model 
from Fig.~\ref{fig:network directed Z2}
decouples into two directed CC models that are related by time reversal.
Panels (a) and (b) are drawn using
the vertex and the brick-wall representations, respectively.
        }
\label{fig:network theta=0}
\end{figure}

We set 
\begin{equation}
\delta^{2}=\theta=0. 
\label{eq: delta=theta=0}
\end{equation}
Since $\theta$ measures in dimensionless units the characteristic
strength of spin-orbit interactions with the convention that
the Rashba-like spin-orbit coupling vanishes when $\theta=0$, 
the spin-directed $\mathbb{Z}^{\,}_{2}$ network model
with $\delta^{2}=\theta=0$ is expected to decouple 
into two two-dimensional directed Chalker-Coddington (CC) models,%
~\cite{Chalker95} 
one for the spin-up and one for the
spin down plane waves, that are related by time-reversal symmetry.
This decoupling is shown in Fig.~\ref{fig:network theta=0}.
In a two-dimensional directed CC network model,%
~\cite{Halperin87,Chalker95,Balents96,Kim96,Mathur97,Gruzberg97,Cho97,Balents97}
propagation is uni-directional along the horizontal direction
and bi-directional along the vertical direction of the 
two-dimensional network. 

To establish this decoupling at $\theta=0$, 
we start from
\begin{subequations}
\label{eq: S matrix theta=0}
\begin{equation}
\begin{pmatrix}
\psi^{(o)}_{1,\uparrow}\\
\psi^{(o)}_{2,\downarrow}\\
\psi^{(o)}_{3,\uparrow}\\
\psi^{(o)}_{4,\downarrow}
\end{pmatrix}
=
S\,
\begin{pmatrix}
\psi^{(i)}_{2,\uparrow} 
\\
\psi^{(i)}_{1,\downarrow}
\\
\psi^{(i)}_{4,\uparrow}
\\
\psi^{(i)}_{3,\downarrow}
\end{pmatrix},
\end{equation}
where $S$ takes the limiting form
\begin{equation}
S=
e^{+{i}\phi^{\,}_{0}}\,
\begin{pmatrix}
r\, 
e^{+{i}\phi^{\,}_{3}}\,
\sigma_{0} 
& 
+t\, 
Q
\\
-t\, 
Q^{\dag}
& 
+r\,
 
e^{-{i}\phi^{\,}_{3}}\, 
\sigma^{\,}_{0}
\end{pmatrix}
\end{equation}
with the $2\times2$ diagonal block
\begin{equation}
Q=
\begin{pmatrix}
+e^{+{i}\phi^{\,}_{1}} 
& 
0  
\\
0 
& 
-e^{-{i}\phi^{\,}_{1}}
\end{pmatrix}.
\end{equation}
\end{subequations}
It is then advantageous for our purpose to 
do a unitary transformation that renders explicit the reducibility
of the scattering matrix~(\ref{eq: S matrix theta=0})
in the spin degrees of freedom,
\begin{subequations}
\label{eq: tilde S matrix theta=0}
\begin{equation}
\begin{pmatrix}
\psi^{(o)}_{1,\uparrow} 
\\
\psi^{(o)}_{3,\uparrow}
\\
\psi^{(o)}_{2,\downarrow}
\\
\psi^{(o)}_{4,\downarrow}
\end{pmatrix}
=
\tilde{S}\,
\begin{pmatrix}
\psi^{(i)}_{2,\uparrow} 
\\
\psi^{(i)}_{4,\uparrow}
\\
\psi^{(i)}_{1,\downarrow}
\\
\psi_{3,\downarrow}^{(i)}
\end{pmatrix},
\end{equation}
where $\tilde{S}$ takes the limiting form
\begin{equation}
\begin{split}
\tilde{S}=
e^{+{i}\phi^{\,}_{0}}\,
\begin{pmatrix}
\tilde{Q}(t,\phi^{\,}_{1},\phi^{\,}_{3})
&
0
\\
0 
& 
\tilde{Q}^{T}(t,\phi^{\,}_{1},\phi^{\,}_{3})
\end{pmatrix}
\end{split}
\end{equation}
with the $2\times2$ block
\begin{equation}
\tilde{Q}(t,\phi^{\,}_{1},\phi^{\,}_{3})=
\begin{pmatrix}
+r\,
e^{+{i}\phi^{\,}_{3}} 
& 
+t\, 
e^{+{i}\phi^{\,}_{1}} 
\\
-t\, 
e^{-{i}\phi^{\,}_{1}} 
& 
+r\, 
e^{-{i}\phi^{\,}_{3}} 
\end{pmatrix}.
\end{equation}
\end{subequations}
The upper-left and lower-right $2\times2$ blocks 
$e^{+{i}\phi^{\,}_{0}}\,\tilde{Q}(t,\phi^{\,}_{1},\phi^{\,}_{3})$
and
$e^{+{i}\phi^{\,}_{0}}\,\tilde{Q}^{T}(t,\phi^{\,}_{1},\phi^{\,}_{3})$, 
respectively, are related by time reversal. 
Each block defines the elementary scattering matrix
of the two-dimensional directed CC network model.%
~\cite{Chalker95} 
It is shown in Ref.~\onlinecite{Chalker95}
(see also Refs.%
~\onlinecite{Halperin87}
and \onlinecite{Balents96,Kim96,Mathur97,Gruzberg97,Cho97,Balents97})
that the localization properties of the two-dimensional directed 
CC network model are the following. Transport is highly anisotropic,
for it is perfect along one direction and critical 
along the orthogonal direction of the two-dimensional network.%
~\cite{footnote critical directed CC}
The fixed point~(\ref{eq: delta=theta=0}) 
is thus the \textit{spin-chiral metallic fixed point}.

\subsubsection{Limit $r^{2}=0$ without dimerization}

We set 
\begin{equation}
\delta^{2}=r^{2}=0.
\label{eq: delta=r=0}
\end{equation}
When $r^{2}=0$, the elementary scattering matrix from Eq.%
~(\ref{eq:S-matrix-dimer})
becomes
\begin{subequations}
\label{eq: S matrix t=1}
\begin{equation}
\begin{pmatrix}
\psi^{(o)}_{1,\uparrow}
\\
\psi^{(o)}_{2,\downarrow}
\\
\psi^{(o)}_{3,\uparrow}
\\
\psi^{(o)}_{4,\downarrow}
\end{pmatrix}
=
S\,
\begin{pmatrix}
\psi^{(i)}_{2,\uparrow} 
\\
\psi^{(i)}_{1,\downarrow}
\\
\psi^{(i)}_{4,\uparrow}
\\
\psi^{(i)}_{3,\downarrow}
\end{pmatrix},
\end{equation}
where $S$ takes the limiting form
\begin{equation}
S=
e^{+{i}\phi^{\,}_{0}}\,
\begin{pmatrix}
0 
& 
+Q(\theta,\phi^{\,}_{1},\phi^{\,}_{2})
\\
-
Q^{\dag}(\theta,\phi^{\,}_{1},\phi^{\,}_{2})
& 
0
\end{pmatrix}
\end{equation}
with the $2\times2$ block
\begin{equation}
Q(\theta,\phi^{\,}_{1},\phi^{\,}_{2})=
\begin{pmatrix}
+e^{+{i}\phi^{\,}_{1}}\, 
\cos\theta
& 
+
e^{+{i}\phi^{\,}_{2}}\,
\sin\theta  
\\
+e^{-{i}\phi^{\,}_{2}}\,
\sin\theta 
& 
-e^{-{i}\phi^{\,}_{1}}\,
\cos\theta 
\end{pmatrix}.
\end{equation}
\end{subequations}
We do the unitary transformation
\begin{subequations}
\label{eq: tilde S-matrix t=1}
\begin{equation}
\begin{pmatrix}
\psi^{(o)}_{1,\uparrow} 
\\
\psi^{(o)}_{2,\downarrow}
\\
\psi^{(o)}_{3,\uparrow}
\\
\psi^{(o)}_{4,\downarrow}
\end{pmatrix}
=
\tilde{S}\,
\begin{pmatrix}
\psi^{(i)}_{3,\downarrow}
\\
\psi^{(i)}_{4,\uparrow}
\\
\psi^{(i)}_{1,\downarrow}
\\
\psi^{(i)}_{2,\uparrow}
\end{pmatrix},
\label{eq: tilde S-matrix t=1 a}
\end{equation}
where $\tilde{S}$ takes the limiting form
\begin{equation}
\tilde{S}=
e^{+{i}\phi^{\,}_{0}}\,
\begin{pmatrix}
\tilde{Q}(\cos\theta,\phi^{\,}_{1},\phi^{\,}_{2})
&
0
\\
0
&
\tilde{Q}^{T}(\cos\theta,\phi^{\,}_{1},\phi^{\,}_{2}+\pi)
\end{pmatrix}
\label{eq: tilde S-matrix t=1 b}
\end{equation}
with the $2\times2$ block
\begin{equation}
\tilde{Q}(\cos\theta,\phi^{\,}_{1},\phi^{\,}_{2})=
\begin{pmatrix}
+e^{+{i}\phi^{\,}_{2}}\,
\sin\theta 
& 
+e^{+{i}\phi^{\,}_{1}}\,
\cos\theta
\\
-e^{-{i}\phi^{\,}_{1}}\, 
\cos\theta 
& 
+e^{-{i}\phi^{\,}_{2}}\, 
\sin\theta
\end{pmatrix}.
\label{eq: tilde S-matrix t=1 c}
\end{equation}
\end{subequations}
The upper-left and lower-right $2\times2$ blocks 
$e^{+{i}\phi^{\,}_{0}}\,
\tilde{Q}(\cos\theta,\phi^{\,}_{1},\phi^{\,}_{2})$
and
$e^{+{i}\phi^{\,}_{0}}\,
\tilde{Q}^{T}(\cos\theta,\phi^{\,}_{1},\phi^{\,}_{2}+\pi)$, 
respectively, are related by time reversal.

\begin{figure}[t]
\centering
\includegraphics[width=0.4\textwidth]{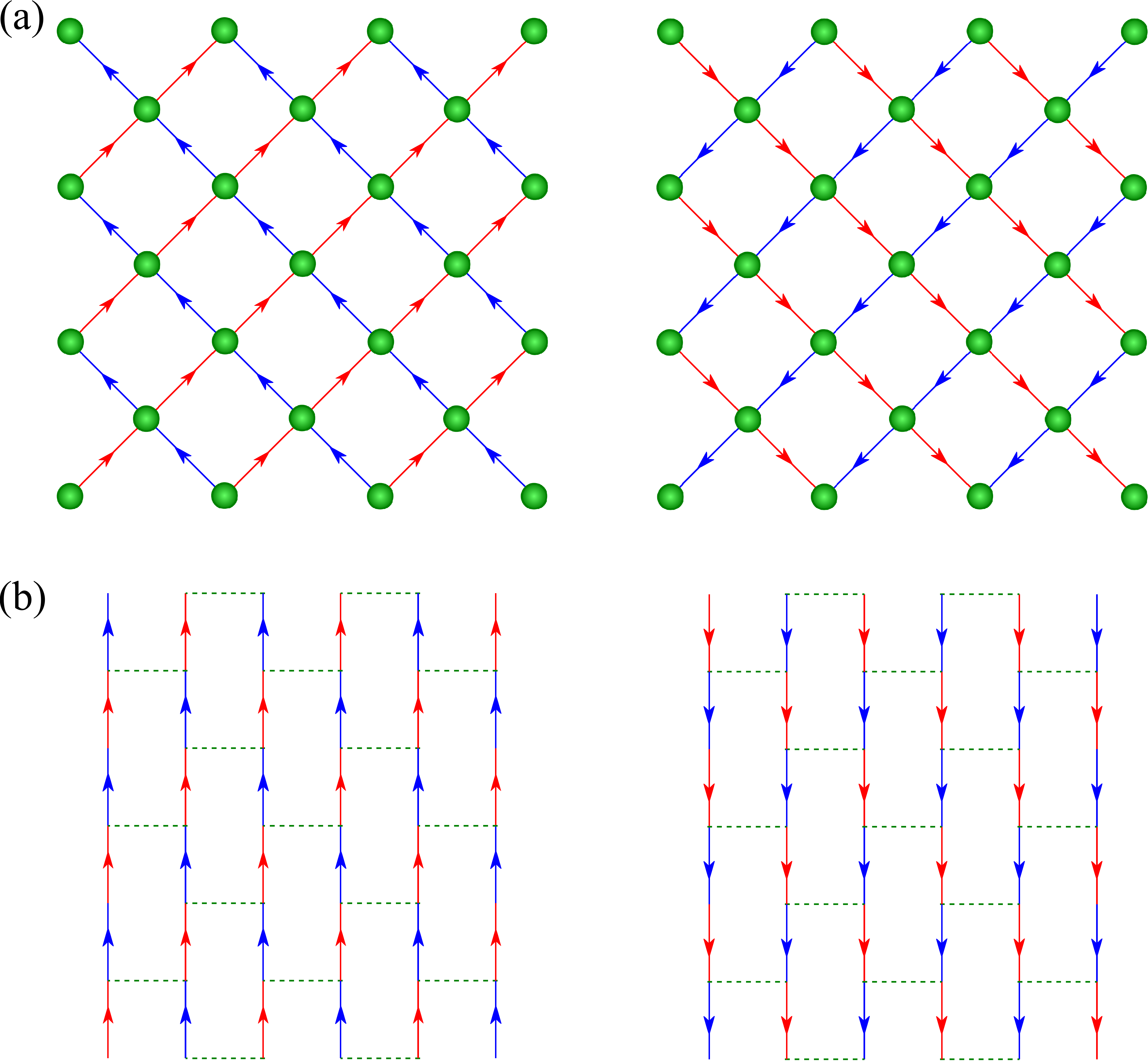}
\caption{
(Color online:)
When $\delta^{2}=r^{2}=0$, 
the two-dimensional spin-directed $\mathbb{Z}^{\,}_{2}$ network model 
from Fig.~\ref{fig:network directed Z2}
decouples into two directed CC models that are related by time reversal.
Panels (a) and (b) are drawn using
the vertex and the brick-wall representations, respectively.
        }
\label{fig: network r=0}
\end{figure}

The elementary scattering matrix~(\ref{eq: tilde S-matrix t=1})
is identical to the elementary scattering matrix%
~(\ref{eq: tilde S matrix theta=0}) 
as implied by the identifications
\begin{equation}
\sin\theta\to r,
\qquad
\cos\theta\to t,
\qquad
\phi^{\,}_{2}\to\phi^{\,}_{3}.
\end{equation}
Consequently,
the two-dimensional directed $\mathbb{Z}^{\,}_{2}$ network model 
from Fig.~\ref{fig:network directed Z2}
decouples into two directed CC network models that are related by time reversal
when $\delta^{2}=r^{2}=0$.
This decoupling is shown in Fig.~\ref{fig: network r=0}.
Transport is highly anisotropic,
for it is perfect along one spanning vector and critical 
along the second spanning vector of the two-dimensional network.
The fixed point~(\ref{eq: delta=r=0}) 
is thus again the \textit{spin-chiral metallic fixed point}.
The relationship between the fixed points
(\ref{eq: delta=r=0}) 
and
(\ref{eq: delta=theta=0}) 
is that the directions for perfect and critical transport
have been interchanged.

\subsubsection{Limit $\theta=\pi/2$ without dimerization}
\label{subsubsec: Limit theta=pi/2 without dimerization}

\begin{figure}[t]
\centering
\includegraphics[width=0.4\textwidth]{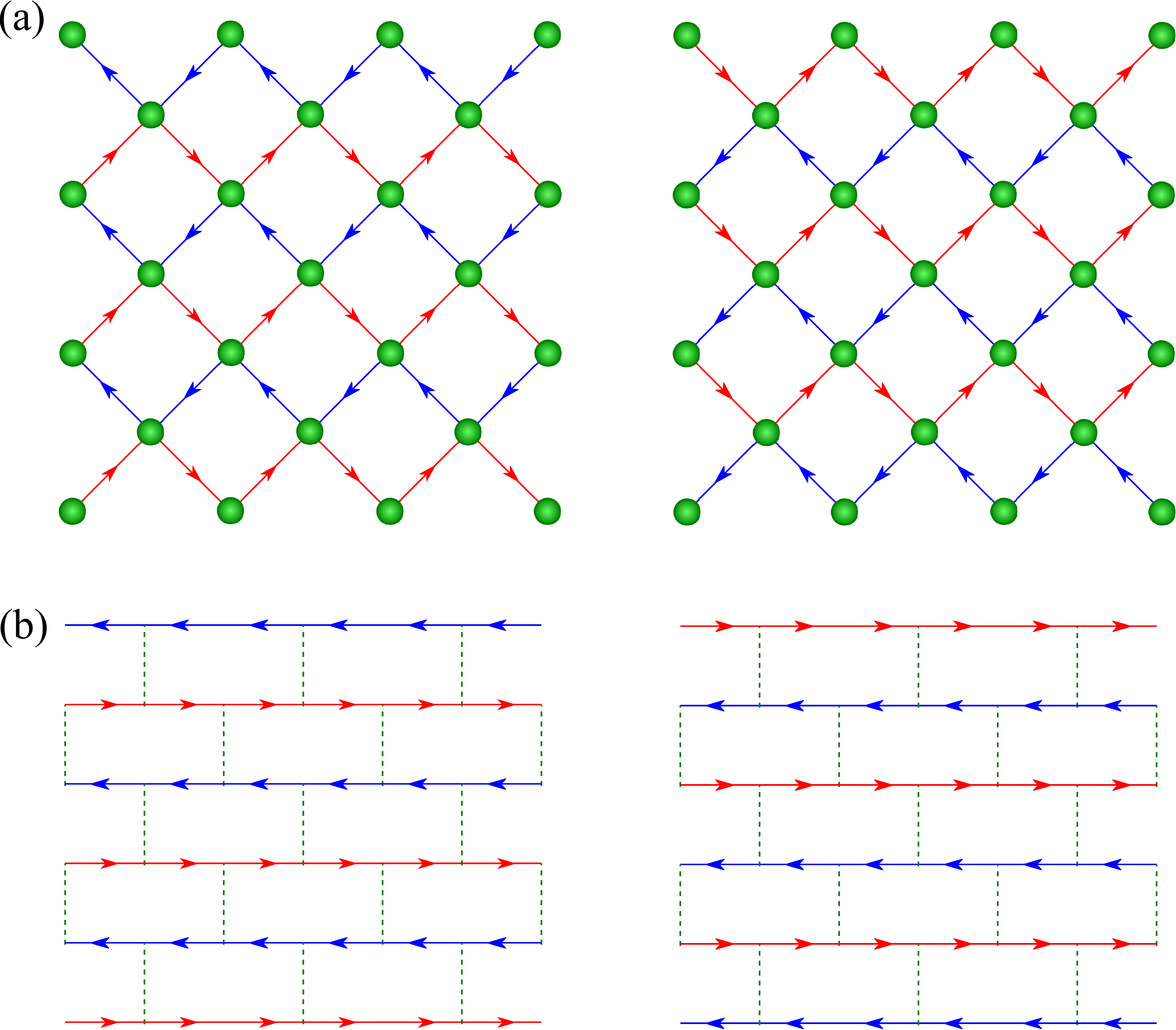}
\caption{
(Color online:)
When $\delta^{2}=\theta-\pi/2=0$, 
the two-dimensional spin-directed $\mathbb{Z}^{\,}_{2}$ network model 
from Fig.~\ref{fig:network directed Z2} decouples into
two directed CC network models that are related by time reversal.
Panels (a) and (b) are drawn using
the vertex and the brick-wall representations, respectively.
Since all elementary scattering $2\times2$ blocks in Eq.%
~(\ref{eq: tilde S matrix theta=pi/2})
are the same, criticality holds for any
$0\leq t^{2}\le1$. At the isotropic point $t^{2}=r^{2}=1/2$, the system
exhibits the criticality of the usual isotropic CC model, otherwise the
system exhibits the critical behavior of the anisotropic CC model.
        }
\label{fig: network theta=pi/2}
\end{figure}

We set 
\begin{equation}
\delta^{2}=\theta-\frac{\pi}{2}=0.
\label{eq: delta=theta-pi/2=0}
\end{equation}
When $\theta=\pi/2$, the elementary scattering matrix from Eq.%
~(\ref{eq:S-matrix-dimer})
becomes
\begin{subequations}
\label{eq: S matrix theta=pi/2}
\begin{equation}
\begin{pmatrix}
\psi^{(o)}_{1,\uparrow} 
\\
\psi^{(o)}_{2,\downarrow}
\\
\psi^{(o)}_{3,\uparrow}
\\
\psi^{(o)}_{4,\downarrow}
\end{pmatrix}
=
S\,
\begin{pmatrix}
\psi^{(i)}_{2,\uparrow} 
\\
\psi^{(i)}_{1,\downarrow}
\\
\psi^{(i)}_{4,\uparrow}
\\
\psi^{(i)}_{3,\downarrow}
\end{pmatrix},
\end{equation}
where $S$ takes the limiting form
\begin{equation}
S=
e^{+{i}\phi^{\,}_{0}}\,
\begin{pmatrix}
+r\, 
e^{+{i}\phi^{\,}_{3}}\,
\sigma^{\,}_{0} 
& 
+t\, 
Q
\\
-t\, 
Q^{\dag}
& 
+r\, 
e^{-{i}\phi^{\,}_{3}}\, 
\sigma^{\,}_{0}
\end{pmatrix}
\end{equation}
with the $2\times2$ block
\begin{equation}
Q=
\begin{pmatrix}
0 
& 
e^{+{i}\phi^{\,}_{2}}  
\\
e^{-{i}\phi^{\,}_{2}} 
& 
0
\end{pmatrix}.
\end{equation}
\end{subequations}
We do the unitary transformation
\begin{subequations}
\label{eq: tilde S matrix theta=pi/2}
\begin{equation}
\begin{pmatrix}
\psi^{(o)}_{1,\uparrow} 
\\
\psi^{(o)}_{4,\downarrow}
\\
\psi^{(o)}_{2,\downarrow}
\\
\psi^{(o)}_{3,\uparrow}
\end{pmatrix}
=
\tilde{S}\,
\begin{pmatrix}
\psi^{(i)}_{2,\uparrow} 
\\
\psi^{(i)}_{3,\downarrow}
\\
\psi^{(i)}_{1,\downarrow}
\\
\psi^{(i)}_{4,\uparrow}
\end{pmatrix},
\label{eq: tilde S matrix theta=pi/2 a}
\end{equation}
where $\tilde{S}$ takes the limiting form
\begin{equation}
\tilde{S}=
e^{+{i}\phi^{\,}_{0}}\,
\begin{pmatrix}
\tilde{Q}(t,\phi^{\,}_{2},\phi^{\,}_{3})
&
0
\\
0
&
\tilde{Q}^{T}(t,\phi^{\,}_{2}+\pi,\phi^{\,}_{3})
\end{pmatrix}
\label{eq: tilde S matrix theta=pi/2 b}
\end{equation}
with
\begin{equation}
\tilde{Q}(t,\phi^{\,}_{2},\phi^{\,}_{3})=
\begin{pmatrix}
+r\, 
e^{+{i}\phi^{\,}_{3}} 
& 
+t\, 
e^{+{i}\phi^{\,}_{2}}
\\
-t\, 
e^{-{i}\phi^{\,}_{2}}
&
+r\, 
e^{-{i}\phi^{\,}_{3}} 
\end{pmatrix}.
\label{eq: tilde S matrix theta=pi/2 c}
\end{equation}
\end{subequations}
The upper-left and lower-right $2\times2$ blocks 
$e^{+{i}\phi^{\,}_{0}}\,
\tilde{Q}(t,\phi^{\,}_{2},\phi^{\,}_{3})$
and
$e^{+{i}\phi^{\,}_{0}}\,
\tilde{Q}^{T}(t,\phi^{\,}_{2}+\pi,\phi^{\,}_{3})$, 
respectively, are related by time reversal. 
Moreover, they are nothing but the elementary 
scattering matrix for the two-dimensional CC
network model that describes the localization properties
of the lowest Landau level perturbed by disorder in the
integer quantum Hall effect (IQHE). Consequently,
the two-dimensional spin-directed $\mathbb{Z}^{\,}_{2}$ network model 
from Fig.~\ref{fig:network directed Z2}
decouples into two CC network models that are related by time reversal
when $\delta^{2}=\theta-\pi/2=0$.
This decoupling is shown in Fig.~\ref{fig: network theta=pi/2}.

Chalker and Coddington showed in Ref.~\onlinecite{Chalker88} that
the CC network model is critical if every node of the network
is described by the scattering matrix 
$e^{+{i}\phi^{\,}_{0}}\,
\tilde{Q}(t,\phi^{\,}_{2},\phi^{\,}_{3})$ 
given in Eq.%
~(\ref{eq: tilde S matrix theta=pi/2 c})
sharing the same tunneling amplitude $t$.
Hence, the two-dimensional spin-directed $\mathbb{Z}^{\,}_{2}$ network model 
with $\delta^{2}=\theta-\pi/2=0$ is always critical. In the special 
isotropic case when $t^{2}=r^{2}=1/2$, this critical point is called
the \textit{isotropic CC critical point}. In the generic anisotropic case when 
$t^{2}=1-r^{2}\neq 1/2$, this critical point is called
the \textit{anisotropic CC critical point}.

\subsubsection{Phase diagram without dimerization}

\begin{figure*}[t]
\centering
\includegraphics[width=\textwidth]{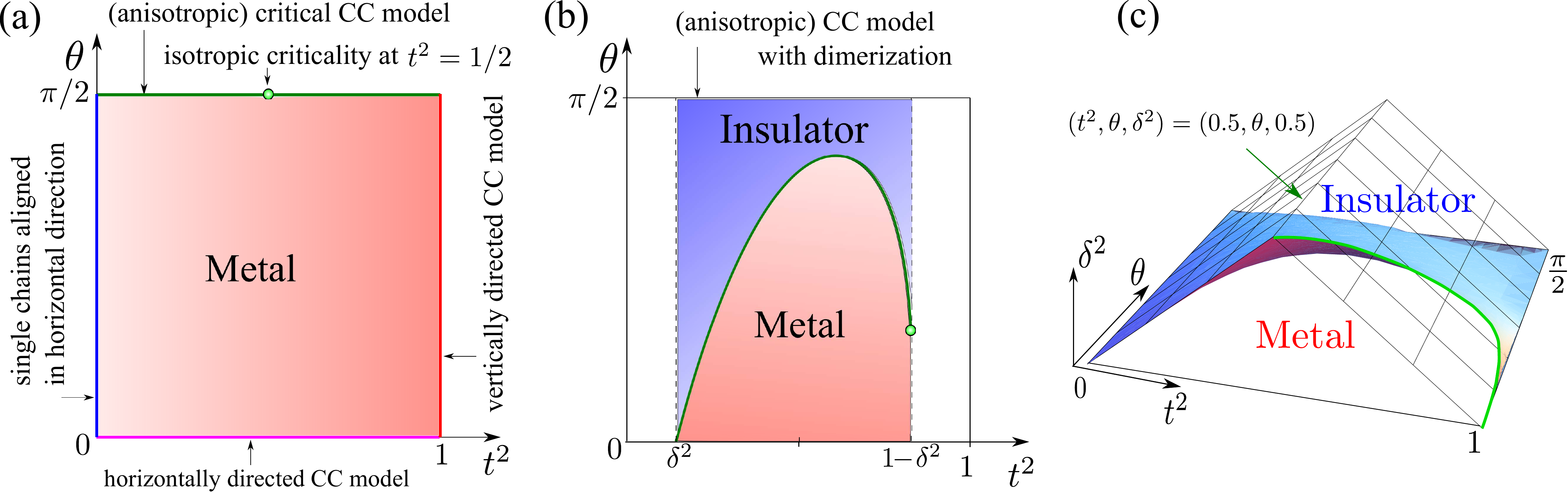}
\caption{
(Color online:)
(a) 
Phase diagram in the two-dimensional parameter space (a square)%
~(\ref{eq: def parameter space no dimerization})
without dimerization of 
the two-dimensional spin-directed $\mathbb{Z}^{\,}_{2}$ network model.
The boundaries of the square corresponds to 
a two-dimensional chiral metal when $\theta=0$,
a two-dimensional chiral metal when $t=1$,
the metal-insulator critical point of the (two-dimensional)
CC model when $\theta=\pi/2$, 
and the quasi-one-dimensional symplectic metallic
fixed point when $t^{2}=0$. Numerics of 
the two-dimensional spin-directed $\mathbb{Z}^{\,}_{2}$ network model 
are consistent with a metallic
phase in the interior of the square except in the region of $t^{2}\ll1$
for which numerics are inconclusive. However,
we show in Appendix A that
the numerical results from Ref.~\onlinecite{Mong12} 
apply to this region, thereby implying that this region is metallic.
(b) 
Two-dimensional cut of the
phase diagram in the three-dimensional parameter space%
~(\ref{eq: def parameter space is 3D})
with \textit{fixed and non-vanishing} dimerization of 
the two-dimensional spin-directed $\mathbb{Z}^{\,}_{2}$ network model. 
The green dot is on the boundary $1/2 \le t^2=1-\delta^2 \le 1$ of the
 three-dimensional parameter space%
~(\ref{eq: def parameter space is 3D}).
The presence of the dimerization
improves the quality of the numerics. Dimerization reduces the area 
of the metallic phase. The divergences of the localization length at
the metal-insulator transitions are consistent with a metal-insulator
critical point belonging to the two-dimensional symplectic class
of Anderson localization. 
(c) 
Qualitative phase diagram in the three-dimensional parameter space%
~(\ref{eq: def parameter space is 3D})
with \textit{fixed and non-vanishing} dimerization.
The phase boundary represented by the green curve is on the boundary $1/2 \le t^2=1-\delta^2 \le 1$ of the
 three-dimensional parameter space%
~(\ref{eq: def parameter space is 3D}).
The phase diagram that includes the effect of the sign of 
the dimerization in Eq.~(\ref{eq: def parameter space is 3D})
can be represented by gluing along the negative $\delta^{2}$
axis the mirror image about the plane $\delta^{2}=0$ 
of the phase diagram for positive $\delta^{2}$.
This delivers two insulating phases separated by a metallic phase.
        }
\label{fig: phase-diagram 3 parameter}
\end{figure*}

All four boundaries of parameter space%
~(\ref{eq: def parameter space no dimerization})
shown as the sides of the square in Fig.%
~\ref{fig: phase-diagram 3 parameter}(a)
evade localization. Three of these boundaries belong to the unitary class,
one to the symplectic class. Any deviation away from these boundaries
puts the two-dimensional spin-directed $\mathbb{Z}^{\,}_{2}$ network model
in the two-dimensional symplectic class of Anderson localization.
The symplectic class for disordered metals is the most robust to the effects 
of disorder in that it displays the phenomenon of weak antilocalization
for any dimensionality, i.e.,
the perturbative effect of disorder in the diffusive metallic regime
is to enhance the conductivity in the symplectic class.%
~\cite{Hikami80}
Hence, the most economical conjecture regarding the nature of the
interior of parameter space%
~(\ref{eq: def parameter space no dimerization})
with regard to the physics of localization is that it is metallic.
This scenario is confirmed by our numerical calculations 
that we present in Sec.~\ref{sec: Numerical data with an even number of channels}
except for the region where $t^{2}\ll1$.
Although the numerical results in this region
are inconclusive, we argue in favor of a metallic phase.
In fact, the effective Dirac Hamiltonian derived for $t^{2}\ll1$
in Appendix A is nothing but the Hamiltonian studied by Mong
\textit{et al.}~\cite{Mong12}
with additional anisotropy in velocities.
Their numerical results imply
that this region of question is metallic.
We can thus conclude that there is only a metallic phase
in the interior of the square in Fig.~\ref{fig: phase-diagram 3 parameter}(a).

\subsubsection{Phase diagram with dimerization}

We close Sec.~\ref{sec: Definitions}
with the summary of our numerical results, to be described in more
detail in Sec.~\ref{sec: Numerical data with an even number of channels}, 
in the form of the cuts of the schematic
three-dimensional phase diagram displayed
in Fig.~\ref{fig: phase-diagram 3 parameter}(c).

Any non-vanishing dimerization $\delta^{2}$
shrinks parameter space%
~(\ref{eq: def parameter space no dimerization})
through the condition
\begin{equation}
\delta^{2}\leq t^{2}\leq 1-\delta^{2}.
\end{equation}
According to Appendix 
\ref{appsec: Dirac Hamiltonian from the directed Z2 network model}, 
the continuum limit of 
the two-dimensional spin-directed ${\mathbb Z}^{\,}_{2}$ network model
in the vicinity of $\theta=0$ is that of a gapless Hamiltonian
for any allowed value of $t^{2}$ and $\delta^{2}$
in the three-dimensional parameter space%
~(\ref{eq: def parameter space is 3D}). 
Hence, we anticipate
a metallic phase when $\theta=0$ 
in the three-dimensional parameter space~(\ref{eq: def parameter space is 3D}).
According to Appendix 
\ref{appsec: Dirac Hamiltonian from the directed Z2 network model}, 
the continuum limit of 
the two-dimensional spin-directed ${\mathbb Z}^{\,}_{2}$ network model
in the vicinity of $\theta=\pi/2$ is that of a massive Dirac equation
for any allowed value of $t^{2}$ and $\delta^{2}\neq0$
in the three-dimensional parameter space~(\ref{eq: def parameter space is 3D}).
Hence, we anticipate an insulating phase when $\theta=\pi/2$ and
$\delta^{2}\ne0$
in the three-dimensional parameter space~(\ref{eq: def parameter space is 3D}).
Incidentally, a numerical study of two-component Dirac fermions
with random mass and random potentials from Ref.%
~\onlinecite{Nomura08}
supports this conclusion.

On the boundary
$0\leq t^{2}=\delta^{2}\leq 1/2$
of the three-dimensional parameter space~(\ref{eq: def parameter space is 3D}),
Eq.~(\ref{eq: t pm})
implies that the two values taken by the dimerized hopping
are $t^{2}_{+}=2t^{2}$ and $t^{2}_{-}=0$, respectively.
Correspondingly (recall Fig.~\ref{fig:layeredz2pancake}),
the network model effectively describes the one-dimensional propagation of 
two Kramers' degenerate pairs of helical edge states on this boundary.
Hence, this boundary, if supplemented by the condition $\theta=0$,
is the end line of a critical surface at which a 
metal-insulator transition takes place 
in the three-dimensional parameter space~(\ref{eq: def parameter space is 3D}).
On the insulating side of this quantum phase transition, dimerization is strong
relative to the disorder strength in that the
density of states is so low in the presence of disorder that localization rules. 
On the metallic side of this quantum phase transition, dimerization is weak
relative to the disorder strength in that the metallic fixed point
in the two-dimensional symplectic symmetry class of Anderson localization
is realized.

On the boundary $1/2\leq t^{2}=1-\delta^{2}\leq1$ 
of the three-dimensional parameter 
space~(\ref{eq: def parameter space is 3D}),
Eq.~(\ref{eq: t pm})
implies that the two values taken by the dimerized hopping
are $t^{2}_{+}=1$ and $t^{2}_{-}=2t^{2}-1$, respectively.
Correspondingly (recall Fig.~\ref{fig:layeredz2pancake}),
the network model effectively describes the one-dimensional propagation of 
two Kramers' degenerate pairs of helical edge states only at the point
$t^{2}=\delta^{2}=1/2$ on this boundary.
Hence, the localization properties of surface states on this boundary 
cannot be deduced by mere inspection.
We expect a metal-insulator transition 
driven by $\theta$, for the surface states 
are extended on the boundary $\theta=0$,
while they are localized on the boundary $\theta=\pi/2$. 
Finally, upon
increasing $\delta^{2}$ towards $1/2$, 
the critical point $\theta^{\,}_{\mathrm{c}}(\delta^{2})$ 
at which the 
metal-insulator transition takes place should
decrease, since transport is effectively one-dimensional at $\delta^{2}=1/2$.
A qualitative numerical study of the localization properties
along the boundary $1/2\leq t^{2}=1-\delta^{2}\leq1$ 
can be found at the end of Sec.~\ref{subsec:theta dependence}. 

From these considerations, 
we conjecture the schematic three-dimensional
phase diagram shown in Fig.~\ref{fig: phase-diagram 3 parameter}(c).
We now present numerical support for the
phase diagram~\ref{fig: phase-diagram 3 parameter}(c).

\medskip

\begin{figure*}[t]
\centering
\includegraphics[width=0.9\textwidth]{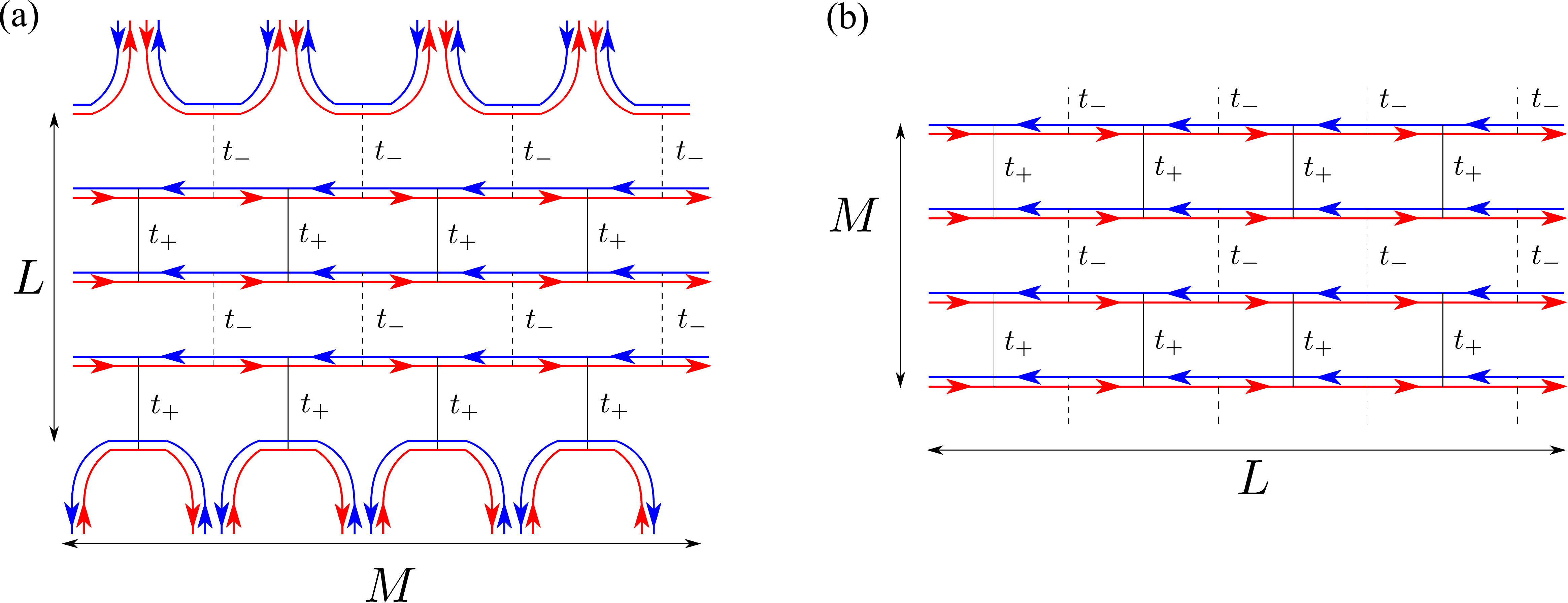}
\caption{
(Color online:)
Two examples in the brick-wall representation
of two-dimensional spin-directed $\mathbb{Z}^{\,}_{2}$ network models 
built out of the elementary scattering processes shown in 
Fig.~\ref{fig:network-dimer}.
The periodic choice of the transmission amplitudes
$t^{\,}_{+}$ and $t^{\,}_{-}$ implements a dimerization pattern.
The red and blue lines represent the flow of electron with up and down
spins, respectively. The continuous (dotted) black line represents 
the scattering with transmission amplitude $t^{\,}_{+}$ $(t^{\,}_{-})$ 
between them. Periodic boundary conditions are imposed along the
horizontal and vertical directions in panels (a) and (b), respectively.
The dimensions of these
two-dimensional spin-directed $\mathbb{Z}^{\,}_{2}$ network models 
are $M\times L=8\times4$ and $L\times M=8\times4$ for examples (a) and (b), 
respectively.
        }
\label{fig:network-staggered}
\end{figure*}

\begin{widetext}

\section{Numerical data for an even number of dimerized channels}
\label{sec: Numerical data with an even number of channels}

\subsection{Transfer matrix}
\label{subsec: Transfer matrix}

In numerical studies of network models, it is convenient to
do a similarity transformation on the space of scattering states.
Instead of defining
a two-dimensional spin-directed $\mathbb{Z}^{\,}_{2}$ network model 
by a unitary scattering matrix that maps $2M$ incoming
plane waves into $2M$ outgoing plane waves,
we can do a similarity transformation
on the scattering states under which a
unitary scattering matrix turns into 
a pseudo-unitary transfer matrix.
The transfer matrix maps
the plane waves at the bottom of the brick wall
to the plane waves at the top of the brick wall
if the boundary conditions are those of
Fig.~\ref{fig:network-staggered}(a).
If the  boundary conditions are those of
Fig.~\ref{fig:network-staggered}(b),
the transfer matrix maps 
the plane waves at the left of the brick wall
to the plane waves at the right of the brick wall.
In the former case, the elementary transfer matrix is defined by
\begin{equation}
\begin{pmatrix}
\psi^{(i)}_{2,\uparrow}\\
\psi^{(i)}_{1,\downarrow}\\
\psi^{(o)}_{1,\uparrow}\\
\psi^{(o)}_{2,\downarrow}
\end{pmatrix}
=
\mathcal{M}^{\,}_{\perp}\,
\begin{pmatrix}
\psi^{(i)}_{4,\uparrow}\\
\psi^{(i)}_{3,\downarrow}\\
\psi^{(o)}_{3,\uparrow}\\
\psi^{(o)}_{4,\downarrow}
\end{pmatrix}, 
\qquad
\mathcal{M}^{\,}_{\perp}:=
\begin{pmatrix}
\displaystyle
+\frac{r}{t}\, 
e^{-{i}\phi^{\,}_{3}}\, Q 
& 
\displaystyle
-\frac{1}{t}\,
e^{-{i}\phi^{\,}_{0}}\,Q
\\&\\
\displaystyle 
+\frac{1}{t}\,
e^{+{i}\phi^{\,}_{0}}\,Q 
&
\displaystyle 
-\frac{r}{t}\,e^{+{i}\phi^{\,}_{3}}\, Q
\end{pmatrix}.
\label{eq: T perp} 
\end{equation}
In the latter case,
the elementary transfer matrix is defined by
\begin{subequations}
\label{eq: T right 1 dimer}
\begin{equation}
\begin{pmatrix}
\psi^{(o)}_{1,\uparrow}\\
\psi^{(o)}_{3,\uparrow}\\
\psi^{(i)}_{1,\downarrow}\\
\psi^{(i)}_{3,\downarrow}
\end{pmatrix}=
\mathcal{M}^{\,}_{\parallel}\,
\begin{pmatrix}
\psi^{(i)}_{2,\uparrow}\\
\psi^{(i)}_{4,\uparrow}\\
\psi^{(o)}_{2,\downarrow}\\
\psi^{(o)}_{4,\downarrow}
\end{pmatrix}, 
\end{equation}
where
\begin{equation}
\mathcal{M}^{\,}_{\parallel}:=
\frac{1}{r^{2}+t^{2}\cos^{2}\theta}
\begin{pmatrix}
r 
e^{+{i}(\phi_{0}+\phi_{3}) } 
& 
t 
e^{+{i}(\phi_{0}+\phi_{1})} \cos \theta 
&
-
t^{2} 
e^{+{i} (\phi_{1}+\phi_{2})} 
\sin\theta 
\cos\theta 
&
r 
t 
e^{+{i}(\phi_{2}+\phi_{3})}
\sin\theta 
\\
-
t 
e^{+{i}(\phi_{0}-\phi_{1})} 
\cos\theta 
& 
r 
e^{+{i}(\phi_{0}-\phi_{3})} 
&
-
r 
t 
e^{+{i}(\phi_{2}-\phi_{3})} 
\sin\theta 
&
-
t^{2} 
e^{-{i}(\phi_{1}-\phi_{2})}
\sin\theta 
\cos\theta 
\\
t^{2} 
e^{-{i}(\phi_{1}+\phi_{2})} 
\sin\theta 
\cos\theta 
&
-
r 
t 
e^{-{i}(\phi_{2}+\phi_{3})} 
\sin\theta 
&
r 
e^{-{i}(\phi_{0}+\phi_{3})} 
& 
t 
e^{-{i}(\phi_{0}+\phi_{1})} 
\cos\theta 
\\
r 
t 
e^{-{i}(\phi_{2}-\phi_{3})} 
\sin\theta 
& 
t^{2} 
e^{+{i}(\phi_{1}-\phi_{2})} 
\sin\theta 
\cos\theta 
&
-t 
e^{-{i}(\phi_{0}-\phi_{1})} 
\cos\theta 
&
r 
e^{-{i}(\phi_{0}-\phi_{3})}
\end{pmatrix}.
\label{eq: T parallel}
\end{equation}
\end{subequations}
We refer the reader to
Ref.~\onlinecite{Obuse07}
for the detailed construction of the total transfer matrix 
$\mathcal{M}^{\,}_{\mathrm{tot}}$
corresponding to Fig.~\ref{fig:network-staggered}
from the elementary $4\times4$ transfer matrix of
Eq.~(\ref{eq: T right 1 dimer}).

\end{widetext}

\subsection{Definition of the normalized localization length}
\label{subsec: Definition of the normalized localization length}

As explained in Ref.~\onlinecite{Obuse07},
the transfer matrix
$\mathcal{M}^{\,}_{\mathrm{tot}}$ 
is pseudo-unitary and belongs to the group $SO^{*}(2M)$
given the definition of $M$ in Fig.~\ref{fig:network-staggered}
($M$ is even).
The matrix
$\mathcal{M}^{\dag}_{\mathrm{tot}}\,\mathcal{M}^{\,}_{\mathrm{tot}}$
built up from $\mathcal{M}^{\,}_{\mathrm{x}}$ 
with $\mathrm{x}=\perp$ or $\parallel$ 
is positive definite and has the doubly degenerate eigenvalues
of the form 
$\exp(\pm2X^{\,}_{\mathrm{x},j})$
with $j=1,\cdots,M/2$,
which can be ordered according to the convention
\begin{equation}
0<X^{\,}_{\mathrm{x},1}<\cdots<X^{\,}_{\mathrm{x},M/2}.
\end{equation}
The ordered numbers $X^{\,}_{\mathrm{x},j}$ with $j=1,\cdots,M/2$
are called Lyapunov exponents. 
They become selfaveraging
random variables as the quasi-one-dimensional limit
$L\to\infty$ for fixed $M$ is taken in 
Fig.~\ref{fig:network-staggered}.%
~\cite{Johnston83}
The quasi-one-dimensional localization length
is defined by
\begin{equation}
\xi^{\,}_{\mathrm{x},M}:=
\lim_{L\to\infty}
\frac{L}{X^{\,}_{\mathrm{x},1}}.
\label{eq: def xi}
\end{equation}
As shown by MacKinnon and Kramer,%
~\cite{MacKinnon83}
criticality in two dimensions can be probed through the dependence of
the normalized localization length
\begin{equation}
\Lambda_{\mathrm{x}}:=
\frac{\xi^{\,}_{\mathrm{x},M}}{M}
\label{eq: def Lambda}
\end{equation}
on the width $M$
of the quasi-one-dimensional spin-directed network model
in Fig.~\ref{fig:network-staggered}.
We are going to study numerically the dependence 
on $M$ of the normalized localization lengths 
$\Lambda^{\,}_{\perp}$ 
and
$\Lambda^{\,}_{\parallel}$ 
defined by the geometries of Fig.%
~\ref{fig:network-staggered}(a)
and%
~\ref{fig:network-staggered}(b),
respectively. To this end, we shall
use the transfer matrix method from Ref.~\onlinecite{MacKinnon83}
in a quasi-one-dimensional geometry with the aspect ratio
$L/M=5\times10^{5}$.
We work in the parameter space%
~(\ref{eq: def parameter space is 3D}).

\subsection{Finite-size scaling analysis}
\label{subsec:finite-size-scaling}

We apply a finite-size scaling analysis to 
the normalized localization lengths 
$\Lambda^{\,}_{\perp}$ and $\Lambda^{\,}_{\parallel}$
with the goal of studying the critical properties 
at the metal-insulator transition of 
the two-dimensional spin-directed $\mathbb{Z}^{\,}_{2}$ network model, 
if any. To this end, the normalized localization lengths
along the vertical $\Lambda^{\,}_{\perp}$ 
and horizontal $\Lambda^{\,}_{\parallel}$ directions 
of the two-dimensional network
are calculated numerically by the transfer matrix method\cite{MacKinnon83}
as a function of $\delta^{2}$ with fixed values of 
the intrinsic parameters $t^{2}$ and $\theta$ and the geometric
parameter $M$, whereby the width $M$ ranges from a minimal value
$M^{\,}_{\mathrm{min}}$
to a maximal value
$M^{\,}_{\mathrm{max}}$.

We extract the critical exponent $\nu$ 
for the power-law divergence of the localization length defined by
\begin{equation}
\xi\propto 
\left|z-z^{\,}_{\mathrm{c}}\right|^{-\nu},
\label{eq:xi delta}
\end{equation} 
where $z$ and $z^{\,}_{\mathrm{c}}$ denote any relevant parameter driving to
the metal-insulator transition and its critical point, respectively.
In the present work, 
we choose the parameter $z$ that controls the distance to the
critical point $z^{\,}_{\mathrm{c}}$ out of
$\delta^{2}$, $\theta$, $\delta\phi$, and $t^{2}$
defined in Eq.~(\ref{eq: def parameter space is 4D}).
It is well known that the normalized localization length near the Anderson
transition obeys a scaling law (not necessarily a power law, though).%
~\cite{MacKinnon83}

We treat the scaling behavior of the inverse
\begin{equation}
\Gamma^{\,}_{\mathrm{x}}:= 
\Lambda^{-1}_{\mathrm{x}},
\qquad
\mathrm{x}\equiv\perp,\parallel,
\end{equation}
of the normalized localization length
$\Lambda^{\,}_{\mathrm{x}}$,
which is proportional
to the Lyapunov exponent $X^{\,}_{\mathrm{x},1}$,
a random variable that is selfaveraging in the thermodynamic limit.%
~\cite{Johnston83}
The error bars on
$\Gamma^{\,}_{\mathrm{x}}$
are easier to obtain than the error bars on $\Lambda^{\,}_{\mathrm{x}}$.

We assume that 
$\Gamma^{\,}_{\mathrm{x}}$
is a scaling function,\cite{Slevin99}
\begin{equation}
\Gamma^{\,}_{\mathrm{x}}= 
F^{\,}_{\mathrm{x}}\left(\eta\,M^{1/\nu},\zeta\,M^{y}\right),
\label{eq:scaling function}
\end{equation}
where $\eta$ and $\zeta$ are the relevant and first leading irrelevant
scaling variables, respectively. The irrelevant exponent $y$ satisfies
$y<0$. In the limit $M\rightarrow \infty$, 
we furthermore assume that the scaling law~(\ref{eq:scaling function})
obeys a Taylor expansion in powers of the scaling fields 
$\eta$ and $\zeta$ about 
$\eta\,M^{1/\nu}=\zeta\,M^{y}=0$
that we truncate to the order
$N^{\,}_{\mathrm{rel}}$ and $N^{\,}_{\mathrm{irr}}(p)$, respectively,
\begin{equation}
\Gamma^{\,}_{\mathrm{x}}\approx
\sum_{p=0}^{N^{\,}_{\mathrm{rel}}} 
\sum_{q=0}^{N^{\,}_{\mathrm{irr}}(p)} 
F^{p,q}_{\mathrm{x}}\,
\eta^{p}\, 
\zeta^{q}\,  
M^{p/\nu+qy},
\label{eq:scaling function expansion 1}
\end{equation}
where $F^{p,q}_{\mathrm{x}}$ is the expansion coefficient at the $p$
and $q$-th order of $\eta M^{1/\nu}$ and $\zeta M^y$, respectively. 
Here, we have allowed for the possibility that
$N^{\,}_{\mathrm{irr}}(p)$ depends on $p$.
If we assume that the expansion coefficients
$F^{p,0}_{\mathrm{x}}$
are all non-vanishing, we may then do the factorization
\begin{equation}
\begin{split}
&
\Gamma^{\,}_{\mathrm{x}}\approx
\sum_{p=0}^{N^{\,}_{\mathrm{rel}}}
F^{p,0}_{\mathrm{x}}\,
\left({\eta}\,M^{1/\nu}\right)^p 
\left[
1
+
\sum_{q=1}^{N^{\,}_{\mathrm{irr}}(p)} 
f^{p,q}_{\mathrm{x}}\,
\left(\zeta\,M^{y}\right)^{q}
\right],
\\
&
f^{p,q}_{\mathrm{x}}:=
\frac{F^{p,q}_{\mathrm{x}}}{F^{p,0}_{\mathrm{x}}}.
\end{split}
\label{eq:scaling function expansion 2}
\end{equation}
It is an empirical fact that the fitting procedure%
~(\ref{eq:scaling function expansion 2})
is more stable than the fitting procedure%
~(\ref{eq:scaling function expansion 1}).
Finally, we assume that
\begin{equation}
\eta:=
z - z^{\,}_{\mathrm{c}},
\label{eq:eta}
\end{equation}
where $z$ and $z^{\,}_{\mathrm{c}}$ 
were introduced in Eq.\ (\ref{eq:xi delta})
and
\begin{equation}
\zeta=1. 
\end{equation}
In other words, we choose as the fitting parameters:
(1) the scaling exponents $\nu$ and $y$,
(2) 
the location of the critical point $z^{\,}_{\mathrm{c}}$,
(3) 
the $(N^{\,}_{\mathrm{rel}}+1)$ 
expansion coefficients for the relevant perturbation
$F^{p,0}_{\mathrm{x}}$ for $p=0,\cdots,N^{\,}_{\mathrm{rel}}$,
and (4)
the $\sum_{p=0}^{N^{\,}_{\mathrm{rel}}} N^{\,}_{\mathrm{irr}}(p)$ 
expansion coefficients for the leading irrelevant perturbation, 
$f^{p,q}_{\mathrm{x}}$
for 
$q=1,\cdots,N^{\,}_{\mathrm{irr}}(p)$
and
$p=0,\cdots,N^{\,}_{\mathrm{rel}}$.
We call
\begin{equation}
\Lambda^{\mathrm{c}}_{\mathrm{x}}:= 
\frac{1}{F^{0,0}_{\mathrm{x}}},
\end{equation}
the universal scaling amplitude of the normalized localization lengths
with $\mathrm{x}=\perp,\parallel$, respectively.
In order to demonstrate a single-parameter scaling law, 
we introduce 
\begin{subequations}
\begin{equation}
\Lambda_{\mathrm{x}}^{\prime}:=
\left(\Gamma^{\prime}_{\mathrm{x}}\right)^{-1}, 
\quad 
\mathrm{x}=\perp,\parallel,
\label{eq:Lambda^{prime}}
\end{equation}
where $\Gamma^{\prime}_{\mathrm{x}}$ is defined by subtracting from
$\Gamma^{\,}_{\mathrm{x}}$ its finite-size correction from the leading
irrelevant exponent $y$, i.e.,
\begin{equation}
\Gamma^{\prime}_{\mathrm{x}}:=
\Gamma^{\,}_{\mathrm{x}} 
-
\sum_{p=0}^{N^{\,}_{\mathrm{rel}}}
\sum_{q=1}^{N^{\,}_{\mathrm{irr}}(p)}
F^{p,0}_{\mathrm{x}}\,
f^{p,q}_{\mathrm{x}}\, 
\left(\eta M^{1/\nu}\right)^p
M^{q y}.
\label{eq:Gamma^{prime}}
\end{equation}
\end{subequations}

The quality of the fit of the data set to the scaling function
is tested as follows. The simplest test for fitting numerical data
is given by
\begin{equation}
\frac{\chi^{2}}{N}:=
\frac{1}{N}
\sum_{j=1}^{N}\frac{(O_j-E_j)^{2}}{\sigma^{2}},
\label{eq: def chi{2}/{N}}
\end{equation} 
where $O_j\,(=\Gamma^{\,}_{\mathrm{x}})$ runs over the $N$ data,
$\sigma^{2}$ is the variance of the data, 
and $E_j$ is the value of the fitting function 
computed from Eq.\ (\ref{eq:scaling function expansion 2})
corresponding to data $O$. 
The values of $\chi^{2}/N$
are distributed in the range $[0,\infty[$, and
a perfect fit gives $\chi^{2}/N=0$.
In practice,
a fit is acceptable if $\chi^{2}/N$ 
is smaller than $1$.
Another measure for the quality of the fitting procedure
is the goodness of fit $\mathcal{G}$ defined by
\begin{subequations}
\label{eq:goodness of fit}
\begin{equation}
\mathcal{G}:= 
\Gamma^{\,}_{\mathrm{ni}}\Big((N-p)/2,\chi^{2}/2\Big),
\label{eq:goodness of fit a}
\end{equation}
where
\begin{equation}
\Gamma^{\,}_{\mathrm{ni}}(a,x):= 
\frac{1}{\Gamma(a)} 
\int\limits_{x}^{\infty}\mathrm{d}y\, 
y^{a-1}\, 
e^{-y}
\label{eq:goodness of fit b}
\end{equation}
\end{subequations}
denotes the normalized incomplete gamma function.
The values of $\mathcal{G}$ 
are distributed in the range $[0,1]$.
A perfect fit gives $\mathcal{G}=1$. 

A consistency check on the fitting procedure is obtained by
verifying that the statistical error bar (one sigma)
on a fitting parameter does not exceed the value of the fitting parameter 
by an order of magnitude.
Error bars of fitting parameters
are themselves estimated from
error-propagation theory given the error bars 
of $\Lambda^{\,}_{\mathrm{x}}$. 

The fitting with the nonlinear function 
(\ref{eq:scaling function expansion 2}) 
strongly depends on the initial values of the fitting parameters. 
For this reason, we perform iteratively the fitting for the
data set with different initial values of the fitting parameters.
We then select the best fitting parameters as the ones with
the smallest value of $\chi^{2}/N$, that we denote
$\chi^{2}_{\mathrm{min}}/N$. 
The best fitting parameters and their
$\chi^{2}_{\mathrm{min}}/N$ define the most reliable fit for a given data set. 
The typical number of iterations done for a given data set
is about $1000$. 

The best fitting parameters should not strongly depend on the given data set.
For this reason, we repeat our scaling analysis for data sets
differing through the choice of their minimum and maximum values
$M^{\,}_{\mathrm{min}}$ and $M^{\,}_{\mathrm{max}}$ for the width of the 
quasi-one-dimensional spin-directed $\mathbb{Z}^{\,}_{2}$ network model.
As we demonstrate later, 
the fitting parameters obtained from data sets 
with different minimum
$M^{\,}_{\mathrm{min}}$ and maximum $M^{\,}_{\mathrm{max}}$ widths  
are not always consistent with each other.
To overcome this difficulty, we estimate the 
error bars on the fitting parameters
by making use of the practical-error-bar procedure
from Ref.~\onlinecite{Obuse13}.

\begin{figure}[t]
\includegraphics[width=0.45\textwidth]{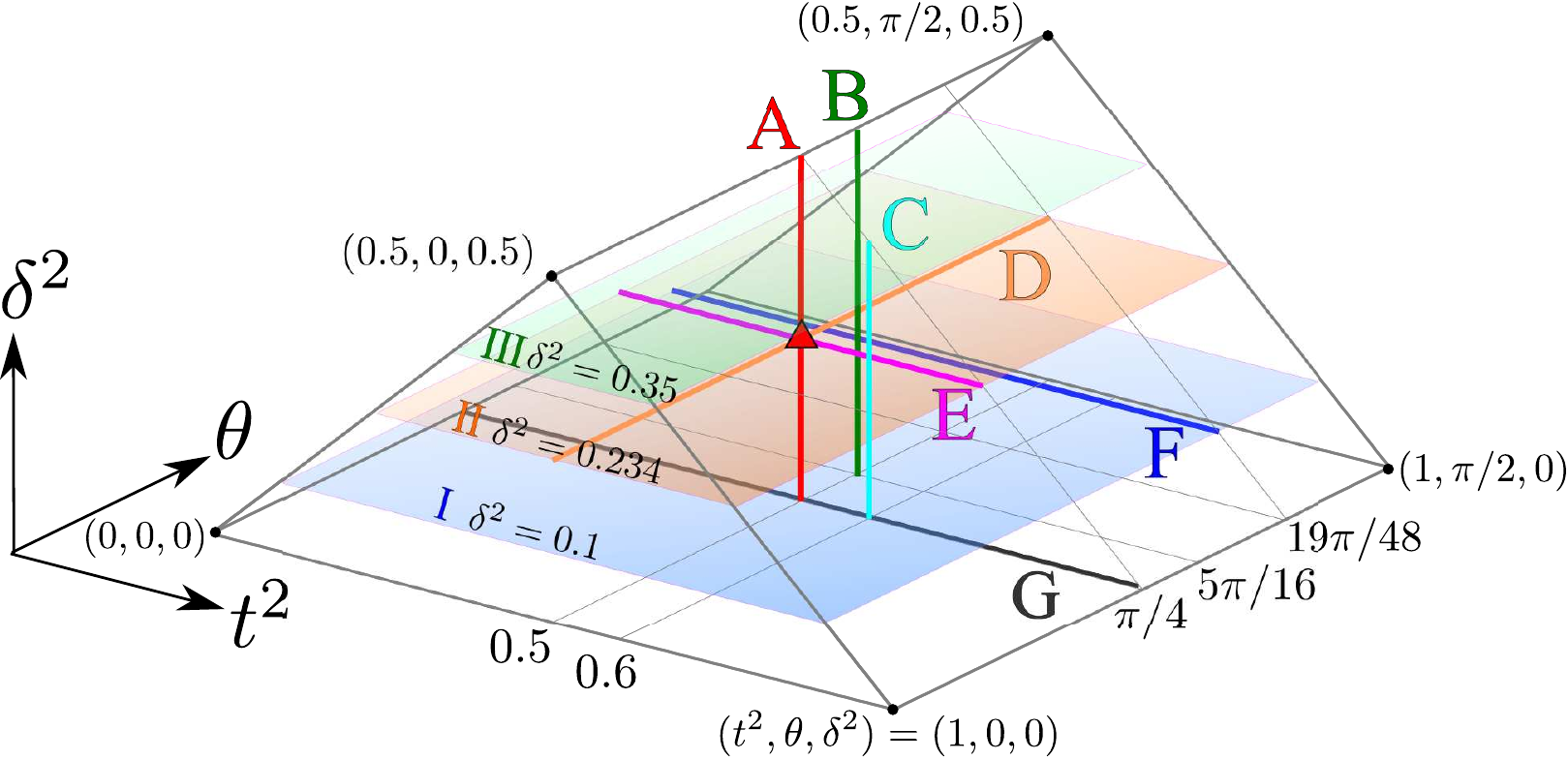}
\caption{
(Color online:)
One and two-dimensional cuts in the parameter space%
~(\ref{eq: def parameter space is 3D}) studied in this paper.
The one-dimensional cuts, 
A defined in Eq.\ (\ref{eq: cut choice 1}),
B in (\ref{eq: cut choice 2}),
C in (\ref{eq: cut choice 3}),
D in (\ref{eq: cut choice D}),
E in (\ref{eq:cut choice E}),
F in (\ref{eq:cut choice F}), and 
G in (\ref{eq:cut choice G}), as well as
the two-dimensional cuts, 
I-III, defined in Eq.\ (\ref{eq:two-dimensional cut}) are shown.
The red triangle 
where the three one-dimensional cuts A,D, and E, cross 
represents the critical point  (\ref{eq: critical point along ct 1}).
}
\label{fig:cut}
\end{figure}

\begin{figure*}[th]
(a-1) \hspace{7cm} (a-2)\\
\includegraphics[width=0.42\textwidth]{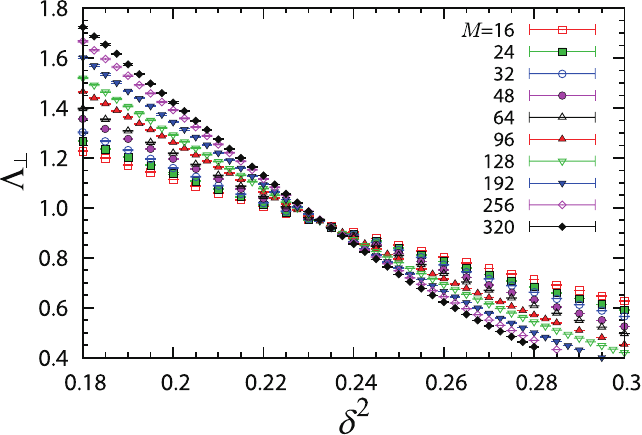}
\includegraphics[width=0.38\textwidth]{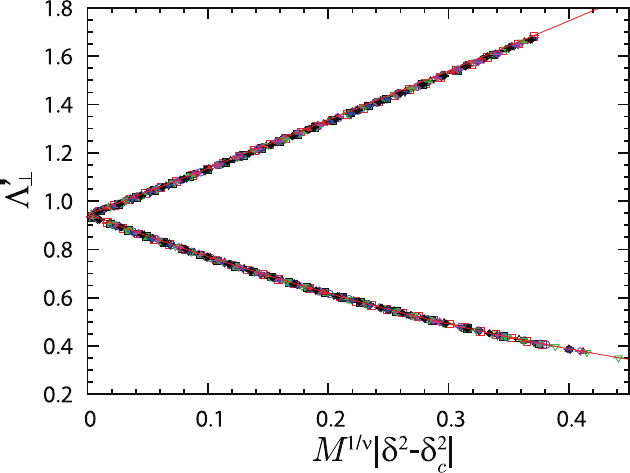}\\
(b-1) \hspace{7cm} (b-2)\\
\includegraphics[width=0.42\textwidth]{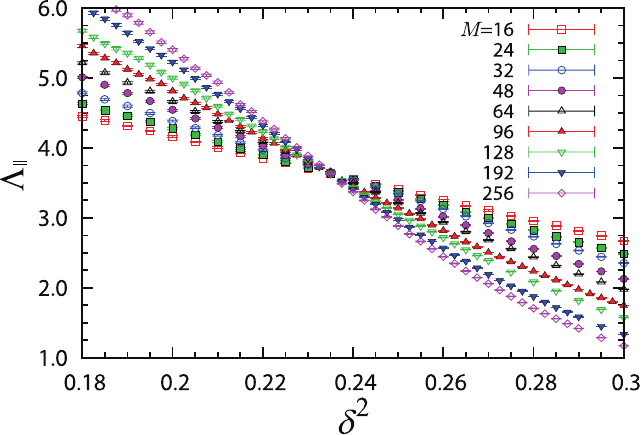}
\includegraphics[width=0.38\textwidth]{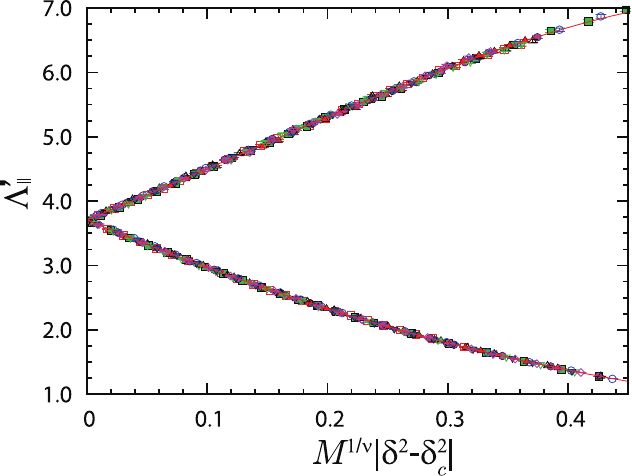}
\caption{
(Color online:)
The two-dimensional spin-directed $\mathbb{Z}^{\,}_{2}$ network model 
is solved numerically along the one-dimensional cut~(\ref{eq: cut choice 1})
in the three-dimensional parameter space%
~(\ref{eq: def parameter space is 3D}).
Panel (a-1) shows
the $\delta^{2}$ dependence of the normalized localization length
$\Lambda^{\,}_{\perp}$ corresponding to the geometry of 
Fig.~\ref{fig:network-staggered}(a)
for several values of $M$.
Panel (b-1) shows
the $\delta^{2}$ dependence of the normalized localization lengths
$\Lambda^{\,}_{\parallel}$ 
corresponding to the geometry of 
Fig.~\ref{fig:network-staggered}(b)
for several values of $M$.
A finite-size scaling analysis of panels
(a-1) and (b-1) is performed in panels (a-2) and (b-2), respectively.
The horizontal axis is $M^{1/\nu}|\delta^{2}-\delta^{2}_{\mathrm{c}}|$
with $\nu$ and $\delta^{2}_{\mathrm{c}}$ given 
in 
Table~\ref{tab:fitting parameters cut 1 perpendicular} and
Table~\ref{tab:fitting parameters cut 1 parallel} in 
Appendix \ref{appsec:finite-size-scaling}.
The vertical axis $\Lambda^{\prime}_{\mathrm{x}}$
with $\mathrm{x}=\perp,\parallel$
is defined by subtracting from the
normalized localization length $\Lambda^{\,}_{\mathrm{x}}$
its finite-size correction from the leading irrelevant exponent $y$
given in Table~S-I and Table~S-II from Ref.\ \onlinecite{suppl}.
The red solid curve demonstrates the quality of the data collapse
onto a one-parameter scaling function.
        }
\label{fig:Lambda dimerization 0.5 PIover4}
\end{figure*}

\subsection{Dependence of the normalized localization length on
$\delta^{2}$}
\label{subsec:dimerization delta dependence}

To begin with, we choose the one-dimensional cut
\begin{equation}
\mathrm A:\quad (0.5,\pi/4,\delta^{2}),
\qquad \delta^{2}\in[0,0.5],
\label{eq: cut choice 1}
\end{equation}
of three-dimensional parameter space%
~(\ref{eq: def parameter space is 3D}).
Cut A is shown in Fig.~\ref{fig:cut} with the label ``A''.
This cut is far from the isotropic CC critical point
\begin{equation}
\left(0.5,\pi/2,0\right).
\label{eq: def isotropic CC point}
\end{equation}

Figures~\ref{fig:Lambda dimerization 0.5 PIover4}(a-1) 
and \ref{fig:Lambda dimerization 0.5 PIover4}(b-1)
show the $\delta^{2}$ dependence of the normalized localization lengths 
$\Lambda^{\,}_{\perp}$ and $\Lambda^{\,}_{\parallel}$ 
defined by the geometries of Fig.%
~\ref{fig:network-staggered}(a)
and%
~\ref{fig:network-staggered}(b),
respectively, along the cut~(\ref{eq: cut choice 1})
as the width $M$ is
increased from $M=16$ to $M=320$. According to 
Fig.~\ref{fig:Lambda dimerization 0.5 PIover4}(a-1),
$\Lambda^{\,}_{\perp}$
and 
$\Lambda^{\,}_{\parallel}$ 
increase (decrease) with increasing $M$ for small (large) values of $\delta^{2}$.
The dependence on $M$ of either
$\Lambda^{\,}_{\perp}$ or $\Lambda^{\,}_{\parallel}$ 
appears to vanish at a value of $\delta^{2}$ approximately given by
$0.23$. This suggests that the point 
\begin{equation}
(t^{2}_{\mathrm{c}},\theta^{\,}_{\mathrm{c}},\delta^{2}_{\mathrm{c}}):=
\left(0.5,\pi/4,0.23\right)
\label{eq: critical point along ct 1}
\end{equation}
in the three-dimensional parameter space%
~(\ref{eq: def parameter space is 3D})
is a critical point at which a metal-insulator transition
takes place.

\subsubsection{Finite-size scaling for the 
normalized localization length with dimerization}

To confirm this interpretation of the point%
~(\ref{eq: critical point along ct 1})
in the three-dimensional parameter space%
~(\ref{eq: def parameter space is 3D}),
we have done a finite-size scaling analysis of the data presented in
Fig.~\ref{fig:Lambda dimerization 0.5 PIover4}(a-1) 
and~\ref{fig:Lambda dimerization 0.5 PIover4}(b-1)
by regarding $\delta^{2}$ as the driving parameter $z$ in 
Eq.\ (\ref{eq:xi delta}),
the details of which are to be found in 
Tables~\ref{tab:summary_FSS},
\ref{tab:fitting parameters cut 1 perpendicular}, and
\ref{tab:fitting parameters cut 1 parallel} 
(from Appendix \ref{appsec:finite-size-scaling} 
for Tables%
~\ref{tab:fitting parameters cut 1 perpendicular} 
and
\ref{tab:fitting parameters cut 1 parallel}).

Figures~\ref{fig:Lambda dimerization 0.5 PIover4}(a-2) 
and~\ref{fig:Lambda dimerization 0.5 PIover4}(b-2)
support the one-parameter scaling obeyed by
$\Lambda^{\prime}_{\perp}$ and $\Lambda^{\prime}_{\parallel}$
in Eqs.~(\ref{eq:Lambda^{prime}}) and (\ref{eq:Gamma^{prime}})
close to the critical point%
~(\ref{eq: critical point along ct 1})
in the three-dimensional parameter space%
~(\ref{eq: def parameter space is 3D}).
We use values of the irrelevant exponent $y$ and the expansion
coefficient $f_{\mathrm{x}}^{p,q}$ given in Tables 
 \ref{tab:fitting parameters cut 1 perpendicular} and
\ref{tab:fitting parameters cut 1 parallel}
to calculate
$\Lambda^{\prime}_{\perp}$ and $\Lambda^{\prime}_{\parallel}$
through 
Eqs.~(\ref{eq:Lambda^{prime}}) and (\ref{eq:Gamma^{prime}}).

As is reported in Tables \ref{tab:fitting parameters cut 1 perpendicular} and
\ref{tab:fitting parameters cut 1 parallel},
the fitting parameters obtained by varying the minimum and
maximum widths $M^{\,}_{\mathrm{min}}$ and $M^{\,}_{\mathrm{max}}$, 
respectively, 
are not always consistent with the statistical error bars, 
even though $\chi^{2}_{\mathrm{min}}/N$ and $\mathcal{G}$ 
are acceptable. On the one hand, 
Table \ref{tab:fitting parameters cut 1 perpendicular}
gives values of $\nu$ from all data set that vary
within the statistical error bars. 
On the other hand,
Table \ref{tab:fitting parameters cut 1 parallel}
gives values of $\nu$ extracted from the data set with 
$M^{\,}_{\mathrm{min}}=16$
and 
$M^{\,}_{\mathrm{max}}=256$ 
that are not within the error bars from 
the values of $\nu$ extracted from the data set with
$M^{\,}_{\mathrm{min}}=32$ 
and
$M^{\,}_{\mathrm{max}}=256$, 
while both set of values are statistically reliable
as measured by $\chi^{2}_{\mathrm{min}}/N$.
We believe that the reason for this inconsistency originates 
from the orders of truncations
$N^{\,}_{\mathrm{rel}}$ and $N^{\,}_{\mathrm{irr}}(p)$ 
in Eq.~(\ref{eq:scaling function expansion 2}),
constrained as they are by the accuracy of our data,
being too small.
To overcome this difficulty, we estimate the 
error bars on the fitting parameters shown in Table~\ref{tab:summary_FSS}
by making use of the practical-error-bar procedure
from Ref.~\onlinecite{Obuse13}.
The first two lines of Table~\ref{tab:summary_FSS} give,
at the critical point~(\ref{eq: critical point along ct 1})
along the cut (\ref{eq: cut choice 1}),
the value of the critical exponent $\nu$ that
controls the power-law divergence of the localization length,
the value $|y|$ for the leading irrelevant scaling exponent $y$, 
the strength of $\delta^{2}_{\mathrm{c}}$, 
the values of the normalized localization length 
$\Lambda^{\mathrm{c}}_{\perp}$ and $\Lambda^{\mathrm{c}}_{\parallel}$,
and their geometrical mean
$\sqrt{\Lambda^{\mathrm{c}}_{\perp}\Lambda^{\mathrm{c}}_{\parallel}}$.

The subsequent two pairs of lines from
Table~\ref{tab:summary_FSS} summarize the results of the
same analysis performed along the one-dimensional cuts
\begin{equation}
\mathrm{B:}\quad \left(0.5,5\pi/16,\delta^{2}\right),
\qquad
\delta^{2}\in[0,0.5],
\label{eq: cut choice 2}
\end{equation} 
and 
\begin{equation}
\mathrm{C:}\quad\left(0.6,\pi/4,\delta^{2}\right),
\qquad
\delta^{2}\in[0,0.4],
\label{eq: cut choice 3}
\end{equation}
in the three-dimensional parameter space%
~(\ref{eq: def parameter space is 3D}).
Cuts B and C are shown
in Fig.~\ref{fig:cut} with the labels ``B'' and ``C''.
The details of the finite-size scaling analysis along the cuts
``B'' and ``C'' are presented in
Tables 
\ref{tab:fitting parameters cut 2 perpendicular}--%
\ref{tab:fitting parameters cut 2 parallel}  
and
\ref{tab:fitting parameters cut 3 perpendicular}--%
\ref{tab:fitting parameters cut 3 parallel} 
from
Appendix \ref{appsec:finite-size-scaling}, respectively.
The $\delta^2$ dependence of the normalized localization lengths
$\Lambda^{\,}_\perp$ and $\Lambda^{\,}_\parallel$ for the
cuts ``B'' and ``C'' are shown in
Figs.\ \ref{fig:Lambda_dimerization_0.5_5PIover16}
and \ref{fig:Lambda_dimerization_0.6_PIover4}
from Appendix \ref{appsec:finite-size-scaling}, respectively.

The value for the critical point $\delta^{\,}_{\mathrm{c}}$ 
along a one-dimensional cut with $t^{{2}}$ and $\theta$ fixed
in the three-dimensional parameter space%
~(\ref{eq: def parameter space is 3D})
obtained from 
$\Lambda^{\,}_{\perp}$ 
agrees within error bars with that obtained from
$\Lambda^{\,}_{\parallel}$. 
This agreement is required if
$(t^{{2}},\theta,\delta^{{2}}_{\mathrm{c}})$
is to be interpreted as a quantum critical point
in Anderson localization.

We find the values of the scaling exponent $\nu$ 
to be distributed around $2.7\pm0.2$ in Table~\ref{tab:summary_FSS}.
For comparison,
the scaling exponent $\nu$ for the ordinary two-dimensional symplectic class 
is $\nu\approx2.7$.%
~\cite{Asada02}

In contrast,
we observe that the normalized localization lengths at the critical point
$\Lambda^{\mathrm{c}}_{\perp}$ and $\Lambda^{\mathrm{c}}_{\parallel}$ 
along a given one-dimensional cut in the three-dimensional parameter space%
~(\ref{eq: def parameter space is 3D}) from Table~\ref{tab:summary_FSS}
differ. These values also differ along
different cuts in the three-dimensional parameter space%
~(\ref{eq: def parameter space is 3D}) 
as well as from the value
$\Lambda^{\mathrm{c}}\approx1.84$
obtained for the ordinary two-dimensional symplectic class
in Ref.~\onlinecite{Asada02}.
However, if we take the geometric average of
$\Lambda^{\mathrm{c}}_{\perp}$ and $\Lambda^{\mathrm{c}}_{\parallel}$,
we find that the value
\begin{equation}
\sqrt{
\Lambda^{\mathrm{c}}_{\perp}\Lambda^{\mathrm{c}}_{\parallel}
     }
\approx 
1.84
\end{equation}
agrees with $\Lambda^{\mathrm{c}}$. 
The result that the geometric average of 
$\Lambda^{\mathrm{c}}_{\perp}$ 
and 
$\Lambda^{\mathrm{c}}_{\parallel}$ 
for the two-dimensional spin-directed $\mathbb{Z}^{\,}_{2}$ network model 
agrees with the value $\Lambda^{\mathrm{c}}$ for
the two-dimensional $\mathbb{Z}^{\,}_{2}$ network model 
from Ref.~\onlinecite{Obuse07} has a counterpart for
the CC network model. In the anisotropic CC network model 
defined by the condition $0\leq t^{2}\neq1/2\leq1$, 
there are two normalized localization lengths 
$\Lambda^{\mathrm{c}}_{\perp}$ 
and 
$\Lambda^{\mathrm{c}}_{\parallel}$ 
whose geometric average equals the
normalized localization length of the 
isotropic CC network model 
defined by the condition $t^{2}=1/2$.%
~\cite{Chalker88}
In either {case}, this relation is a manifestation of
two-dimensional conformal invariance at a critical point.%
~\cite{Cardy87}

The difference between 
$\Lambda^{\mathrm{c}}_{\perp}$ and $\Lambda^{\mathrm{c}}_{\parallel}$ 
along the cut~(\ref{eq: cut choice 2})
is smaller than that 
along the cut~(\ref{eq: cut choice 1}).
This observation is consistent with the fact that
the former cut is closer than the latter cut
to the isotropic CC critical point~(\ref{eq: def isotropic CC point}).

The difference between 
$\Lambda^{\mathrm{c}}_{\perp}$ and $\Lambda^{\mathrm{c}}_{\parallel}$ 
along the cut~(\ref{eq: cut choice 3})
is smaller than that 
along the cut~(\ref{eq: cut choice 1}).
We attribute this fact to the property that
increasing small $t$ increases $\Lambda^{\,}_{\perp}$, 
while it decreases $\Lambda^{\,}_{\parallel}$. 

From these observations,
we conjecture that the surface states of 
weak three-dimensional $\mathbb{Z}^{\,}_{2}$ topological insulators 
undergo a metal-insulator transition as a result of the
competition between disorder and dimerization that
belongs to the ordinarily two-dimensional symplectic universality class.

\begin{table*}[t]
\caption{
Summary of the finite-size scaling analysis for
$\Gamma^{\,}_{\mathrm{x}}(=1/\Lambda^{\,}_{\mathrm{x}})$
with $\mathrm{x}=\perp,\parallel$
for the three cuts
(\ref{eq: cut choice 1}),
(\ref{eq: cut choice 2}),
and
(\ref{eq: cut choice 3})
from the three-dimensional parameter space%
~(\ref{eq: def parameter space is 3D}).
Only the most important fitting parameters, namely 
$\nu$, 
$y$, 
$\delta^{2}_{\mathrm{c}}$, 
$\Lambda^{\mathrm{c}}_{\mathrm{x}}$,
and
$\sqrt{\Lambda^{\mathrm{c}}_{\perp}\Lambda^{\mathrm{c}}_{\parallel}}$
are shown. More details on this finite-size scaling analysis can be
found in 
Tables \ref{tab:fitting parameters cut 1 perpendicular} -
 \ref{tab:fitting parameters cut 3 parallel}
in
Appendix \ref{appsec:finite-size-scaling}.
The expected value and its error bar are estimated by employing
the practical error bar procedure.~\cite{Obuse13}
        }
\label{tab:summary_FSS}
\begin{tabular}{c c c|r l r l r l r l r}
\hline\hline
$\text{x}$ 
& 
$\theta$ 
& 
$t^{2}$ 
& 
$\nu$ 
&& 
$|y|$ 
&& 
$\delta^{2}_{\mathrm{c}}$ 
&&
$\Lambda^{\mathrm{c}}_{\mathrm{x}}$ 
&& 
$\sqrt{
\Lambda^{\mathrm{c}}_{\perp}
\Lambda^{\mathrm{c}}_{\parallel}
       }$
\\
\hline
$\perp$ 
& 
$\pi/4$ 
& 
$0.5$ 
& 
$2.89$ 
& 
$[2.75:2.98]\quad$ 
& 
$0.72$ 
&
$[0.59:0.90]\quad$ 
& 
$0.2340$ 
& 
$[0.2334:0.2349]\quad$ 
&
$0.935$ 
& 
$[0.919:0.944]\quad$ 
& 
$1.849$
\\
$\parallel$ 
& 
$\pi/4$ 
& 
$0.5$ 
& 
$2.71$ 
& 
$[2.53:2.85]$ 
& 
$0.47$ 
&
$[0.30:0.81]$ 
& 
$0.2343$ 
& 
$[0.2332:0.2355]$ 
&
$3.657$ 
& 
$[3.557:3.734]$
\\
\hline
$\perp$ 
& 
$5\pi/16$ 
& 
$0.5$ 
& 
$2.80$ 
& 
$[2.64:2.91]$ 
& 
$0.91$ 
&
$[0.55:1.49]$ 
& 
$0.1591$ 
& 
$[0.1579:0.1616]$ 
&
$1.290$ 
& 
$[1.234:1.310]$ 
& 
$1.850$
\\
$\parallel$ 
& 
$5\pi/16$ 
& 
$0.5$ 
& 
$2.63$ 
& 
$[2.55:2.79]$ 
& 
$1.14$ 
&
$[0.55:1.54]$ 
& 
$0.1590$ 
& 
$[0.1575:0.1598]$ 
&
$2.652$ 
& 
$[2.622:2.716]$
\\
\hline
$\perp$ 
& 
$\pi/4$ 
& 
$0.6$ 
& 
$2.84$ 
& 
$[2.70:2.92]$ 
& 
$0.81$ 
&
$[0.51:1.21]$ 
& 
$0.2765$ 
& 
$[0.2752:0.2784]$ 
&
$1.225$ 
& 
$[1.193:1.249]$ 
& 
$1.834$
\\
$\parallel$ 
& 
$\pi/4$ 
& 
$0.6$ 
& 
$2.57$ 
& 
$[2.42:2.74]$ 
& 
$0.85$ 
&
$[0.26:1.72]$ 
& 
$0.2764$ 
& 
$[0.2754:0.2772]$ 
&
$2.746$ 
& 
$[2.710:2.783]$
\\
\hline\hline
\end{tabular} 
\end{table*}

\subsection{Dependence of the normalized localization length on
$\theta$}
\label{subsec:theta dependence}

We are going to study the dependence on $\theta$
of the normalized localization
lengths $\Lambda^{\,}_{\perp}$ and $\Lambda^{\,}_{\parallel}$,
whereby we recall that $\theta$ encodes
the strength of the spin-orbit couplings.

\begin{figure}[t]
\centering
\includegraphics[width=0.45\textwidth]{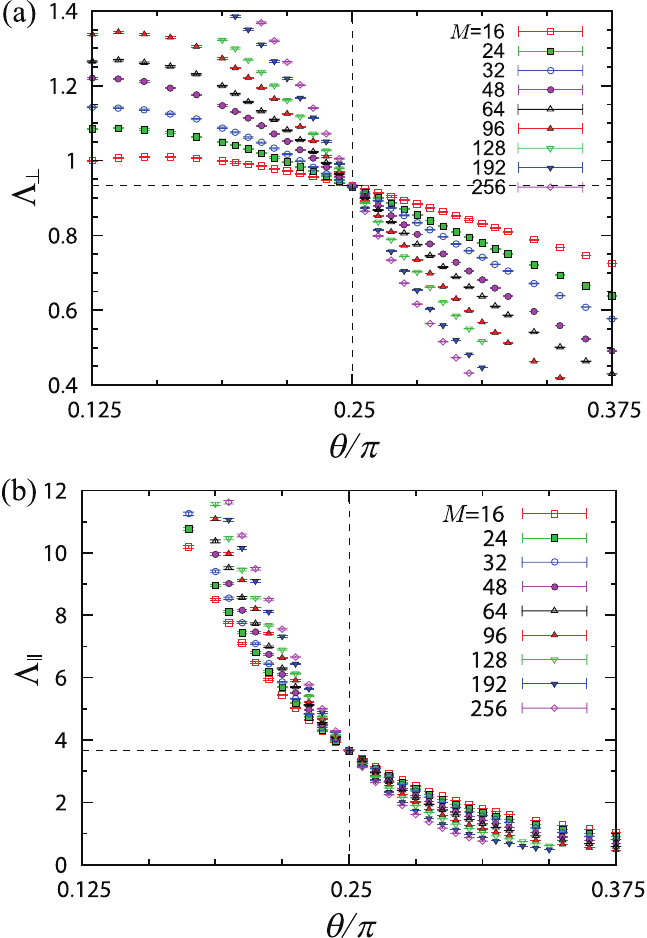}
\caption{
(Color online:)
The $\theta$ dependence of the normalized localization lengths
$\Lambda^{\,}_{\perp}$ in panel (a) 
and $\Lambda^{\,}_{\parallel}$ in panel (b) 
of the two-dimensional spin-directed $\mathbb{Z}^{\,}_{2}$ network model with
dimerization along the one-dimensional cut (\ref{eq: cut choice D}).
The strength of the spin-orbit coupling $\theta$ is varied
given the values $t^{2}=0.5$ and $\delta^{2}=0.234$ 
for fixed $M$ ranging from $16$ to $256$.
The vertical and horizontal dashed lines represent 
$\theta^{\,}_{\mathrm{c}}$ 
and
$\Lambda^{\,}_{\mathrm{x,c}}$, respectively,
as deduced from the finite-size scaling analysis summarized 
in Table.\ \ref{tab:summary_FSS}.
}
\label{fig:Lambda theta dependence}
\end{figure}

We first choose the one-dimensional cut
\begin{equation}
\mathrm{D:}\quad (0.5, \theta, 0.234), \quad \theta \in [0,\pi/2],
\label{eq: cut choice D}
\end{equation}
of the three-dimensional parameter space (\ref{eq: def parameter space
is 3D}). 
Cut D is shown in Fig.~\ref{fig:cut} with the label ``D''.
Figures \ref{fig:Lambda theta dependence}(a) and 
\ref{fig:Lambda theta dependence}(b) show 
the normalized localization lengths
$\Lambda^{\,}_{\perp}$ and
$\Lambda^{\,}_{\parallel}$, respectively, as a function of $\theta$  along the
cut (\ref{eq: cut choice D}).
We note that, along the cut (\ref{eq: cut choice D}),
there is the critical point $(0.5,\pi/4,0.234)$ of 
Eq.\ (\ref{eq: critical point along ct 1}) identified in Sec.\
\ref{subsec:dimerization delta dependence}.
The dependence of the normalized localization lengths
$\Lambda^{\,}_{\perp}$ 
and
$\Lambda^{\,}_{\parallel}$ 
on the width $M$ depicted
in Fig.~\ref{fig:Lambda theta dependence} 
show the expected metallic behavior for $\theta<\pi/4$ 
and the expected insulating behavior for $\theta>\pi/4$
from the phase diagram~\ref{fig: phase-diagram 3 parameter}(c).
We apply the finite-size scaling analysis by using the data points
presented in Fig. \ref{fig:Lambda theta dependence}(b)
by choosing $\theta$ as the driving parameter $z$ in Eq.\ (\ref{eq:xi delta}).
To improve the stability of the finite-size scaling analysis 
with the number of data points at our disposal,
we have used $\theta^{\,}_{\mathrm{c}}=\pi/4$ and $\Lambda^{\,}_{\mathrm{c}}=3.657$ 
as given in the scaling function Eq.~(\ref{eq:scaling function expansion 2}).
The parameter sets used in the finite-size scaling analysis
is shown in Table%
~\ref{tab:fitting parameters cut D parallel} from Appendix%
~\ref{appsec:finite-size-scaling}. Applying the practical error bar
procedure,\cite{Obuse13} 
we obtain 
\begin{equation}
\nu=2.88
\quad [2.81:2.98].
\label{eq:nu from theta}
\end{equation}
This result suggests
that the strength of the spin-orbit couplings is also a control
parameter that drives the surface states of
weak three-dimensional $\mathbb{Z}^{\,}_{2}$ topological insulators 
through the metal-insulator transition belonging to
the ordinary two-dimensional symplectic universality 
class among the Anderson transitions.

We proceed with the numerical exploration of 
the three-dimensional parameter space 
(\ref{eq: def parameter space is 3D}) 
depicted in
Fig.~\ref{fig: phase-diagram 3 parameter}(c)
with the three two-dimensional cuts
\begin{subequations}
\label{eq:two-dimensional cut}
\begin{eqnarray}
\mathrm{I}&:&\quad (t^{2},\theta, 0.1), 
\label{eq:two-dimensional cut I}\\
&&\quad\quad t^{2} \in [0.1,0.9] ,\quad \theta \in [0,\pi/2],
\nonumber\\
\mathrm{II}&:&\quad (t^{2},\theta, 0.234),
\label{eq:two-dimensional cut II}\\
&&\quad\quad t^{2} \in [0.234,0.766],
\quad \theta \in [0,\pi/2],
\nonumber\\
\mathrm{III}&:&\quad (t^{2},\theta, 0.35),
\label{eq:two-dimensional cut III}\\
&&\quad\quad t^{2} \in [0.35,0.65],
\quad \theta \in [0,\pi/2],\nonumber
\end{eqnarray}
\end{subequations}
which are shown in Fig.~{\ref{fig:cut} with the labels
``I'', ``II'', and ``III'', respectively.

We have calculated the normalized localization length 
$\Lambda^{\,}_{\parallel}$ 
on any one of the two-dimensional cuts I, II, and III.
We have estimated the positions 
$(t^{2}_{\mathrm{c}},\theta^{\,}_{\mathrm{c}})$
of the critical points from the $\theta$
dependence of the normalized localization length $\Lambda^{\,}_{\parallel}$ 
at fixed $t^{2}$ by using the scaling function
(\ref{eq:scaling function expansion 2}), but neglecting corrections involving
the irrelevant exponent $y$. Because of this simplification, the goodness of
fit $\mathcal{G}$ defined in Eq.~(\ref{eq:goodness of fit}) is 
almost zero so that the obtained values of $\nu$
are not reliable. Nevertheless, we believe that 
the accuracy of the positions of the 
critical point $(t^{2}_{\mathrm{c}},\theta^{\,}_{\mathrm{c}})$ 
are sufficient to confirm the phase diagram%
~\ref{fig: phase-diagram 3 parameter}(c).
Figures \ref{fig:phase diagram theta dependence}(a),
\ref{fig:phase diagram theta dependence}(b), 
and 
\ref{fig:phase diagram theta dependence}(c) show
the phase diagrams on the two-dimensional cuts I, II, and III, respectively.
We find that the metallic phase survives 
along the boundary $t^{2}=1-\delta^{2}$ 
for a window of values of $\theta$'s not too large.
This metallic phase undergoes
the transition to the insulating phase upon increasing $\theta$.
We also observe the reentrance driven by $t^{2}$ near 
$\theta\approx3\pi/8$ 
on the two-dimensional cut (\ref{eq:two-dimensional cut I})
in Fig.~\ref{fig:phase diagram theta dependence}(a).
The lower upper bound on the allowed values of $t^{2}$ 
resulting from the larger fixed values for $\delta^{2}$
preempts any reentrance driven by $t^{2}$
for the two-dimensional cuts II and III
in Figs.~\ref{fig:phase diagram theta dependence}(b) 
and 
\ref{fig:phase diagram theta dependence}(c),
respectively.

\begin{figure*}[t]
\centering
\includegraphics[width=0.95\textwidth]{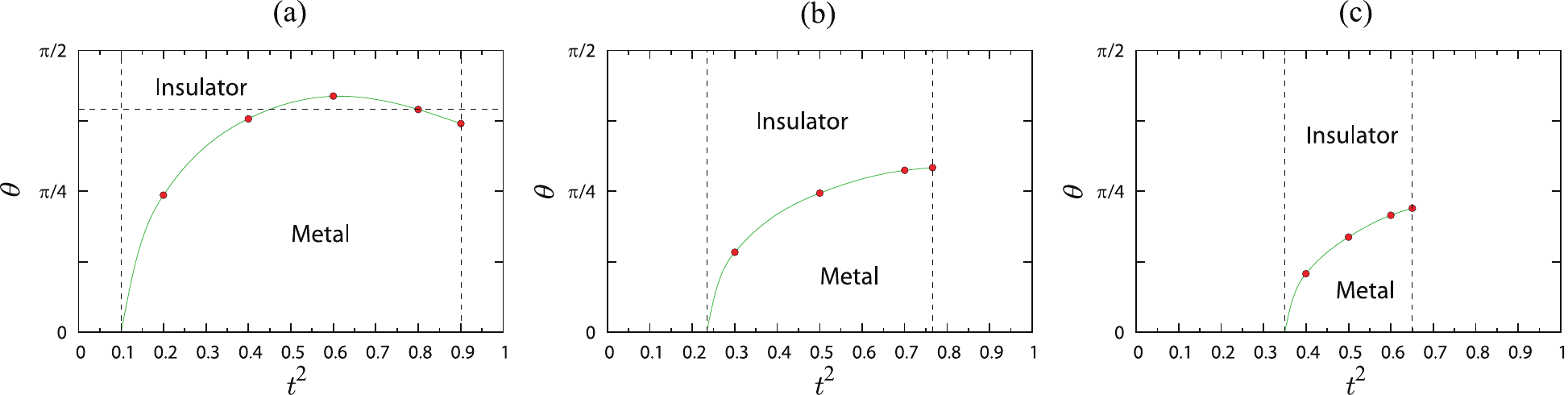}
\caption{
(Color online:)
The phase diagrams obtained from the $\theta$ dependence of
the normalized localization length $\Lambda^{\,}_{\parallel}$ 
on the two-dimensional cuts 
(a) I ($\delta^{2}=0.1$), 
(b) II ($\delta^{2}=0.234$), 
and (c) III ($\delta^{2}=0.35$). 
The critical points obtained by the finite-size scaling
analysis are marked by red dots. 
The connecting green lines are guides to the eyes.
The point 
$t^{2}=\delta^{2}$ and $\theta=0$ 
is critical by construction.
The two vertical dashed lines in each panel indicate 
the allowed minimum and maximum values of $t^{2}$. 
The horizontal dashed line in panel (a) represents
the one-dimensional cut ``F'' ($\theta=19\pi/48$) defined in
Eq.\ (\ref{eq:cut choice F}).}
\label{fig:phase diagram theta dependence}
\end{figure*}

\subsection{Dependence of the normalized localization length on
$t^{2}$}

We continue exploring the phase diagram%
~\ref{fig: phase-diagram 3 parameter}(c)
by studying the $t^{2}$ dependence of the normalized localization lengths
$\Lambda^{\,}_{\perp}$ and $\Lambda^{\,}_{\parallel}$.

\subsubsection{Finite-size scaling for the 
normalized localization length with dimerization}

To begin with, we consider the
one-dimensional cut
\begin{equation}
 \mathrm{E:}\quad (t^{2},\pi/4,0.234),\quad t^{2} \in [0.234,0.766],
\label{eq:cut choice E}
\end{equation}
of the three-dimensional parameter space (\ref{eq: def parameter space
is 3D}). 
Cut E is shown in Fig.~\ref{fig:cut} with the label ``E''.
We note that this one-dimensional cut also contains
the critical point $(0.5,\pi/4,0.234)$ in Eq.%
~(\ref{eq: critical point along ct 1}).
Figures \ref{fig:Lambda dimerization t^{2} dependence}(a) 
and 
\ref{fig:Lambda dimerization t^{2} dependence}(b) show 
the normalized localization lengths
$\Lambda^{\,}_{\perp}$ and
$\Lambda^{\,}_{\parallel}$, respectively, as a function of $t^{2}$  along the
cut (\ref{eq:cut choice E}).
The dependence of 
$\Lambda^{\,}_{\perp}$ 
and
$\Lambda^{\,}_{\parallel}$ 
in Fig.~\ref{fig:Lambda dimerization t^{2} dependence} 
on the width $M$ 
is reversed compared to the one displayed 
in Fig.~\ref{fig:Lambda theta dependence}.
This is so because increasing $t^{2}$ along the cut E 
in Fig.~\ref{fig:cut}
drives a transition from the insulating to the metallic phase,
while increasing $\theta$ along the cut D
in Fig.~\ref{fig:cut}
drives a transition from the metallic to the insulating phase.
We also note for Fig.~\ref{fig:Lambda dimerization t^{2} dependence}(b) 
that, when the width $M$ is sufficiently small, 
the normalized localization length $\Lambda^{\,}_{\parallel}$ for $t^{2}<0.5$
is larger than that for $t^{2}>0.5$, while this relation
is inverted when $M$ is sufficiently large.
This observation is a manifestation of finite-size corrections to scaling.

\begin{figure}[t]
\centering
\includegraphics[width=0.45\textwidth]{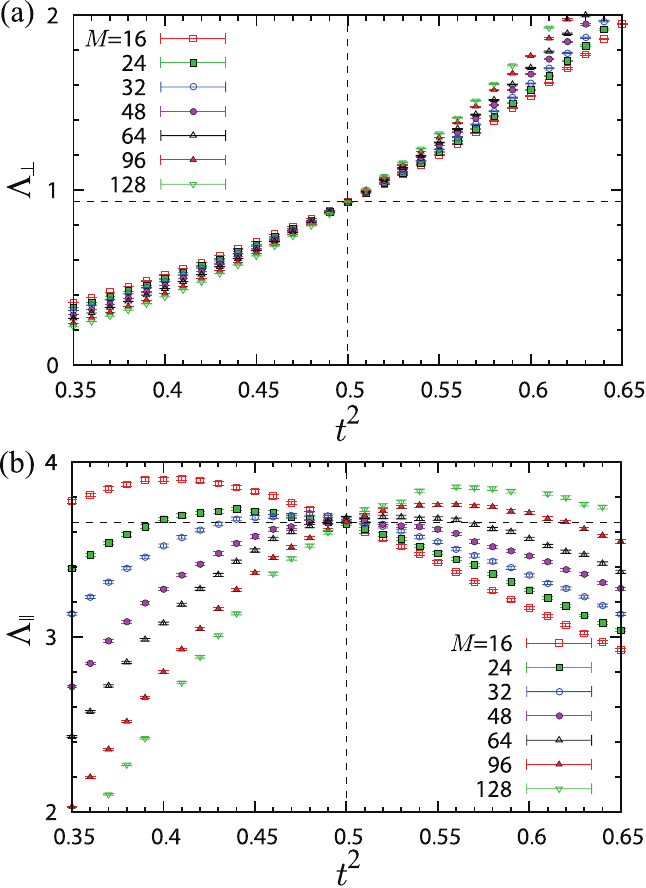}
\caption{
(Color online:)
The $t^{2}$ dependence of the normalized localization lengths
$\Lambda^{\,}_{\perp}$ in panel (a) and $\Lambda^{\,}_{\parallel}$ in panel (b) 
of the two-dimensional spin-directed $\mathbb{Z}^{\,}_{2}$ network model with
dimerization along the one-dimensional cut (\ref{eq:cut choice E})
at $\theta=\pi/4$ and $\delta^{2}=0.234$ for fixed $M$ ranging from 
$16$ to $128$.
The vertical and horizontal dashed lines represent 
$t^{2}_{\mathrm{c}}$ 
and
$\Lambda^{\,}_{\mathrm{x,c}}$ 
as deduced from the finite-size scaling analysis summarized 
in Table.~\ref{tab:summary_FSS}, respectively.
}
\label{fig:Lambda dimerization t^{2} dependence}
\end{figure}

In Fig.~\ref{fig:phase diagram theta dependence}(a), 
we have observed reentrance near $\theta\approx3\pi/8$ 
in the $t^2$ dependence of the
normalized localization lengths
$\Lambda^{\,}_{\perp}$ and $\Lambda^{\,}_{\parallel}$.
To examine in more detail reentrance driven by $t^{2}$,
we choose the one-dimensional cut
\begin{equation}
 \mathrm{F:}\quad (t^{2}, 19\pi/48,0.1),\quad t^{2} \in [0.1,0.9],
\label{eq:cut choice F}
\end{equation}
of the three-dimensional parameter space 
(\ref{eq: def parameter space is 3D}).
Cut F is shown in Fig.~\ref{fig:cut} with the label ``F''.
Cut F is also shown in Fig.~\ref{fig:phase diagram theta dependence}(a)
by the horizontal dashed line.

Figures \ref{fig:Lambda dimerization t^{2} dependence reentrant}(a) and
\ref{fig:Lambda dimerization t^{2} dependence reentrant}(b)
show the normalized localization lengths $\Lambda^{\,}_{\perp}$ and
$\Lambda^{\,}_{\parallel}$, respectively, along the cut 
(\ref{eq:cut choice F}).
Inspection of Fig.~\ref{fig:phase diagram theta dependence}(a)
shows that cut F comes close to an extended segment of
the boundary between the metallic and insulating phases
when $0.45\leq t^{2}\leq0.8$.
Hence, the proximity to the phase boundary of
cut F when $0.45\leq t^{2}\leq0.8$
implies that the dependence of the normalized
localization lengths on the width $M$ is weak
along this portion of cut F in Fig.~\ref{fig:cut}.
Nevertheless, the signature of a metal-to-insulator transition
upon increasing $t^{2}$ through $t^{2}\approx0.75$
and the signature of an insulator-to-metal transition
upon increasing $t^{2}$ through $t^{2}\approx 0.45$ 
are visible in Fig.%
~\ref{fig:Lambda dimerization t^{2} dependence reentrant}(a)
and in its inset, respectively. Reentrance is also visible
in Fig.~\ref{fig:Lambda dimerization t^{2} dependence reentrant}(b)
and in its inset.
We have also studied the normalized localization lengths along
other one-dimensional cuts parametrized by $t^{2}$ 
in the planes I, II, and III defined by
Eqs.\ (\ref{eq:two-dimensional cut I}),
(\ref{eq:two-dimensional cut II}),
and
(\ref{eq:two-dimensional cut III}),
respectively. We have opted not to present these results for lack of space.
It suffices to say that the dependence on $t^{2}$ along these cuts 
is compatible with the phase diagram
in Fig.\ \ref{fig:phase diagram theta dependence}.

\begin{figure}[t]
\centering
\includegraphics[width=0.45\textwidth]{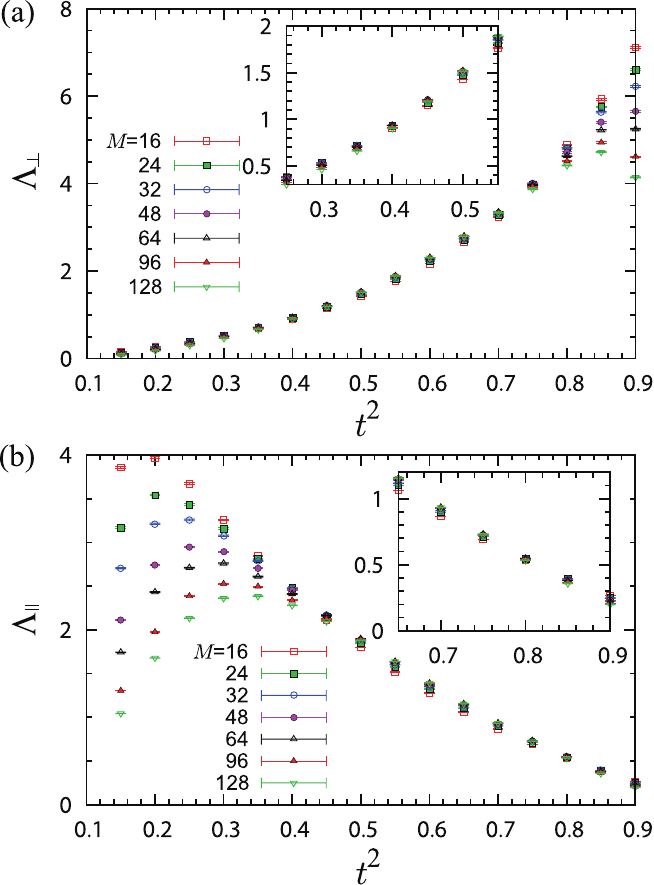}
\caption{
(Color online:)
The $t^{2}$ dependence of the normalized localization lengths
$\Lambda^{\,}_{\perp}$ in panel (a) and $\Lambda^{\,}_{\parallel}$ in panel (b) 
of the two-dimensional spin-directed $\mathbb{Z}^{\,}_{2}$ network model with
dimerization along the one-dimensional cut (\ref{eq:cut choice F})
at $\theta=19\pi/48$ and $\delta^{2}=0.1$ for fixed $M$ ranging from $16$
 to $128$.
Inset: Dependence on $t^{2}$ near $t^{2}\approx0.4$ for panel (a)
and near $t^{2}\approx0.8$ for panel (b).
}
\label{fig:Lambda dimerization t^{2} dependence reentrant}
\end{figure}

\begin{figure}[t]
\centering
\includegraphics[width=0.45\textwidth]{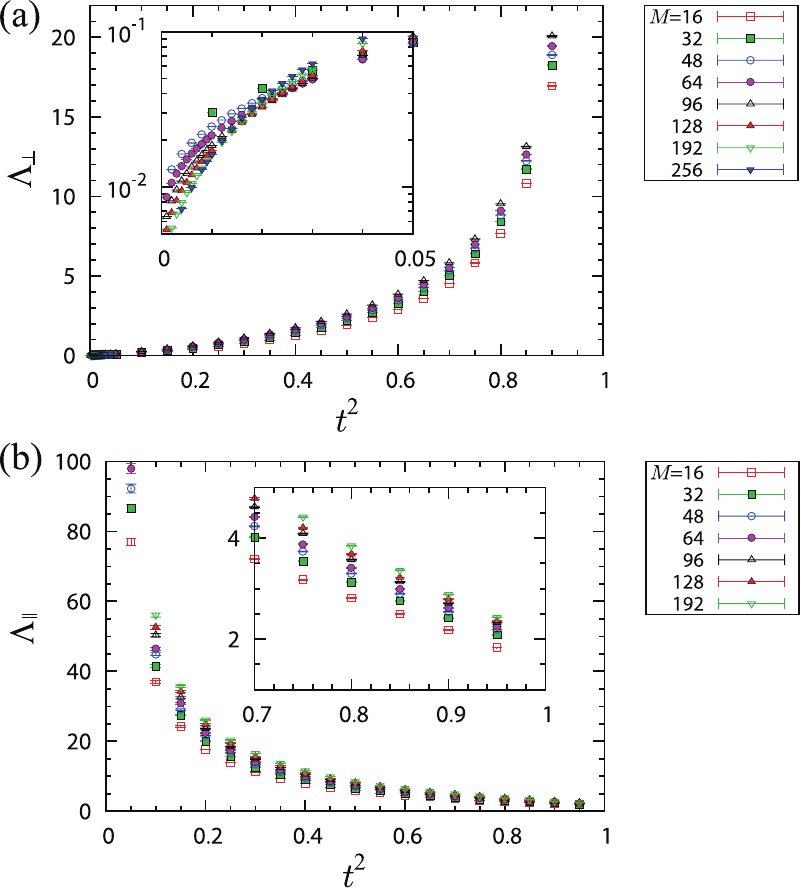}
\caption{ 
(Color online:)
The $t^{2}$ dependence of the normalized localization lengths
$\Lambda^{\,}_{\perp}$ in panel (a)
and $\Lambda^{\,}_{\parallel}$  in panel (b)
of the two-dimensional spin-directed $\mathbb{Z}^{\,}_{2}$  
network model without dimerization, at $\theta=\pi/4$,
and for fixed $M$ ranging from $16$ to $256$.
Inset: Dependence on $t^{2}$ near $t^{2}=0$ for panel (a) 
and near $t^{2}=1$ for panel (b).
         }
\label{fig:Lambda_no-staggering_PIover4}
\end{figure}

\subsubsection{Finite-size scaling for the 
normalized localization length without dimerization}
\label{sec:absence of dimerization}

We continue by studying surface states of a weak three-dimensional
$\mathbb{Z}^{\,}_{2}$ topological insulators without
dimerization, i.e., $\delta^{2}=0$ and we fix
$\theta$ to the value $\pi/4$.
This defines the one-dimensional cut
\begin{equation}
 \mathrm{G:}\quad (t^{2},\pi/4,0),\quad t^{2} \in [0,1],
\label{eq:cut choice G}
\end{equation}
of the three-dimensional parameter space 
~(\ref{eq: def parameter space is 3D}).
Cut G is shown in Fig.~\ref{fig:cut} with the label ``G''.
The dependence on $M$ of the normalized localization length
$\Lambda^{\,}_{\perp}$
and
$\Lambda^{\,}_{\parallel}$
corresponding to the geometries of Figs.%
~\ref{fig:network-staggered}(a)
and%
~\ref{fig:network-staggered}(b),
respectively,
as $t^{2}$ is increased along the interval $[0,1]$
is the following.

Figure~\ref{fig:Lambda_no-staggering_PIover4} 
shows the normalized localization length $\Lambda^{\,}_{\perp}$
in panel (a) and $\Lambda^{\,}_{\parallel}$ in panel (b)
as a function of $t^{2}$. 

On the one hand,
according to Fig.~\ref{fig:Lambda_no-staggering_PIover4}(b),
$\Lambda^{\,}_{\parallel}$ increases with increasing $M$ for 
values of $t^{2}$ ranging from 0.05 to 0.95.
This would be the signature for a metallic phase for these values
of $t^{2}$ if we could show that $\Lambda^{\,}_{\perp}$ 
also diverges in the limit $M\to\infty$. 
According to Fig.~\ref{fig:Lambda_no-staggering_PIover4}(a),
$\Lambda^{\,}_{\perp}$ also increases with increasing $M$ for 
$t^{2}$ larger than 0.05. We conclude that the
phase at $\delta^{2}=0$, $\theta=\pi/4$, and
for $0.05<t^{2}<0.95$ is metallic. 

For values of $t^{2}$ smaller than 0.05,
$\Lambda^{\,}_{\perp}$ decreases with increasing $M$
very close to $t^{2}=0$ according to the inset of
Fig.~\ref{fig:Lambda_no-staggering_PIover4}(a).
However, this apparent insulating dependence on $M$
of $\Lambda^{\,}_{\perp}$ for $t^{2}<0.05$ 
might be a finite-size artifact due to the fact that
$\Lambda^{\,}_{\perp}$ must vanish at 
$t^{2}=0$. 
In this scenario, the value of $M$ beyond which metallic dependence 
of $\Lambda^{\,}_{\perp}$ on $M$ is the rule diverges 
as $t^{2}\to0$. 
This scenario is consistent 
with the observation that the goodness of fit ${\mathcal{G}}$ in
Eq.~(\ref{eq:goodness of fit})
is the poorest
for $t^{2}<0.05$.
Moreover, the following argument supports this scenario
and the conclusion that the phase is
metallic inside the two-dimensional cut 
of parameter space at $\delta^{2}=0$.

When $t^{2}=0$, 
the two-dimensional spin-directed $\mathbb{Z}^{\,}_{2}$ 
network model realizes a unidirectional metal 
that consists of $M$ (even) independent
pairs of helical edge states
supporting the dimensionless conductance $M$ 
along their direction of propagation 
[the direction $\parallel$ in the geometry of 
Fig.~\ref{fig:Lambda_no-staggering_PIover4}(b)]
and a vanishing conductance in the orthogonal direction
[the direction $\perp$ in the geometry of 
Fig.~\ref{fig:Lambda_no-staggering_PIover4}(a)].
We first assume that 
$\lim_{M\to\infty}\Lambda^{\,}_{\perp}=0$,
where the limit $M\to\infty$ is taken with $M$ even, 
persists away from $t^{2}=0$ for sufficiently 
small values of $t^{2}$. We are going to show that 
this first assumption contradicts the second assumption
$\lim_{M\to\infty}\Lambda^{\,}_{\parallel}=\infty$,
where the limit $M\to\infty$ is taken with $M$ even,
for all values of $t^{2}$.
[The second assumption is supported empirically for
all the values of $t^{2}$ shown in Fig.%
~\ref{fig:Lambda_no-staggering_PIover4}(b),
whereas the first assumption is not unambiguously
supported by Fig.%
~\ref{fig:Lambda_no-staggering_PIover4}(a).]
These two assumptions are in mutual contradiction,
for the first assumption implies that 
the two-dimensional spin-directed $\mathbb{Z}^{\,}_{2}$ 
network model realizes a quasi-one-dimensional
quantum wire with an even number $2M$ of channels
in the symplectic symmetry class. Such a 
quasi-one-dimensional wire is
necessarily localized, i.e., 
$\lim_{M\to\infty}\Lambda^{\,}_{\parallel}=0$,
in contradiction with the second assumption.
Since our finite-size scaling analysis puts the
second assumption on firmer ground
than the first assumption, we conclude that
a two-dimensional metallic phase in the symplectic
symmetry class is established for any non-vanishing
$t^{2}$. 

\subsection{Dependence of the normalized localization length on $\delta \phi$}
\label{subsec:delta phi dependence}

So far, we have focused on the maximally disordered case by setting 
$\delta\phi=2\pi$ in Eq.\ (\ref{eq: def parameter space is 3D}).
It is time to investigate how the normalized localization lengths
$\Lambda^{\,}_{\perp}$ 
and
$\Lambda^{\,}_{\parallel}$ 
at the point~(\ref{eq: critical point along ct 1}) 
marked by the triangle in Fig.~\ref{fig:cut}
depends on the width $\delta\phi$ of the random phases
defined in Eq.~(\ref{eq: def parameter space is 4D}). 
We recall that the triangle in Fig.~\ref{fig:cut}
is the critical point defined
by the intersection of the cuts A, D, and E
when $\delta\phi=2\pi$.

The dependence on $0\leq\delta\phi\leq2\pi$ 
of the normalized localization lengths $\Lambda^{\,}_{\perp}$  
and $\Lambda^{\,}_{\parallel}$, respectively,
at the point~(\ref{eq: critical point along ct 1})
is shown in 
Figs.~\ref{fig:Lambda dimerization delta phi dependence}(a) 
and 
\ref{fig:Lambda dimerization delta phi dependence}(b).
We observe that
$\Lambda^{\,}_{\perp}$ and $\Lambda^{\,}_{\parallel}$ 
are both increasing functions of $0\leq\delta\phi<2\pi$
that appear to converge to the critical values of 
$\Lambda^{\mathrm{c}}_{\perp}$ 
and
$\Lambda^{\mathrm{c}}_{\parallel}$, respectively,
at the critical point~(\ref{eq: critical point along ct 1})
when $\delta\phi=2\pi$.
Hence, disorder favors delocalization over localization
for the surface states of weak
three-dimensional $\mathbb{Z}^{\,}_{2}$ topological insulators.
For any fixed $0\leq\delta\phi<2\pi$, we also observe that
the normalized localization lengths 
$\Lambda^{\,}_{\perp}$  and $\Lambda^{\,}_{\parallel}$
both decrease with increasing the width $M$.
This insulating behavior is caused by the finite dimerization
$\delta^2=0.23$.
Hence, we deduce that the cut
\begin{equation}
\mathrm{H}:\quad(t^{2},\theta,\delta\phi,\delta^{2}):=(0.5,\pi/4,\delta\phi,0.23),
\quad
0\leq\delta\phi<2\pi
\end{equation}
realizes an insulating phase.
Along the same lines, we expect that
the phase diagrams presented in 
Fig.~\ref{fig:phase diagram theta dependence}
are qualitatively correct for a weaker disorder $0<\delta\phi<2\pi$
than $\delta\phi=2\pi$, 
albeit with a smaller metallic region.

\begin{figure}[t]
\centering
\includegraphics[width=0.45\textwidth]{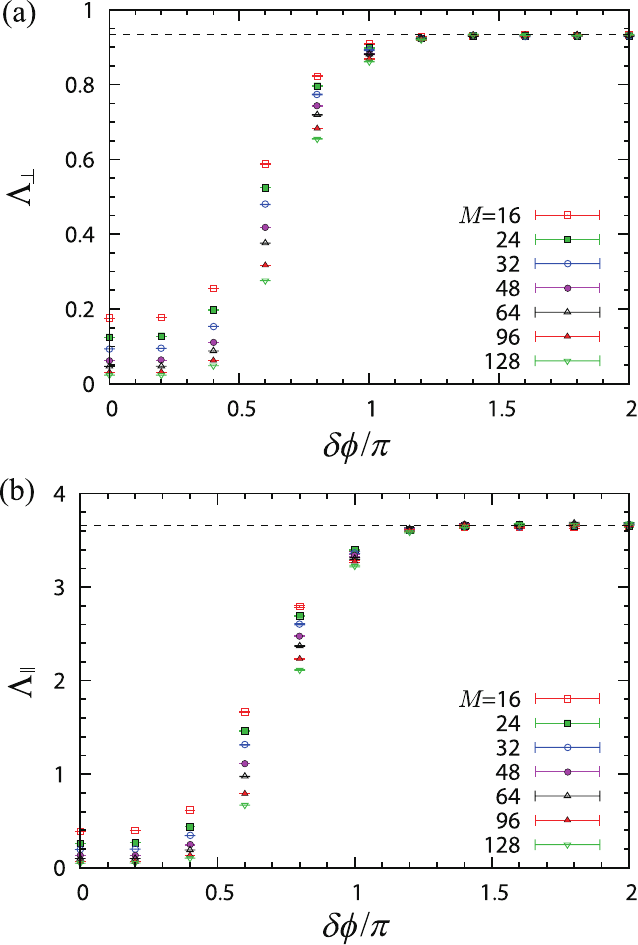}\\
\caption{
(Color online:)
The $\delta \phi$ dependence of the normalized localization lengths
$\Lambda^{\,}_{\perp}$ in panel (a) and $\Lambda^{\,}_{\parallel}$ in panel (b) 
of the two-dimensional spin-directed $\mathbb{Z}^{\,}_{2}$ network model
at $t^{2}=0.5$, $\theta=\pi/4$ and $\delta^{2}=0.234$
for fixed $M$ ranging from $16$ to $128$.
The horizontal dashed lines in panels (a) and (b) represent 
$\Lambda^{\mathrm{c}}_{\perp}=0.935$ 
and
$\Lambda^{\mathrm{c}}_{\parallel}=3.657$, respectively.
}
\label{fig:Lambda dimerization delta phi dependence}
\end{figure}

\section{Numerical data for an odd number of dimerized channels}
\label{Numerical data with an odd number of channels}

As was emphasized in Refs.%
~\onlinecite{Ringel11},%
~\onlinecite{Imura12},
and~\onlinecite{Yoshimura13},
weak three-dimensional $\mathbb{Z}^{\,}_{2}$ topological insulators
are characterized by a dependence on the parity in the stacking
number of strong two-dimensional $\mathbb{Z}^{\,}_{2}$ topological insulators.
This parity effect can be illustrated in the context of
the two-dimensional spin-directed $\mathbb{Z}^{\,}_{2}$ network model
by weakly perturbing the quasi-one-dimensional symplectic metallic fixed point%
~(\ref{eq: delta=t=0})
from Sec.~\ref{subsubsec: Limit t2=0 without dimerization}
in the limit $M$ fixed and $L\to\infty$.
In this quasi-one-dimensional limit, the localization properties of 
the two-dimensional spin-directed $\mathbb{Z}^{\,}_{2}$ network model
depend on the parity of $M$. If $M=2M'$ is even, as we have
assumed all along so far, then the transfer matrix belongs
to the Lie group $SO^{*}(4M')$ and exponential localization
is the rule.%
~\cite{Brouwer96}
If $M=2M'+1$ is odd, the transfer matrix belongs
to the Lie group $SO^{*}(4M'+2)$ and there is one 
pair of Kramers degenerate helical quasi-one-dimensional channels 
that is perfectly conducting.%
~\cite{Takane04,Suzuura02,Ando02}

\begin{figure}[t]
\centering
\includegraphics[width=0.3\textwidth]{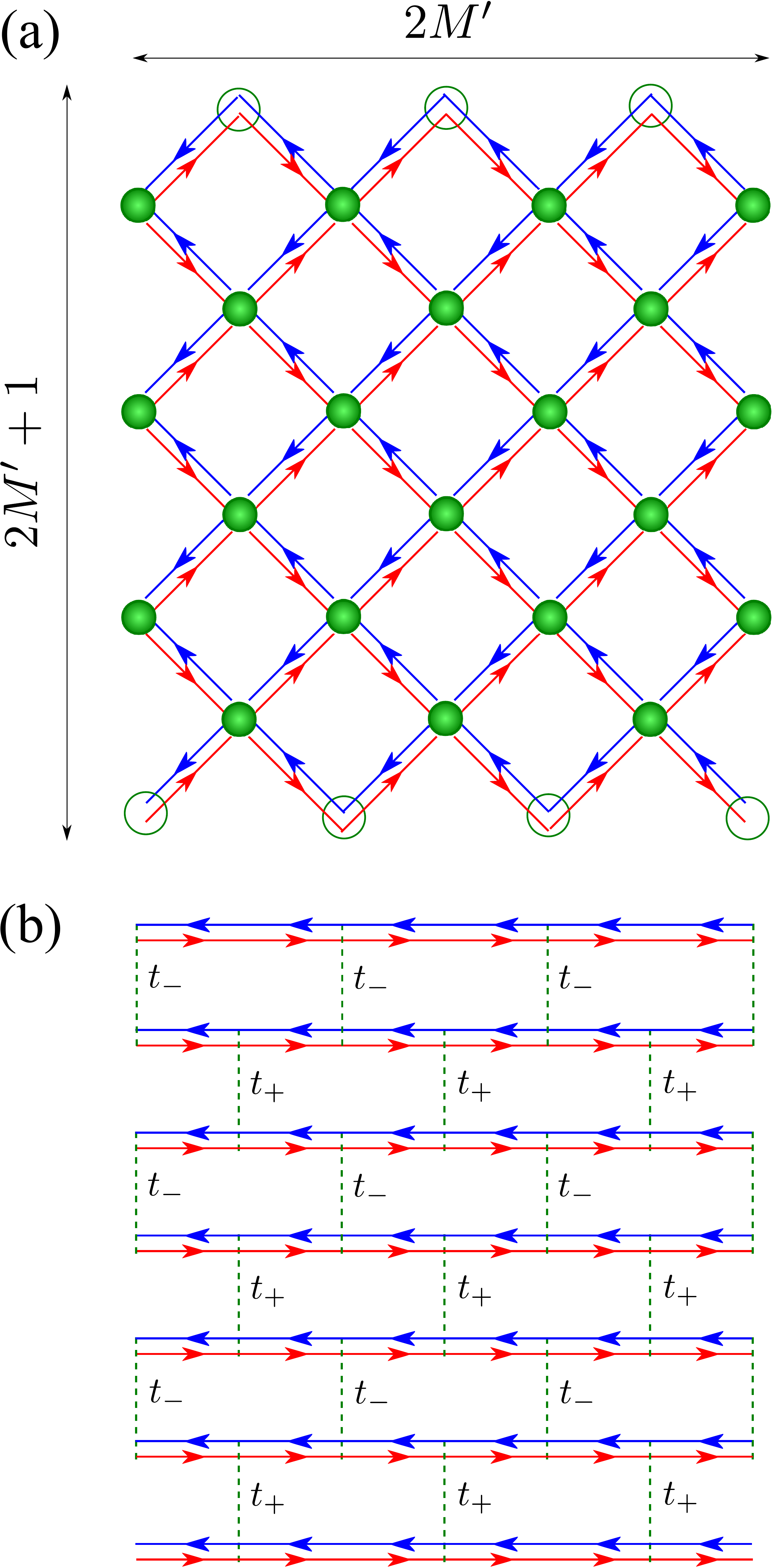}
\caption{
(Color online:)
A two-dimensional spin-directed $\mathbb{Z}^{\,}_{2}$ network model with
the dimensions $(2M^{\prime}+1)\times 2M^{\prime}$ 
where $M^{\prime}=3$ 
in the vertex (a) and the brick-wall representations (b),
respectively. 
The difference with the
two-dimensional spin-directed $\mathbb{Z}^{\,}_{2}$ network model 
shown in Fig.~\ref{fig:network directed Z2}(a) 
is the addition of a bottom and a top row of vertices represented by
empty circles at which perfectly reflecting boundary conditions
must be imposed in panel (a).
        }
\label{fig:network directed Z2 odd}
\end{figure}

We are going to derive this parity effect
for the two-dimensional spin-directed $\mathbb{Z}^{\,}_{2}$ network model
shown in Fig.~\ref{fig:network directed Z2 odd} that represents
surface states from stacked layers of
an odd number of strong two-dimensional $\mathbb{Z}^{\,}_{2}$
topological insulators in Fig.~\ref{fig:layeredz2pancake}.
By comparing
Fig.~\ref{fig:network directed Z2}(a)
with
Fig.~\ref{fig:network directed Z2 odd}(a),
we observe that two rows of vertices denoted by empty circles were
added in Fig.~\ref{fig:network directed Z2 odd}(a)
at the bottom and at the top of
Fig.~\ref{fig:network directed Z2}(a).
Correspondingly, all vertices represented by filled circles colored in green in
Fig.~\ref{fig:network directed Z2 odd}(a)
correspond to the elementary scattering process 
between four incoming and four outgoing plane waves 
defined in Eq.~(\ref{eq:S-matrix-dimer}),
while all vertices represented by empty circles in
Fig.~\ref{fig:network directed Z2 odd}(a)
correspond to the perfectly reflecting boundary condition 
\begin{equation}
\begin{pmatrix}
\psi^{(o)}_{\uparrow} \\ \psi^{(o)}_{\downarrow}
\end{pmatrix}
=
S^{\,}_{\mathrm{ref}}\,
\begin{pmatrix}
\psi^{(i)}_{\uparrow} \\ \psi^{(i)}_{\downarrow}
\end{pmatrix},
\qquad
S^{\,}_\mathrm{ref}:=
e^{{i}\phi}\,
\sigma^{\,}_{0},
\label{eq:reflecting boundary condition}
\end{equation}
where $0\le\phi\le2\pi$.
Fig.~\ref{fig:network directed Z2 odd}(a).

The total transfer matrix $\mathcal{M}^{\,}_{\parallel,\mathrm{tot}}$ 
that defines a two-dimensional spin-directed 
$\mathbb{Z}^{\,}_{2}$ network model 
with $2M^{\prime}+1$ Kramers' pairs of conducting channels
is built from the elementary 
$4\times4$ transfer matrix $\mathcal{M}^{\,}_{\parallel}$ 
defined in Eq.\ (\ref{eq: T parallel}) and the elementary 
$2\times2$ transfer matrix defined by
\begin{equation}
\begin{pmatrix}
\psi^{(o)}_{\uparrow} \\ \psi^{(i)}_{\downarrow}
\end{pmatrix}
= 
\mathcal{M}^{\,}_{\mathrm{ref}}
\begin{pmatrix}
\psi^{(i)}_{\uparrow} \\ \psi^{(o)}_{\downarrow}
\end{pmatrix},
\qquad
\mathcal{M}^{\,}_{\mathrm{ref}}
:=
\begin{pmatrix}
e^{+{i}\phi} & 0 \\
0 & e^{-{i}\phi}
\end{pmatrix}.
\label{eq:transfer matrix reflecting boundary condition}
\end{equation}
The matrix $\mathcal{M}^{\,}_{\parallel,\mathrm{tot}}$
belongs to the Lie group $SO^*(4M^{\prime}+2)$.
Hence, the matrix 
$\mathcal{M}^{\dag}_{\parallel,\mathrm{tot}}\,\mathcal{M}^{\,}_{\parallel,\mathrm{tot}}$ 
has doubly degenerate eigenvalues of the form 
$\exp(\pm 2X^{\,}_{\parallel,j})$ with $j=1,\cdots,M/2$ 
as well as doubly degenerate
eigenvalues of the form
$\exp(2X^{\,}_{\parallel,0})=1$, 
i.e., $X^{\,}_{\parallel,0}=0$.
We shall assume the convention
\begin{equation}
 0=X^{\,}_{\parallel,0}<X^{\,}_{\parallel,1}<\cdots<X^{\,}_{\parallel,M/2}
\end{equation}
when ordering the Lyapunov exponents.

We have computed numerically
the Lyapunov exponents $X^{\,}_{\parallel,j}$ by the transfer matrix
method.~\cite{MacKinnon83} 
In the quasi-one-dimensional limit, as anticipated, 
we have found the doubly degenerate eigenvalues $X^{\,}_{\parallel,0}=0$ for 
a sampling of points from the 
three-dimensional parameter space 
(\ref{eq: def parameter space is 3D}). 
Since the largest localization length is 
nothing but the inverse of
the Lyapunov exponent $X^{\,}_{\parallel,0}$, 
the fact $X^{\,}_{\parallel,0}=0$
is interpreted as the existence of 
a perfectly conducting pair 
of Kramers' degenerate quasi-one-dimensional channels
associated to an infinite localization length anywhere in the 
three-dimensional parameter space
(\ref{eq: def parameter space is 3D}).
In other words, 
stacking an odd number of strong two-dimensional $\mathbb{Z}^{\,}_{2}$
topological insulators in Fig.~\ref{fig:layeredz2pancake}
always delivers 
a single perfectly conducting pair
of Kramers' degenerate quasi-one-dimensional channels 
for arbitrary dimerization and arbitrary local disorder 
that preserve time-reversal symmetry.
The same result was also obtained in Ref.~\onlinecite{Ringel11}.
This implies that, even in the insulating phase in the
phase diagram of Fig.~\ref{fig: phase-diagram 3 parameter},
there exists a single pair
of Kramers' degenerate quasi-one-dimensional channels 
of perfect conduction as long as $M$ is odd. Hereby, 
surface states of a weak three-dimensional $\mathbb{Z}^{\,}_{2}$ 
topological insulator with an odd number of stacking layers 
support, in the quasi-one-dimensional limit,
a perfectly conducting pair 
of Kramers' degenerate quasi-one-dimensional channels, 
even in the presence of dimerizations. 

We have also applied the finite-size scaling analysis 
encoded by Eqs.\ (\ref{eq: def xi}) and (\ref{eq: def Lambda})
to the normalized localization length $\Lambda^{\,}_{\parallel}$ obtained from
the \textit{second} smallest Lyapunov exponent $X^{\,}_{\parallel,1}$.
The $\delta^{2}$ dependence of the normalized localization length 
$\Lambda^{\,}_{\parallel}$ 
along the one-dimensional cut ``A'' defined in 
Eq.\ (\ref{eq: cut choice 1}) for various width $M$
is summarized in Fig.~\ref{fig:Lambda odd dimerization}.
Although finite-size corrections are more pronounced 
for an odd stacking than for an even stacking 
in Fig.~\ref{fig:layeredz2pancake},
the first subleading normalized localization length 
$\Lambda^{\,}_{\parallel}$ 
undergoes a metal-insulator transition at 
$\delta^{2}_{\mathrm{c}}\approx0.234$
belonging to the same universality class as  that occurring
for an even stacking.
This suggests that a ``two-fluid picture'' applies when stacking
an odd number of strong two-dimensional $\mathbb{Z}^{\,}_{2}$
topological insulators in Fig.~\ref{fig:layeredz2pancake}
in the quasi-one-dimensional limit.
While there exists a perfectly conducting 
quasi-one-dimensional channel,
the localization properties
of all remaining quasi-one-dimensional channels are those of 
an even number of stacked strong two-dimensional $\mathbb{Z}^{\,}_{2}$
topological insulators in Fig.~\ref{fig:layeredz2pancake}.

\begin{figure}[t]
\centering
\includegraphics[width=0.4\textwidth]{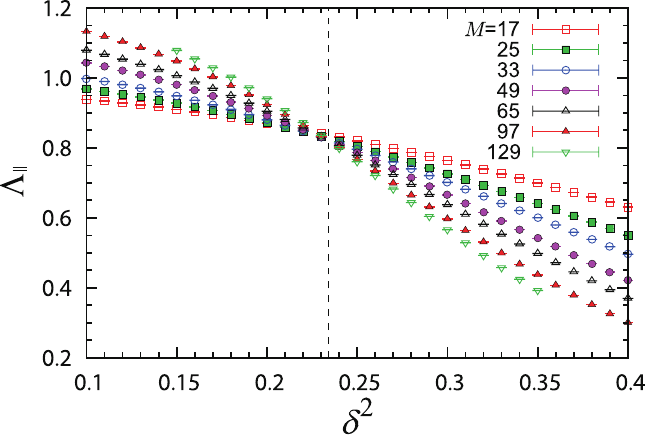}
\caption{
(Color online:)
The $\delta^{2}$ dependence of the normalized localization length 
$\Lambda^{\,}_{\parallel}$
calculated from the second smallest positive Lyapunov exponent 
$X^{\,}_{\parallel,1}$ for the spin-directed 
$\mathbb{Z}^{\,}_{2}$ network model with an odd number channels. 
The parameter $t^{2}$ and $\theta$ are fixed to $t^{2}=0.5$ and
$\theta=\pi/4$, so as to facilitate comparison with
Fig.~\ref{fig:Lambda dimerization 0.5 PIover4} 
when the number of channels is even, while all other
parameters are the same. 
The vertical dashed line represents $\delta^{2}_\mathrm{c}=0.234$,
as deduced from the finite-size scaling analysis summarized 
in Table.~\ref{tab:summary_FSS}.
}
\label{fig:Lambda odd dimerization}
\end{figure}

\section{Numerical data with trimerization}
\label{sec: Numerical data with trimerization}

The parity effect discussed in
Sec.~\ref{Numerical data with an odd number of channels} 
is also the reason for which 
the spin-directed $\mathbb{Z}^{\,}_{2}$ network model
with trimerization shown in
Fig.~\ref{fig:network-trimer} 
is always delivering a metallic phase.
We first defend this assertion using a qualitative argument.
We then present numerical results in support of this assertion.

\subsection{Qualitative argument}

In the limit of $t^{\,}_{+}\gg t^{\,}_{-}$,
we may replace the three pairs of helical modes in a trimer
by a single effective pair of helical modes, 
which is then coupled by $t^{\,}_{-}$ to its neighboring effective pairs of
helical modes. Thus, the 
two-dimensional trimerized spin-directed $\mathbb{Z}^{\,}_{2}$ network model
reduces to a two-dimensional spin-directed $\mathbb{Z}^{\,}_{2}$ network model
without polymerization in this limit.
In the opposite limit $t^{\,}_{+}\ll t^{\,}_{-}$, any two pairs of 
helical modes coupled by $t^{\,}_{-}$ become inert (localized) 
and the remaining pairs of helical modes are weakly coupled
without polymerization. The same conclusion follows from
the point of view of surface states realizing an even number of
Dirac cones. Scattering matrix elements are needed that couple pairwise the 
surface Dirac cones in order to localize the Dirac modes.
Trimerization does not deliver such matrix elements.

From the examples of dimerization and trimerization,
we conjecture the following parity effect.
The combined effects of polymerization and disorder for
the two-dimensional spin-directed $\mathbb{Z}^{\,}_{2}$ network model
produces a phase diagram with either (i) two insulating phases separated by
a metallic phase when the breaking of translation symmetry
involves a repeat unit cell consisting of an even number of helical modes 
in the clean limit,
(ii) or a single metallic phase when the breaking of translation symmetry
involves a repeat unit cell consisting of an odd number of helical modes 
in the clean limit.

\begin{figure}[t]
\centering
\includegraphics[width=0.4\textwidth]{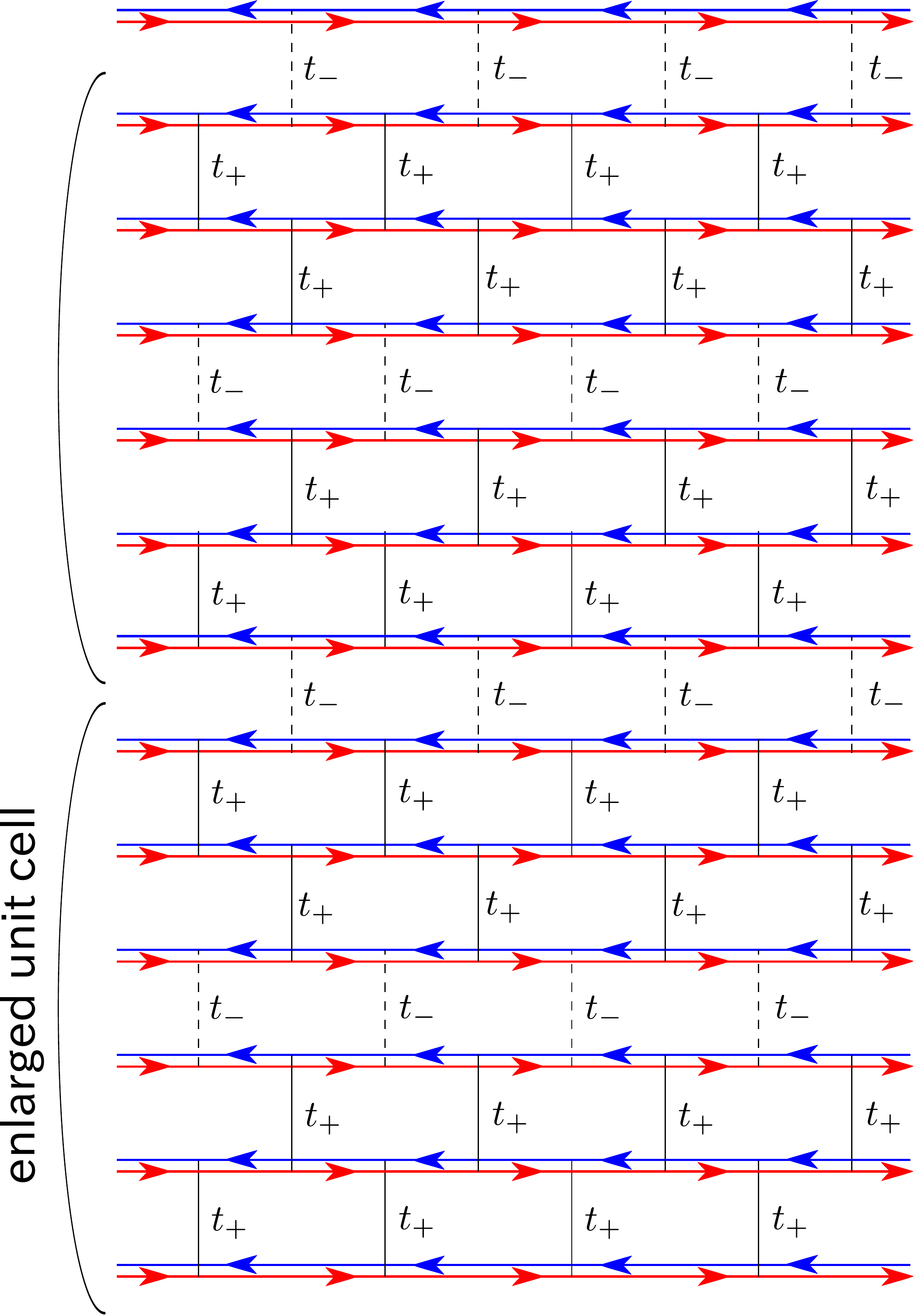}
\caption{
(Color online:)
Two-dimensional spin-directed $\mathbb{Z}^{\,}_{2}$ network model
with $M=6M'$ an even integer multiple of 3. The periodic pattern for
the transmission amplitudes implements a trimerization, as is
indicated by the enlarged unit cell.
        }
\label{fig:network-trimer}
\end{figure}

\subsection{Numerics}

\subsubsection{Transfer matrix}

Two transfer matrices 
$\mathcal{M}^{\,}_{\perp}$
and
$\mathcal{M}^{\,}_{\parallel}$
are defined as in Sec.%
~\ref{subsec: Transfer matrix},
except for the pattern of trimerization
shown in Fig.~\ref{fig:network-trimer}
for the transmission amplitude.

\subsubsection{Definition of the normalized localization length}
\label{subsec: Definition of the normalized localization length trimerization}

The normalized localization lengths
$\Lambda^{\,}_{\perp}$
and
$\Lambda^{\,}_{\parallel}$
are defined as in Sec.%
~\ref{subsec: Definition of the normalized localization length}, 
except for the pattern of trimerization
shown in Fig.~\ref{fig:network-trimer}
for the transmission amplitude.

\begin{figure}[t]
\centering
\includegraphics[width=0.45\textwidth]{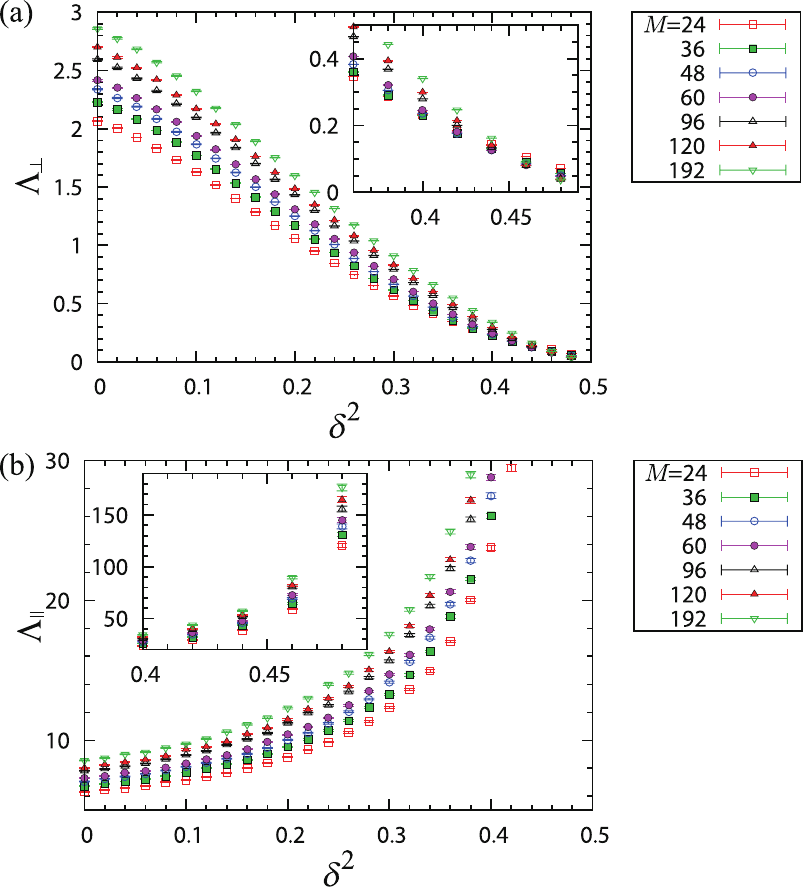}
\caption{
(Color online:) 
Combined effects of trimerization and disorder for a
two-dimensional spin-directed $\mathbb{Z}^{\,}_{2}$ network model.
The $\delta^{2}$ dependence of the normalized localization length
at $t^{2}=0.5$ and $\theta=\pi/4$
for fixed $M$ ranging from $24$ to $192$
is shown in panel (a) for $\Lambda^{\,}_{\perp}$ 
and in panel (b) for $\Lambda^{\,}_{\parallel}$.
Inset: Dependence on $\delta^{2}$ near $\delta^{2}=0.5$ for panel (a)
and for panel (b).
       }
\label{fig:Lambda_trimerization_0.5_PIover4}
\end{figure}

\subsubsection{Finite-size scaling for the 
normalized localization length in the presence of trimerization}
\label{subsec: Finite-size scaling for the ... trimerization}

Figure \ref{fig:Lambda_trimerization_0.5_PIover4} 
shows the $\delta^{2}$ dependence of the
normalized localization lengths
$\Lambda^{\,}_{\perp}$
and
$\Lambda^{\,}_{\parallel}$
along the one-dimensional cut (\ref{eq: cut choice 1})
with $t^{2}_{\pm}=0.5\pm\delta^{2}$ and $\theta=\pi/4$
in the two-dimensional trimerized spin-directed 
$\mathbb{Z}^{\,}_{2}$ network model.

According to Fig.%
~\ref{fig:Lambda_trimerization_0.5_PIover4}(a),
the normalized localization length $\Lambda^{\,}_{\perp}$
at fixed $t^{2}_{\pm}=0.5\pm\delta^{2}$, $\theta=\pi/4$, and $\delta^{2}$
increases with increasing $M$, whereas it is a decreasing function
with increasing $\delta^{2}$ 
at fixed $t^{2}_{\pm}=0.5\pm\delta^{2}$, $\theta=\pi/4$, and $M$.
The latter decrease of $\Lambda^{\,}_{\perp}$ with increasing
$\delta^{2}$ is expected since 
$\Lambda^{\,}_{\perp}=0$ because of
$t^{2}_{-}=0$ at $\delta^{2}=0.5$.
A signature of the point $t^{2}_{-}=0$
is visible in the inset of Fig.%
~\ref{fig:Lambda_trimerization_0.5_PIover4}(a)
in which the dependence of $\Lambda^{\,}_{\perp}$ 
on $0.25<\delta^{2}<0.5$ for fixed $M$ ranging from 24 to 192 is shown.
Indeed, upon approaching from below $\delta^{2}=0.5$,
the dependence of $\Lambda^{\,}_{\perp}$ on $M$ undergoes a crossover from
monotonically increasing to monotonically decreasing. We attribute this
fact to the same finite-size artifact discussed in the last paragraph of
Sec.~\ref{sec:absence of dimerization} that is responsible
for a similar crossover of the  dependence of $\Lambda^{\,}_{\perp}$ on $M$
in the inset of Fig.~\ref{fig:Lambda_no-staggering_PIover4}(a)
upon approaching from above $t^{2}=0$.

According to Fig.%
~\ref{fig:Lambda_trimerization_0.5_PIover4}(b),
the normalized localization length $\Lambda^{\,}_{\parallel}$
at fixed $t^{2}_{\pm}=0.5\pm\delta^{2}$, $\theta=\pi/4$, and $\delta^{2}$
increases with increasing $M$. Moreover, it is an increasing function
of $\delta^{2}$ at fixed $t^{2}_{\pm}=0.5\pm\delta^{2}$, $\theta=\pi/4$,
and $M$.

Thus, there is no sign of a transition from a metallic to an insulating phase
in the two-dimensional trimerized spin-directed $\mathbb{Z}^{\,}_{2}$ network model
upon increasing the value of $\delta^{2}$, as was the case for
the two-dimensional dimerized spin-directed $\mathbb{Z}^{\,}_{2}$ network model
at the critical point~(\ref{eq: critical point along ct 1}),
see Fig.~\ref{fig:Lambda dimerization 0.5 PIover4}.
The absence of the insulating phase in the two-dimensional spin-directed
$\mathbb{Z}^{\,}_{2}$ network model with trimerizations 
has also been confirmed by studying the 
$\theta$ and $t^{2}$ dependences along the cuts (\ref{eq: cut choice D}) and
(\ref{eq:cut choice E}), respectively, of the normalized localization lengths
$\Lambda^{\,}_{\perp}$ and $\Lambda^{\,}_{\parallel}$.

\section{Summary and discussion}
\label{Sec: discussion} 

We have shown that the surface states of
a weak three-dimensional $\mathbb{Z}^{\,}_{2}$ topological insulator
can be modeled by 
a two-dimensional spin-directed $\mathbb{Z}^{\,}_{2}$ network model. 
In other words, a two-dimensional spin-directed $\mathbb{Z}^{\,}_{2}$ network model
can be interpreted as an effective lattice regularization
for the surface states of a weak three-dimensional 
$\mathbb{Z}^{\,}_{2}$ topological insulator.
The qualifier $\mathbb{Z}^{\,}_{2}$ indicates here 
that time-reversal symmetry is present, but $SU(2)$
spin-rotation symmetry is broken. 

We have studied the combined effects of polymerization and disorder
in a two-dimensional spin-directed $\mathbb{Z}^{\,}_{2}$ network model.
Polymerization implies that the lattice symmetry group
$\mathfrak{G}$ of 
the two-dimensional spin-directed $\mathbb{Z}^{\,}_{2}$ network model
is reduced to a nontrivial subgroup $\mathfrak{G}'$
in the clean limit. 

On the one hand,
if the polymerization opens a spectral gap in the clean limit
and if the disorder strength is weak relative to the polymerization gap,
single-particle states are localized. As the ratio
of the disorder strength to the polymerization gap is increased,
a quantum phase transition from an insulating to a metallic phase
takes place. This transition is smooth and we have shown that it
belongs to the two-dimensional symplectic universality class
in the theory of Anderson localization, as measured by the 
power-law divergence of the localization length with the
scaling exponent $\nu\approx2.7$. This metallic phase
is connected to the critical point that separates the two
polymerized-gapped phases in the clean limit.

On the other hand,
if the polymerization does not open a spectral gap in the clean limit,
i.e., if the pattern of symmetry breaking
$\mathfrak{G}\to\mathfrak{G}^{\prime}$ is associated to an enlarged unit cell
of the two-dimensional spin-directed $\mathbb{Z}^{\,}_{2}$ network model
that is built out of an odd integer number of the unit cell prior 
to switching on polymerization,
then the metallic phase is robust to any short-range correlated disorder.

The two-dimensional spin-directed $\mathbb{Z}^{\,}_{2}$ network model 
studied in this paper is the second example of a 
two-dimensional directed network model
after that of the directed CC network model.
Similarly to the directed CC network model, it is
an effective lattice model that captures some
low-energy and long-wave length properties of
surface states of weak three-dimensional topological insulators,
such as the universal properties of a quantum phase
transition from the theory of Anderson localization.
There is an important difference with regard to
the long-distance physics of the two classes of network models, however.
Charge transport for the directed CC network model is
intrinsically anisotropic:
ballistic chiral transport in one
direction and critical (diffusive) transport in the other direction.
By contrast, charge transport for
the spin-directed $\mathbb{Z}^{\,}_{2}$ network model 
is effectively isotropic at long wave lengths, as is illustrated
after a rescaling of velocities in the limiting Dirac Hamiltonians.

There exists a network model for
each of the ten symmetry classes in the theory of 
Anderson localization.%
~\cite{Kramer05}
Regardless of the dimensionality of space,
five of these network models encode the effects of
disorder on strong topological insulators, 
noninteracting insulators with response functions whose
topological character are protected by symmetry.%
~\cite{Schnyder08,Kitaev09,Ryu10b}
By stacking and weakly coupling a family of
strong two-dimensional topological insulators
from a given symmetry class, one obtains
a weak three-dimensional topological insulator.
We conjecture that the combined effects of breaking the stacking
symmetry in a periodic way (polymerization)
and disorder on the phase diagram of
a weak three-dimensional topological insulator from a given
symmetry class are captured by
a two-dimensional ``directed'' network model built out of an
elementary scattering matrix within this symmetry class.
As we have seen for 
the two-dimensional spin-directed $\mathbb{Z}^{\,}_{2}$ network model
of this paper, and as expected from theoretical considerations,%
~\cite{Nomura08,Morimoto13,Yao13,Chiu13,Fulga12}
unpolymerized network models are not expected to support insulating phases.
Furthermore, the dichotomy between anisotropic versus isotropic
transport is expected to apply to 
the three symmetry classes with a $\mathbb{Z}$ index 
(the symmetry classes A, C, and D) and
the two symmetry classes with a $\mathbb{Z}^{\,}_{2}$ index 
(the symmetry classes AII and CII),
respectively.%
~\cite{Morimoto13}

An outstanding open problem is the interplay of disorder
and interactions for the surface states of 
three-dimensional weak $\mathbb{Z}^{\,}_{2}$ topological
insulators, given the fact that interactions can stabilize
states of matter that fall outside the classification
of noninteracting topological insulators.%
~\cite{Wu06,Xu06,Levin09,Fidkowski10,Fidkowski11,Turner11,Wang11,Neupert11,%
Ryu12a,Ryu12b,Chen13,Gu12,Qi13,Yao13}

\section*{Acknowledgments}

We thank Jens Bardarson for helpful discussions.
S. R., A. F., and C. M. thank the
KITP program Topological Insulators and Superconductors 
for hospitality where this work was initiated.
H. O. was supported by Grant-in-Aid (Nos.\ 25800213 and 25390113) from
the Japan Society for Promotion of Science.
The work of A. F. was partly supported by a Grant-in-Aid
(No.\ 24540338) from the Japan Society for Promotion of Science
and by the RIKEN iTHES project.


\begin{appendices}

\renewcommand{\theequation}{A\arabic{equation}}
\section{
Quasi-one-dimensional model for the surface states of a
weak three-dimensional $\mathbb{Z}^{\,}_{2}$ topological insulator
        }
\label{appsec: Wire construction for the surface states of a 3D Z2 WTI}

Figure~\ref{fig:z2pancake} 
depicts a two-dimensional $\mathbb{Z}^{\,}_{2}$
topological \textit{band} insulator. 
The interior of the ellipse shown
in Fig.~\ref{fig:z2pancake}(a) 
is called the bulk.
The boundary of the ellipse shown
in Fig.~\ref{fig:z2pancake}(a) 
is the edge. 
Figure~\ref{fig:z2pancake}
represents a model of noninteracting electrons such that
(i) the single-particle eigenstates with support in the bulk 
display a spectral gap $\Delta$
as is indicated in  Fig.~\ref{fig:z2pancake}(b), 
while (ii) the single-particle eigenstates with support on the edge
realize a two-fold degenerate dispersion that crosses the spectral
gap of the bulk states as is indicated in Fig.~\ref{fig:z2pancake}(b), 
are extended along the edge, but are exponentially localized in the direction
perpendicular to the edge. These edge states represent a single pair of
Kramers' degenerate electrons propagating with opposite velocities.
These edge states are also called helical states as the expectation values
of the electronic spins are opposite for each electron forming the 
Kramers' degenerate pair and change with the momenta $k$
of the electrons parallel to the edge.
The low-energy and long-wave-length effective Hamiltonian of the single pair
of helical states depicted in Fig.~\ref{fig:z2pancake}(a) is
\begin{subequations}
\begin{equation}
\hat{H}^{\,}_{\mathrm{Helical}}:=
\int\limits_{\mathrm{edge}}\mathrm{{d}}x\,
\left(
\hat{\Psi}^{\dag}
(-{i})\,
v\,
\sigma^{\,}_{3}\,
\partial^{\,}_{x}\,
\hat{\Psi}
\right)(x).
\label{eq: def single z2pancake hamiltonian}
\end{equation}
Units are chosen so that $\hbar=1$. The speed $v$ is positive
by convention. The operators
$\hat{\Psi}^{\dag}_{\alpha}(x)$
and
$\hat{\Psi}^{\,}_{\alpha}(x)$
create and destroy at the position $x$ along the edge
an electron with the projection
$\alpha=\uparrow,\downarrow$
of its spin along the spin quantization axis, respectively.
They make up the doublet of operators
$\hat{\Psi}^{\dag}(x)$
and
$\hat{\Psi}(x)$,
respectively. The Pauli matrices 
$\sigma^{\,}_{1}$,
$\sigma^{\,}_{2}$,
and
$\sigma^{\,}_{3}$
act on the spin components of the electrons.
The unit $2\times2$ matrix in spin space is denoted $\sigma^{\,}_{0}$.
Hamiltonian~(\ref{eq: def single z2pancake hamiltonian})
is invariant under the operation of time reversal defined by
\begin{equation}
\hat{\Psi}^{\dag}(x)=
\hat{\Psi}^{\prime\dag}(x)\,
K\,
\sigma^{\,}_{2},
\qquad
\hat{\Psi}(x)=
\sigma^{\,}_{2}\,K\,
\hat{\Psi}^{\prime}(x),
\end{equation}
\end{subequations}
where $K$ denotes the operation of complex conjugation.

A layered microscopic model that captures the tunneling of
helical edge states between adjacent layers for energy scales below
the bulk gap $\Delta$ 
is defined by the Hamiltonian
\begin{subequations}
\label{eq: def layered microscopic model}
\begin{widetext}
\begin{equation}
\begin{split}
\hat{H}^{\,}_{\mathrm{layered}}:=&\,
\int\limits_{\mathrm{edge}}\mathrm{{d}}x\,
\sum_{n=1}^{2N}
\left[
\left(
\hat{\Psi}^{\dag}_{n}
(-{i})\,
v^{\,}_{n}\,
\sigma^{\,}_{3}\,
\partial^{\,}_{x}\,
\hat{\Psi}^{\,}_{n}
\right)(x)
+
\hat{\Psi}^{\dag}_{n}(x)\,
\mu^{\,}_{n}(x)\,
\sigma^{\,}_{0}\,
\hat{\Psi}^{\,}_{n}(x)
\right]
\\
&\,
+
\int\limits_{\mathrm{edge}}\mathrm{{d}}x\,
\sum_{n=1}^{2N-1}
\left[
\hat{\Psi}^{\dag}_{n+1}(x)
\left(
\frac{\lambda^{\,}_{n,0}}{2}
\sigma^{\,}_{0}
+
{i}
\sum_{j=1}^{3}
\frac{\lambda^{\,}_{n,j}}{2}
\sigma^{\,}_{j}
\right)
\hat{\Psi}^{\,}_{n}(x)
+
\mathrm{H.c.}
\right].
\end{split}
\label{eq: def layered z2pancake hamiltonian}
\end{equation}
\end{widetext}

There is an even number of layers $2N$.
Each layer $n$ with $n=1,\cdots,2N$
has its own Fermi velocity $v^{\,}_{n}>0$
and chemical potential $\mu^{\,}_{n}\in\mathbb{R}$.
Any two consecutive layers are coupled by hopping matrix
elements parametrized by the four independent real-valued
couplings $\lambda^{\,}_{n,\mu}$ with $\mu=0,1,2,3$ and 
$n=1,\cdots,2N-1$. Open boundary conditions are chosen
along the layering axis.
Hamiltonian~(\ref{eq: def layered z2pancake hamiltonian})
is invariant under the operation of time reversal defined by
\begin{equation}
\hat{\Psi}^{\dag}_{n}(x)=
\hat{\Psi}^{\prime\dag}_{n}(x)\,
K\,
\sigma^{\,}_{2},
\qquad
\hat{\Psi}^{\,}_{n}(x)=
\sigma^{\,}_{2}\,K\,
\hat{\Psi}^{\prime}_{n}(x),
\label{eq: def time reversal}
\end{equation}
\end{subequations}
where $K$ denotes the operation of complex conjugation
and $n=1,\cdots,2N$. 
Hamiltonian~(\ref{eq: def layered z2pancake hamiltonian})
is depicted in Fig.~\ref{fig:layeredz2pancake}.

We can turn the layered model%
~(\ref{eq: def layered microscopic model}) 
into a layered microscopic model of a weak three-dimensional 
$\mathbb{Z}^{\,}_{2}$ topological \textit{band} insulator by demanding that
\begin{align}
&
v^{\,}_{n}=
v^{\,}_{\mathrm{u},x}
+
(-1)^{n}\,
v^{\,}_{\mathrm{s}},
\qquad
\mu^{\,}_{n}=
\mu^{\,}_{\mathrm{u}}
+
(-1)^{n}\,
\mu^{\,}_{\mathrm{s}},
\nonumber\\
&
\lambda^{\,}_{n,\mu}=
\left(1-\delta^{\,}_{n,2N}\right)
\left[
\lambda^{\,}_{\mathrm{u},\mu}
+
(-1)^{n}\,
\lambda^{\,}_{\mathrm{s},\mu}
\right],
\quad
\mu=0,1,2,3,
\end{align}
for $n=1,\cdots, 2N$.

It is shown in the supplementary material~\cite{suppl} 
how to construct a  continuum limit of $\hat{H}_{\mathrm{layered}}$
that delivers the single-particle Dirac Hamiltonian
\begin{align}
\mathcal{H}^{\mathrm{AII}}_{\mathrm{{d}}}\equiv&\,
v^{\,}_{\mathrm{s},x}\,
\sigma^{\,}_{3}\otimes\tau^{\,}_{3}\,
(-{i})\partial^{\,}_{x}
+
v^{\,}_{\mathrm{u},y}\,
\sigma^{\,}_{0}\otimes\tau^{\,}_{1}\,
(-{i})\partial^{\,}_{y}
\nonumber\\
&\,
+
\bar{u}^{\,}_{+,0}\,
\sigma^{\,}_{0}\otimes\tau^{\,}_{0}
+
\bar{u}^{\,}_{-,0}\,
\sigma^{\,}_{0}\otimes\tau^{\,}_{3}
\nonumber\\
&\,
+
\sum^3_{j=1}
2\,
\bar{w}^{\prime\prime}_{+,j}\,
\sigma^{\,}_{j}\otimes\tau^{\,}_{1}
+
2\,
\bar{w}^{\prime}_{-,0}\,
\sigma^{\,}_{0}\otimes\tau^{\,}_{2}.
\end{align}
The parameters $v^{\,}_{\mathrm{s},x}$ and $v^{\,}_{\mathrm{u},y}$
enter as anisotropic Dirac velocities.
The parameter $\bar{u}^{\,}_{+,0}$ enters as a chemical potential.
The parameter $2\,\bar{w}^{\prime}_{-,0}$ enters as a mass.
The mass term anticommutes with all terms except for
the chemical potential.
The remaining 4 terms with the parameters $\bar{u}^{\,}_{-,0}$
and $\bar{w}^{\prime\prime}_{+,j}$ ($j=1,2,3$)
do not anticommute with all the gamma matrices multiplying the
first derivatives in position space. 

\renewcommand{\theequation}{B\arabic{equation}}

\section{
Dirac Hamiltonian from the two-dimensional 
spin-directed $\mathbb{Z}^{\,}_{2}$ network model
        }
\label{appsec: Dirac Hamiltonian from the directed Z2 network model}

Starting from the two-dimensional CC network model,
Ho and Chalker derived in Ref.~\onlinecite{Ho96}
the random Dirac Hamiltonian studied in Ref.~\onlinecite{Ludwig96}
on its own merits. The two-dimensional $\mathbb{Z}^{\,}_{2}$
network model for a strong two-dimensional $\mathbb{Z}^{\,}_{2}$
topological insulator was related to a random Dirac Hamiltonian
in Ref.~\onlinecite{Ryu10}. 
As we show in the supplementary material, it is possible to
take the continuum limit of 
the two-dimensional spin-directed $\mathbb{Z}^{\,}_{2}$ network model
by performing a gradient expansion of the random phases entering
the network model and by performing an expansion in the deviations
about the points
\begin{equation}
(t^{2},\theta,\delta^{2})=(t^{2},\pi/2,0),
\label{appeq: critical line delta-pi/2=delta2=0}
\end{equation}
and
\begin{equation}
(t^{2},\theta,\delta^{2})=(t^{2},0,0),
\label{appeq: line at theta=delta2=0}
\end{equation}
respectively,
in the three-dimensional parameter space%
~(\ref{eq: def parameter space is 3D}).

As derived in the supplemental material\cite{suppl},
close to the point~(\ref{appeq: critical line delta-pi/2=delta2=0}),
the continuum limit of
the two-dimensional spin-directed $\mathbb{Z}^{\,}_{2}$ network model
is captured by the single-particle Dirac Hamiltonian
\begin{subequations}
\begin{align}
\mathcal{H}^{\,}_{\theta=\pi/2}=&\,
\left[
t\,
r
\left(
p^{\,}_{x}
-
p^{\,}_{y}
\right) 
\sigma^{\,}_{1}
+ 
\left( 
r^{2}\,p^{\,}_{x} 
+ 
t^{2}\, p^{\,}_{y}
\right) 
\sigma^{\,}_{2}
\right] 
\otimes 
\tau^{\,}_{0}
\nonumber\\
&\, 
-
\left[
t\,
r 
\left(
A^{\,}_{x}
-
A^{\,}_{y}
\right) 
\sigma^{\,}_{1}
+ 
\left(
r^{2}\, 
A^{\,}_{x} 
+ 
t^{2}\, 
A^{\,}_{y}
\right) 
\sigma^{\,}_{2} 
\right] 
\otimes 
\tau^{\,}_{3} 
\nonumber\\
&\,
- 
m\,\sigma^{\,}_{3}\otimes\tau^{\,}_{3} 
+ 
V^{\,}_{0}\,
\sigma^{\,}_{0} 
\otimes 
\tau^{\,}_{0}
\nonumber\\
&\,
- 
\theta^{\prime}\,
t\,
\left(
\lambda^{\,}_{\phi}\,
\sigma^{\,}_0 \otimes \tau^{\,}_{1}
+ 
2\,
\sigma^{\,}_{0} 
\otimes
\tau^{\,}_{2}
\right).
\label{appeq: H+- if theta=pi/2 b}
\end{align}
Here, a second set of Pauli matrices
$\tau^{\,}_{1}$, $\tau^{\,}_{2}$, and $\tau^{\,}_{3}$, 
together with the unit
$2\times2$ matrix $\tau^{\,}_{0}$
has been introduced.
There appear the two-dimensional momentum operators, $p_x$ and
$p_y$, and a mass, $m$. 
This Hamiltonian is invariant under reversal of time, i.e.,
\begin{equation}
\mathcal{T}\,
\mathcal{H}^{\,}_{\theta=\pi/2}\,
\mathcal{T}^{-1}= 
\mathcal{H}^{\,}_{\theta=\pi/2}, 
\qquad 
\mathcal{T}:= 
{i}\sigma^{\,}_{2}\otimes\tau^{\,}_{1}\,{K},
\end{equation}
\end{subequations}
where ${K}$ denotes the operation of complex conjugation.
The mass $m$ that encodes the dimerization 
($m\propto\delta^{2}$)
multiplies the matrix
$\sigma^{\,}_{3}\otimes\tau^{\,}_{3}$
that anticommutes with all other contributions to the 
continuum limit~(\ref{appeq: H+- if theta=pi/2 b})
with $V^{\,}_{0}=0$. Hence, dimerization opens a spectral gap in the
spectrum of Hamiltonian~(\ref{appeq: H+- if theta=pi/2 b}).
The other parameters $A_x$, $A_y$, $V_0$, $\lambda_\phi$, and
$\theta^\prime\equiv \pi/2-\theta$ commute with all other terms, except
for the term with $m$.
At the isotropic point defined by 
$t^{2}=r^{2}=1/2$,
Hamiltonian~(\ref{appeq: H+- if theta=pi/2 b}) 
becomes the Dirac Hamiltonian
\begin{subequations}
\begin{align}
\mathcal{H}^{\,}_{\theta=\pi/2}=&\,
\frac{1}{\sqrt{2}}
\left(
p^{\prime}_{x}\,\sigma^{\,}_{1} 
+ 
p^{\prime}_{y}\,\sigma^{\,}_{2}
\right)
\otimes 
\tau^{\,}_{0}
+
V^{\,}_{0} 
\sigma^{\,}_{0} 
\otimes 
\tau^{\,}_{0}
\nonumber\\
&\,
-
\frac{1}{\sqrt{2}}
\left(
A^{\prime}_{x}\,\sigma^{\,}_{1} 
+ 
A^{\prime}_{y}\,\sigma^{\,}_{2}
\right) 
\otimes 
\tau^{\,}_{3}
- 
m\, 
\sigma^{\,}_{3} \otimes \tau^{\,}_{3} 
\nonumber\\
&\,
- 
\frac{\theta^{\prime}}{\sqrt{2}}
\left(
\lambda^{\,}_{\phi}\,
\sigma^{\,}_0 \otimes \tau^{\,}_{1}
+ 
2\,
\sigma^{\,}_{0}\otimes\tau^{\,}_{2} 
\right)
\end{align}
where 
\begin{align}
&
p^{\prime}_{x}\equiv 
\frac{
p^{\,}_{x}-p^{\,}_{y}
     }
     {
\sqrt{2}
     },
\qquad
p^{\prime}_{y}\equiv 
\frac{
p^{\,}_{x}+p^{\,}_{y}
     }
     {
\sqrt{2}
     }, 
\nonumber\\
&
A^{\prime}_{x}\equiv 
\frac{
A^{\,}_{x}-A^{\,}_{y}
     }
     {
\sqrt{2}
     }, 
\qquad
A^{\prime}_{y}\equiv 
\frac{
A^{\,}_{x}+A^{\,}_{y}
     }
     {
\sqrt{2}
     }.
\end{align}
\end{subequations}

Close to the point~(\ref{appeq: line at theta=delta2=0}),
the continuum limit of
the two-dimensional spin-directed $\mathbb{Z}^{\,}_{2}$ network model
is captured by the single-particle Dirac-like Hamiltonian
\begin{subequations}
\begin{align}
\mathcal{H}^{\,}_{\theta=0}=&\,
p^{\,}_{x}\,
\sigma^{\,}_{0}
\otimes
\tau^{\,}_{3}
+
\left(
t\,
r\,
p^{\,}_{y}\,
\sigma^{\,}_{1}
+
t^{2}\,
p^{\,}_{y}\,
\sigma^{\,}_{2}
\right)
\otimes
\tau^{\,}_{0}
\nonumber\\
&\,
+
\left[
t\,
r
\left(
A^{\,}_{x}
-
A^{\,}_{y}
\right) 
\sigma^{\,}_{1}
- 
\left( 
r^{2}\,
A^{\,}_{x} 
+ 
t^{2}\,
A^{\,}_{y}
\right) 
\sigma^{\,}_{2}
\right] 
\otimes 
\tau^{\,}_{3} 
\nonumber\\
&\,
+ 
m\,
\sigma^{\,}_{3}
\otimes
\tau^{\,}_{3} 
+
\theta\,
t\,
\left(
\lambda^{\,}_{\phi}\,
\sigma^{\,}_{0} \otimes \tau^{\,}_{1}
+ 
2\,
\sigma^{\,}_{0} 
\otimes
\tau^{\,}_{2}
\right)
\nonumber\\
&\,
+ 
V^{\,}_{0}\,
\sigma^{\,}_{0} 
\otimes 
\tau^{\,}_{0}.
\label{appeq: H+- if theta=0 b}
\end{align}
This Hamiltonian is invariant under reversal of time, i.e.,
\begin{equation}
\mathcal{T}\,
\mathcal{H}^{\,}_{\theta=0}\,
\mathcal{T}^{-1}= 
\mathcal{H}^{\,}_{\theta=0}, 
\qquad 
\mathcal{T}:= 
{i}\sigma^{\,}_{2}\otimes\tau^{\,}_{1}\,{K},
\end{equation}
\end{subequations}
where ${K}$ denotes the operation of complex conjugation.
As was the case for
the continuum limit~(\ref{appeq: H+- if theta=pi/2 b}),
the term
$V^{\,}_{0}\,\sigma^{\,}_{0}\otimes\tau^{\,}_{0}$
acts as a chemical potential, 
for it commutes with all contributions
to the continuum limit~(\ref{appeq: H+- if theta=0 b}).
We shall set $V^{\,}_{0}=0$ when deciding if a gap at energy 0 
is opened by dimerization.
In comparison to the continuum limit~(\ref{appeq: H+- if theta=pi/2 b}),
the term $p^{\,}_{x}\,\sigma_{0}\otimes\tau^{\,}_{3}$
has appeared that commutes with all contributions
to the continuum limit~(\ref{appeq: H+- if theta=0 b})
except for the term
$
\theta\,
t\,
\left(
\lambda^{\,}_{\phi}\,
\sigma^{\,}_{0} \otimes \tau^{\,}_{1}
+ 
2\,
\sigma^{\,}_{0} 
\otimes
\tau^{\,}_{2}
\right)$.
If we set  $t=A^{\,}_{x}=A^{\,}_{y}=V^{\,}_{0}=0$,
we find the two (two-fold degenerate) gapless dispersions $|p^{\,}_{x}\pm m|$.
More generally, a branch of excitation is expected to cross
the energy 0 at some $m$-dependent value of the momentum 
when $\theta=V^{\,}_{0}=0$.
As the coupling $m$ is caused by dimerization,
dimerization thus fails to open a gap if we set $\theta=V^{\,}_{0}=0$.
On the other hand, because the term
$\theta\,
t\,
\left(
\lambda^{\,}_{\phi}\,
\sigma^{\,}_{0} \otimes \tau^{\,}_{1}
+ 
2\,
\sigma^{\,}_{0} 
\otimes
\tau^{\,}_{2}
\right)$
anticommutes with both 
$m\,\sigma^{\,}_{3}\otimes\tau^{\,}_{3}$ 
and
$p^{\,}_{x}\,\sigma_{0}\otimes\tau^{\,}_{3}$,
we expect that a sufficiently large $\theta$ opens a gap
for a given $m$.

\medskip

\renewcommand{\theequation}{C\arabic{equation}}

\section{Details of Finite-Size Scaling Analysis}
\label{appsec:finite-size-scaling}

Figures \ref{fig:Lambda_dimerization_0.5_5PIover16} and
\ref{fig:Lambda_dimerization_0.6_PIover4} show $\delta^2$ dependence
of the normalized localization lengths $\Lambda^{\,}_\perp$ and
$\Lambda^{\,}_\parallel$ for the one-dimensional cut ``B'' and ``C'',
respectively.
Tables II-VIII present the values of the sets of parameters used in
finite-size scaling analysis.

\begin{widetext}

\begin{figure*}[t]
(a-1) \hspace{7cm} (a-2)\\
\includegraphics[width=0.42\textwidth]{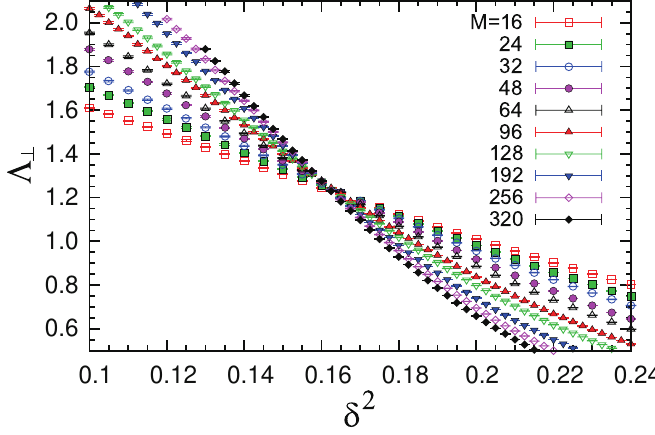}
\includegraphics[width=0.38\textwidth]{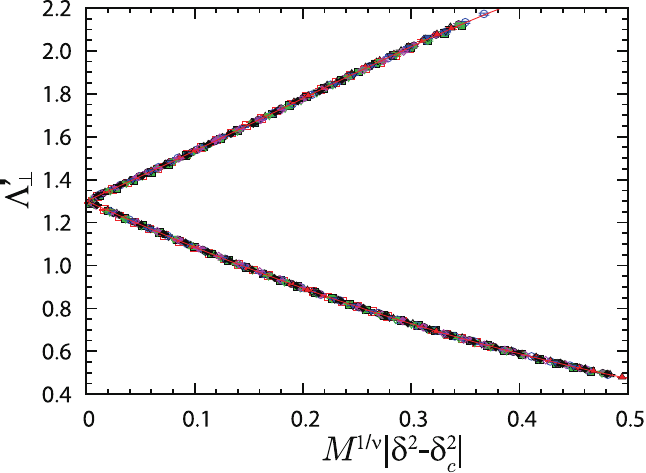}\\
(b-1) \hspace{7cm} (b-2)\\
\includegraphics[width=0.42\textwidth]{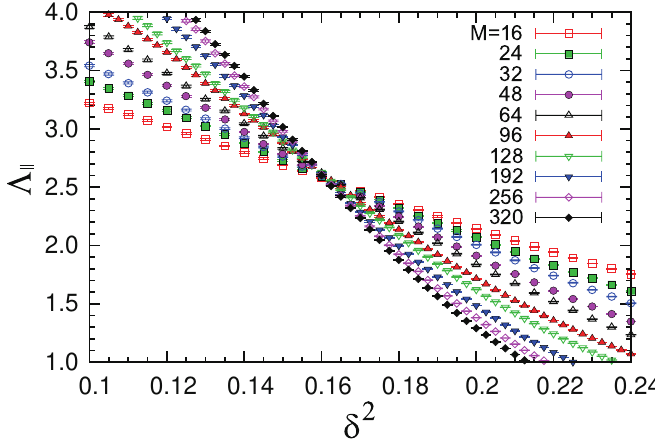}
\includegraphics[width=0.38\textwidth]{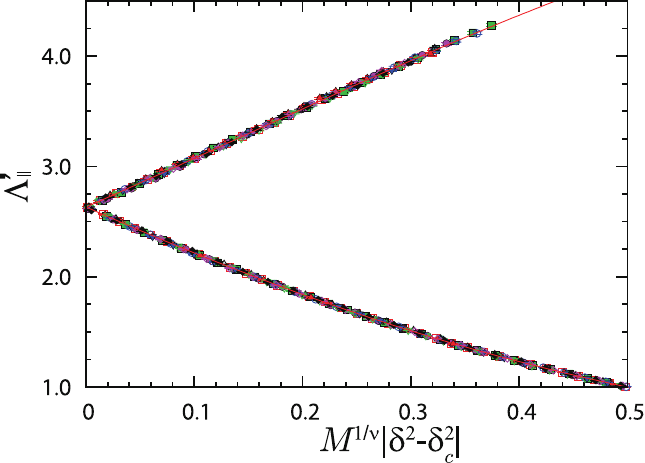}
\caption{
(Color online:)
The two-dimensional spin-directed $\mathbb{Z}^{\,}_{2}$ network model 
is solved numerically along the one-dimensional cut~(\ref{eq: cut choice 2})
in the three-dimensional parameter space%
~(\ref{eq: def parameter space is 3D}).
Panel (a-1) shows
the $\delta^{2}$ dependence of the normalized localization length
$\Lambda^{\,}_{\perp}$ corresponding to the geometry of 
Fig.~\ref{fig:network-staggered}(a)
for several values of $M$.
Panel (b-1) shows
the $\delta^{2}$ dependence of the normalized localization lengths
$\Lambda^{\,}_{\parallel}$ 
corresponding to the geometry of 
Fig.~\ref{fig:network-staggered}(b)
for several values of $M$.
A finite-size scaling analysis of panels
(a-1) and (b-1) is performed in panels (a-2) and (b-2), respectively.
The horizontal axis is $M^{1/\nu}|\delta^{2}-\delta^{2}_{\mathrm{c}}|$
with $\nu$ and $\delta^{2}_{\mathrm{c}}$ given in 
Table~\ref{tab:fitting parameters cut 2 perpendicular} and 
Table~\ref{tab:fitting parameters cut 2 parallel}.
The vertical axis $\Lambda^{\prime}_{\mathrm{x}}$
with $\mathrm{x}=\perp,\parallel$
is defined by subtracting from the
normalized localization length $\Lambda^{\,}_{\mathrm{x}}$
its finite-size correction from the leading irrelevant exponent $y$
given in 
Table~\ref{tab:fitting parameters cut 2 perpendicular} and 
Table~\ref{tab:fitting parameters cut 2 parallel}.
The red solid curve demonstrates the quality of the data collapse
onto a one-parameter scaling function.
        }
\label{fig:Lambda_dimerization_0.5_5PIover16}
\end{figure*}

\begin{figure*}[t]
(a-1) \hspace{7cm} (a-2)\\
\includegraphics[width=0.42\textwidth]{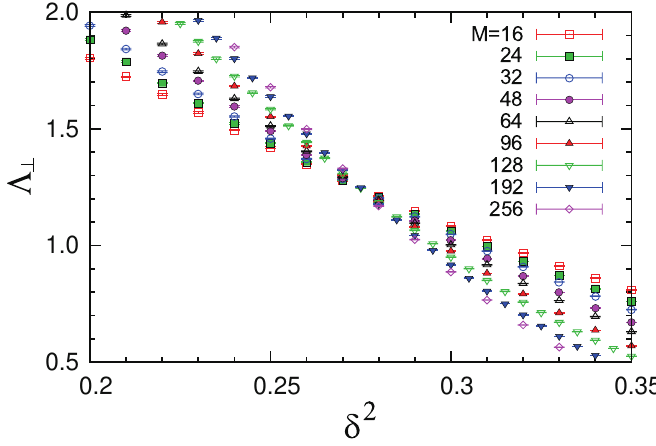}
\includegraphics[width=0.38\textwidth]{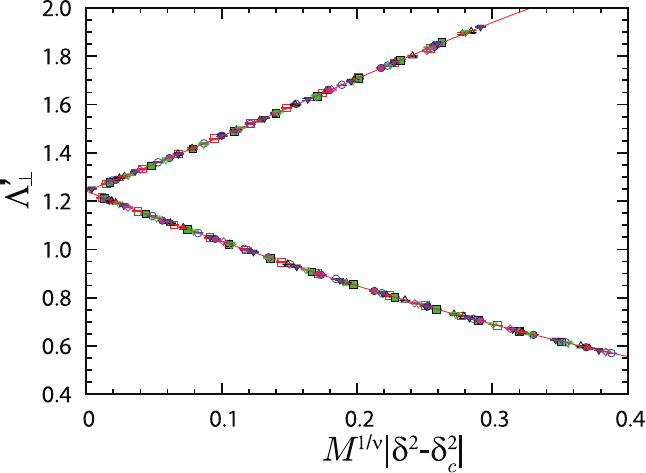}\\
(b-1) \hspace{7cm} (b-2)\\
\includegraphics[width=0.42\textwidth]{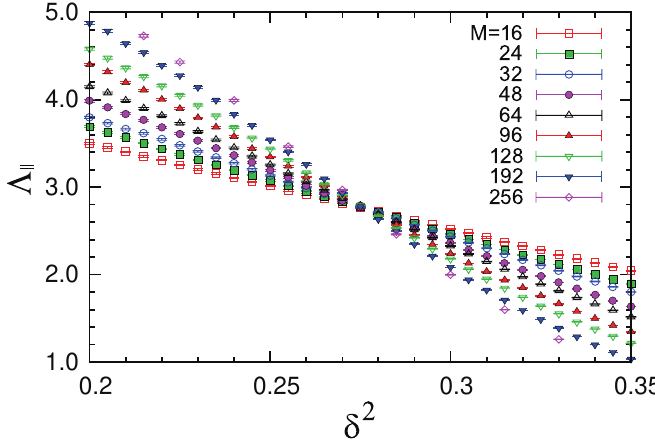}
\includegraphics[width=0.38\textwidth]{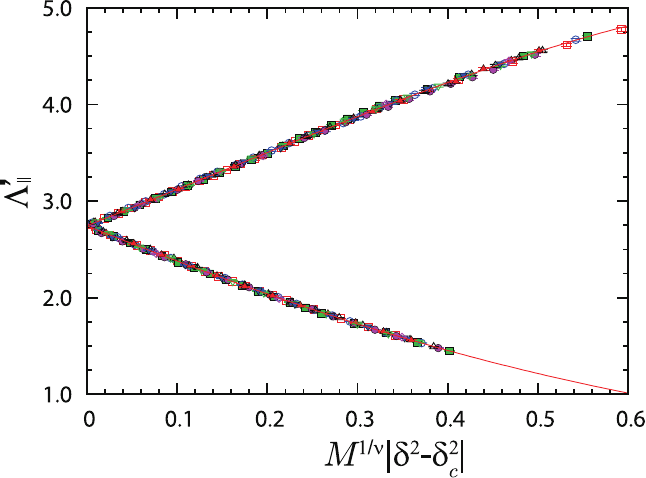}
\caption{
(Color online:)
The two-dimensional spin-directed $\mathbb{Z}^{\,}_{2}$ network model 
is solved numerically along the one-dimensional cut~(\ref{eq: cut choice 3})
in the three-dimensional parameter space%
~(\ref{eq: def parameter space is 3D}).
Panel (a-1) shows
the $\delta^{2}$ dependence of the normalized localization length
$\Lambda^{\,}_{\perp}$ corresponding to the geometry of 
Fig.~\ref{fig:network-staggered}(a)
for several  values of $M$.
Panel (b-1) shows
the $\delta^{2}$ dependence of the normalized localization lengths
$\Lambda^{\,}_{\parallel}$ 
corresponding to the geometry of 
Fig.~\ref{fig:network-staggered}(b)
for several values of $M$.
A finite-size scaling analysis of panels
(a-1) and (b-1) is performed in panels (a-2) and (b-2), respectively.
The horizontal axis is $M^{1/\nu}|\delta^{2}-\delta^{2}_{\mathrm{c}}|$
with $\nu$ and $\delta^{2}_{\mathrm{c}}$ given in 
Table~\ref{tab:fitting parameters cut 3 perpendicular} and 
Table~\ref{tab:fitting parameters cut 3 parallel}.
The vertical axis $\Lambda^{\prime}_{\mathrm{x}}$
with $\mathrm{x}=\perp,\parallel$
is defined by subtracting from the
normalized localization length $\Lambda^{\,}_{\mathrm{x}}$
its finite-size correction from the leading irrelevant exponent $y$
given in 
Table~\ref{tab:fitting parameters cut 3 perpendicular} and 
Table~\ref{tab:fitting parameters cut 3 parallel}.
The red solid curve demonstrates the quality of the data collapse
onto a one-parameter scaling function.
         }
\label{fig:Lambda_dimerization_0.6_PIover4}
\end{figure*}

\begin{table*}
\rotatebox{90}{
\begin{minipage}[t]{1.0\textheight}
\vspace{-1.5cm}
\caption{
Finite-size scaling analysis of $\Lambda^{\,}_{\perp}$ 
for the two-dimensional spin-directed $\mathbb{Z}^{\,}_{2}$ network model 
along the cut (\ref{eq: cut choice 1}) 
[A: $(0.5,\pi/4,\delta^{2})$ for $\delta^{2}\in[0,0.5]$].
Only the values of $\Lambda^{\,}_{\perp}$ satisfying 
$0.35<\Lambda^{\,}_{\perp}<1.7$ 
enter the data set. Different data sets are chosen by varying the
minimum $M^{\,}_{\mathrm{min}}$ and maximum $M^{\,}_{\mathrm{max}}$
taken by the width $M$ 
of the two-dimensional spin-directed $\mathbb{Z}^{\,}_{2}$ network model. 
The quality of fit is measured by
$\chi^{2}_{\mathrm{min}}/N$ and $\mathcal{G}$.
The values $\nu$, $y$, $\delta^{2}_{\mathrm{c}}$, $\Lambda^{\mathrm{c}}_{\perp}$,
$F^{p,0}_{\perp}$, and $f^{p,q}_{\perp}$ for the fitting parameters 
and their $\chi^{2}_{\mathrm{min}}/N$ are shown.
The numbers with $\pm$ are the 
statistical error bars (one sigma).
\label{tab:fitting parameters cut 1 perpendicular}
        }
\begin{tabular}{c c c|c c |r r|r| r r r|r r r| r r r| r r| r r}
\hline\hline
$M^{\,}_{\mathrm{min}}$ 
&  
$M^{\,}_{\mathrm{max}}$ 
& 
$N$ 
& 
$\chi^{2}_{\mathrm{min}}/N$ 
& 
$\mathcal{G}$ 
&
$\nu$ 
& 
$|y|$ 
&
$\delta^{2}_{\mathrm{c}}$
& 
$\Lambda^{\mathrm{c}}_{\perp}\equiv (1/F^{0,0}_{\perp})$ 
& 
$f^{0,1}_{\perp}$ 
& 
$f^{0,2}_{\perp}$
& 
$F^{1,0}_{\perp}$ 
& 
$f^{1,1}_{\perp}$  
& 
$f^{1,2}_{\perp}$
& 
$F^{2,0}_{\perp}$ 
& 
$f^{2,1}_{\perp}$  
& 
$f^{2,2}_{\perp}$  
& 
$F^{3,0}_{\perp}$ 
& 
$f^{3,1}_{\perp}$
& 
$F^{4,0}_{\perp}$ 
& 
$f^{4,1}_{\perp}$\\ 
\hline
$16$ & $320$ & $461$ & $1.009$ & $0.246$ & $2.922$ & $0.712$ & $0.2336$  & $0.94135$ & $0.2081$ & $-1.132$ & $2.060$ & $1.567$ & $-6.563$ & $3.079$& $1.644$ & $-2.246$ & $2.803$ & $5.169$ & $0.993$& $31.085$\\ 
 &  &  &  & & $\pm 0.037$ & $\pm 0.054$ &$\pm 0.0002$  & $\pm 0.00256$ & $\pm 0.0323$ &$\pm 0.302$ & $\pm 0.065$ & $\pm 0.108$ & $\pm 1.397$ & $\pm 0.175$& $\pm 0.210$ & $\pm 0.918$ & $\pm 0.272$ & $\pm 0.279$ & $\pm 0.257$ & $\pm 4.513$\\ 
$24$ & $320$ & $402$ & $0.920$ & $0.714$ & $2.926$ & $0.788$ & $0.2340$  & $0.93539$ & $0.1443$ & $-0.937$ & $2.090$ & $1.883$ & $-11.030$ & $3.131$& $2.053$ & $-5.646$ & $2.903$ & $6.115$ & $1.102$& $36.256$\\ 
 &  &  &  & & $\pm 0.053$ & $\pm 0.107$ &$\pm 0.0002$  & $\pm 0.00299$ & $\pm 0.0766$ &$\pm 0.803$ & $\pm 0.090$ & $\pm 0.283$ & $\pm 6.088$ & $\pm 0.251$& $\pm 0.406$ & $\pm 4.032$ & $\pm 0.392$ & $\pm 0.876$ & $\pm 0.364$ & $\pm 5.269$\\ 
$32$ & $320$ & $361$ & $0.946$ & $0.543$ & $2.850$ & $0.728$ & $0.2345$  & $0.92583$ & $-0.0657$ & $0.683$ & $1.989$ & $2.028$ & $-9.519$ & $2.802$& $2.768$ & $-6.529$ & $2.434$ & $6.600$ & $0.845$& $33.365$\\ 
 &  &  &  & & $\pm 0.102$ & $\pm 0.159$ &$\pm 0.0004$  & $\pm 0.00636$ & $\pm 0.0955$ &$\pm 0.670$ & $\pm 0.174$ & $\pm 0.425$ & $\pm 8.209$ & $\pm 0.481$& $\pm 0.630$ & $\pm 6.497$ & $\pm 0.698$ & $\pm 1.015$ & $\pm 0.544$ & $\pm 6.811$\\
\hline
$16$ & $256$ & $421$ & $1.043$ & $0.118$ & $2.880$ & $0.675$ & $0.2337$  & $0.94084$ & $0.1907$ & $-0.948$ & $1.994$ & $1.647$ & $-5.943$ & $2.892$& $1.805$ & $-2.000$ & $2.510$ & $5.481$ & $0.789$& $34.324$\\ 
 &  &  &  & & $\pm 0.052$ & $\pm 0.057$ &$\pm 0.0002$  & $\pm 0.00344$ & $\pm 0.0353$ &$\pm 0.260$ & $\pm 0.092$ & $\pm 0.144$ & $\pm 1.183$ & $\pm 0.239$& $\pm 0.271$ & $\pm 0.848$ & $\pm 0.359$ & $\pm 0.395$ & $\pm 0.320$ & $\pm 8.998$\\ 
$24$ & $256$ & $362$ & $0.947$ & $0.539$ & $2.845$ & $0.701$ & $0.2343$  & $0.93049$ & $0.0569$ & $-0.314$ & $1.970$ & $1.903$ & $-7.757$ & $2.777$& $2.356$ & $-4.043$ & $2.361$ & $6.403$ & $0.725$& $39.007$\\ 
 &  &  &  & & $\pm 0.089$ & $\pm 0.108$ &$\pm 0.0003$  & $\pm 0.00487$ & $\pm 0.0728$ &$\pm 0.441$ & $\pm 0.153$ & $\pm 0.242$ & $\pm 3.745$ & $\pm 0.407$& $\pm 0.490$ & $\pm 2.723$ & $\pm 0.591$ & $\pm 0.621$ & $\pm 0.477$ & $\pm 12.878$\\ 
\hline\hline
\end{tabular}%
\end{minipage}
             }
\end{table*}

\begin{table*}
\rotatebox{90}{
\begin{minipage}[t]{1.0\textheight}
\vspace{-1.5cm}
\caption{
Finite-size scaling analysis of $\Lambda^{\,}_{\perp}$ 
for the two-dimensional spin-directed $\mathbb{Z}^{\,}_{2}$ network model 
along the cut (\ref{eq: cut choice 1}) 
[A: $(0.5,\pi/4,\delta^{2})$ for $\delta^{2}\in[0,0.5]$].
Only the values of $\Lambda^{\,}_{\parallel}$ satisfying 
$1.4<\Lambda^{\,}_{\perp}<6.0$  
enter the data set. Different data sets are chosen by varying the
minimum $M^{\,}_{\mathrm{min}}$ and maximum $M^{\,}_{\mathrm{max}}$
taken by the width $M$ 
of the two-dimensional spin-directed $\mathbb{Z}^{\,}_{2}$ network model. 
The quality of fit is measured by
$\chi^{2}_{\mathrm{min}}/N$ and $\mathcal{G}$.
The values $\nu$, $y$, $\delta^{2}_{\mathrm{c}}$, $\Lambda^{\mathrm{c}}_{\parallel}$,
$F^{p,0}_{\parallel}$, and $f^{p,q}_{\parallel}$ for the fitting parameters 
and their $\chi^{2}_{\mathrm{min}}/N$ are shown.
The numbers with $\pm$ are the 
statistical error bars (one sigma).
\label{tab:fitting parameters cut 1 parallel}
        }
\begin{tabular}{c c c|c c | r r|r| r r|r r| r r| r r| r r}
\hline\hline
$M^{\,}_{\mathrm{min}}$ 
&  
$M^{\,}_{\mathrm{max}}$ 
& 
$N$ 
& 
$\chi^{2}_{\mathrm{min}}/N$ 
& 
$\mathcal{G}$ 
&
$\nu$ 
& 
$|y|$ 
&
$\delta^{2}_{\mathrm{c}}$
& 
$\Lambda^{\mathrm{c}}_{\parallel}\equiv(1/F^{0,0}_{\parallel})$ 
& 
$f^{0,1}_{\parallel}$
& 
$F^{1,0}_{\parallel}$ 
& 
$f^{1,1}_{\parallel}$ 
& 
$F^{2,0}_{\parallel}$ 
& 
$f^{2,1}_{\parallel}$  
& 
$F^{3,0}_{\parallel}$ 
& 
$f^{3,1}_{\parallel}$
& 
$F^{4,0}_{\parallel}$ 
& 
$f^{4,1}_{\parallel}$\\ 
\hline
$16$ & $256$ & $403$ & $0.941$ & $0.641$ & $2.793 $ & $0.404 $  & $0.2336 $ & $3.70707 $ & $0.041 $ & $0.562 $ & $-0.789 $ & $0.954 $& $-0.967 $ & $1.004 $ & $-0.974 $ & $0.626 $ & $-0.735 $ \\ 
 & & & & & $\pm 0.052$ & $\pm 0.053$  & $\pm 0.0003$ & $\pm 0.01882$ & $\pm 0.010$ & $\pm 0.031$ & $\pm 0.034$ & $\pm 0.096$& $\pm 0.041$ & $\pm 0.147$ & $\pm 0.094$ & $\pm 0.135$ & $\pm 0.442$ \\ 
$24$ & $256$ & $352$ & $0.970$ & $0.450$ & $2.758 $ & $0.387 $  & $0.2340 $ & $3.67677 $ & $0.013 $ & $0.556 $ & $-0.745 $ & $0.903 $& $-0.807 $ & $0.920 $ & $-0.774 $ & $0.615 $ & $-0.972 $ \\ 
 & & & & & $\pm 0.067$ & $\pm 0.083$  & $\pm 0.0003$ & $\pm 0.02158$ & $\pm 0.014$ & $\pm 0.045$ & $\pm 0.055$ & $\pm 0.129$& $\pm 0.065$ & $\pm 0.186$ & $\pm 0.149$ & $\pm 0.168$ & $\pm 0.397$ \\ 
$32$ & $256$ & $307$ & $0.875$ & $0.852$ & $2.653 $ & $0.473 $  & $0.2352 $ & $3.58849 $ & $-0.088 $ & $0.511 $ & $-0.801 $ & $0.747 $& $-0.580 $ & $0.709 $ & $-0.547 $ & $0.444 $ & $-1.156 $ \\ 
 & & & & & $\pm 0.059$ & $\pm 0.125$  & $\pm 0.0003$ & $\pm 0.02051$ & $\pm 0.030$ & $\pm 0.039$ & $\pm 0.146$ & $\pm 0.093$& $\pm 0.130$ & $\pm 0.127$ & $\pm 0.257$ & $\pm 0.117$ & $\pm 0.740$ \\
\hline
$16$ & $192$ & $361$ & $0.943$ & $0.606$ & $2.772 $ & $0.449 $  & $0.2335 $ & $3.71150 $ & $0.049 $ & $0.543 $ & $-0.828 $ & $0.904 $& $-1.008 $ & $0.924 $ & $-0.948 $ & $0.576 $ & $-0.659 $ \\ 
 & & & & & $\pm 0.063$ & $\pm 0.068$  & $\pm 0.0003$ & $\pm 0.02286$ & $\pm 0.012$ & $\pm 0.034$ & $\pm 0.052$ & $\pm 0.104$& $\pm 0.055$ & $\pm 0.153$ & $\pm 0.128$ & $\pm 0.140$ & $\pm 0.586$ \\ 
$24$ & $192$ & $310$ & $0.980$ & $0.381$ & $2.702 $ & $0.476 $  & $0.2340 $ & $3.67609 $ & $0.016 $ & $0.517 $ & $-0.818 $ & $0.799 $& $-0.803 $ & $0.767 $ & $-0.595 $ & $0.514 $ & $-0.980 $ \\ 
 & & & & & $\pm 0.073$ & $\pm 0.114$  & $\pm 0.0004$ & $\pm 0.02398$ & $\pm 0.019$ & $\pm 0.041$ & $\pm 0.115$ & $\pm 0.113$& $\pm 0.094$ & $\pm 0.152$ & $\pm 0.249$ & $\pm 0.143$ & $\pm 0.595$ \\ 
$32$ & $192$ & $265$ & $0.866$ & $0.842$ & $2.578 $ & $0.632 $  & $0.2356 $ & $3.57956 $ & $-0.157 $ & $0.472 $ & $-1.038 $ & $0.646 $& $-0.394 $ & $0.573 $ & $-0.045 $ & $0.362 $ & $-1.722 $ \\ 
 & & & & & $\pm 0.051$ & $\pm 0.175$  & $\pm 0.0003$ & $\pm 0.02268$ & $\pm 0.073$ & $\pm 0.029$ & $\pm 0.361$ & $\pm 0.063$& $\pm 0.250$ & $\pm 0.080$ & $\pm 0.566$ & $\pm 0.082$ & $\pm 1.415$ \\ 
\hline\hline
\end{tabular}
\end{minipage}
              }
\end{table*}

\begin{table*}
\rotatebox{90}{
\begin{minipage}[t]{1.0\textheight}
\caption{
Finite-size scaling analysis of $\Lambda^{\,}_{\perp}$ 
for the two-dimensional spin-directed $\mathbb{Z}^{\,}_{2}$ network model 
along the cut (\ref{eq: cut choice 2}) 
[B: $(0.5,5\pi/16,\delta^{2})$ for $\delta^{2}\in[0,0.5]$].
Only the values of $\Lambda^{\,}_{\perp}$ satisfying 
$0.45<\Lambda^{\,}_{\perp}<2.15$ 
enter the data set. Different data sets are chosen by varying the
minimum $M^{\,}_{\mathrm{min}}$ and maximum $M^{\,}_{\mathrm{max}}$
taken by the width $M$ 
of the two-dimensional spin-directed $\mathbb{Z}^{\,}_{2}$ network model. 
The quality of fit is measured by
$\chi^{2}_{\mathrm{min}}/N$ and $\mathcal{G}$.
The values $\nu$, $y$, $\delta^{2}_{\mathrm{c}}$, $\Lambda^{\mathrm{c}}_{\perp}$,
$F^{p,0}_{\perp}$, and $f^{p,q}_{\perp}$ for the fitting parameters 
and their $\chi^{2}_{\mathrm{min}}/N$ are shown.
The numbers with $\pm$ are the statistical error bars (one sigma).
\label{tab:fitting parameters cut 2 perpendicular}
        }
\begin{tabular}{c c c|c c |r r|r| r r r|r r r| r r| r r| r }
\hline\hline
$M^{\,}_{\mathrm{min}}$ 
&  
$M^{\,}_{\mathrm{max}}$ 
& 
$N$ 
& 
$\chi^{2}_{\mathrm{min}}/N$ 
& 
$\mathcal{G}$ 
&
$\nu$ 
& 
$|y|$ 
&
$\delta^{2}_{\mathrm{c}}$
& 
$\Lambda^{\mathrm{c}}_{\perp}\equiv (1/F^{0,0}_{\perp})$ 
& 
$f^{0,1}_{\perp}$ 
& 
$f^{0,2}_{\perp}$
& 
$F^{1,0}_{\perp}$ 
& 
$f^{1,1}_{\perp}$  
& 
$f^{1,2}_{\perp}$
& 
$F^{2,0}_{\perp}$ 
& 
$f^{2,1}_{\perp}$ 
& 
$F^{3,0}_{\perp}$ 
& 
$f^{3,1}_{\perp}$
& 
$F^{4,0}_{\perp}$\\ 
\hline
$12$ & $320$ & $528$ & $1.067$ & $0.064$ & $2.743$ & $0.580$ & $0.1588$  & $1.30050$ & $0.0919$ & $0.446$ & $1.332$ & $1.204$ & $-4.489$ & $1.741$& $1.515$ & $1.439$ & $0.887$ & $0.999$\\
 &  &  &  & & $\pm 0.021$ & $\pm 0.032$ &$\pm 0.0002$  & $\pm 0.00381$ & $\pm 0.0354$ &$\pm 0.051$ & $\pm 0.030$ & $\pm 0.083$ & $\pm 0.344$ & $\pm 0.072$& $\pm 0.072$ & $\pm 0.086$ & $\pm 0.190$ & $\pm 0.046$ \\
$16$ & $320$ & $477$ & $1.019$ & $0.223$ & $2.731$ & $0.625$ & $0.1595$  & $1.28399$ & $-0.0079$ & $0.911$ & $1.350$ & $1.162$ & $-5.039$ & $1.743$& $1.722$ & $1.428$ & $1.015$ & $0.960$\\ 
 &  &  &  & & $\pm 0.025$ & $\pm 0.045$ &$\pm 0.0002$  & $\pm 0.00500$ & $\pm 0.0497$ &$\pm 0.109$ & $\pm 0.033$ & $\pm 0.080$ & $\pm 0.735$ & $\pm 0.085$& $\pm 0.079$ & $\pm 0.098$ & $\pm 0.215$ & $\pm 0.054$ \\ 
$32$ & $320$ & $373$ & $0.751$ & $0.999$ & $2.688$ & $0.729$ & $0.1611$  & $1.24759$ & $-0.4620$ & $4.587$ & $1.362$ & $1.361$ & $-8.908$ & $1.722$& $2.422$ & $1.364$ & $1.643$ & $0.717$\\ 
 &  &  &  & & $\pm 0.045$ & $\pm 0.111$ &$\pm 0.0005$  & $\pm 0.01359$ & $\pm 0.0893$ &$\pm 2.293$ & $\pm 0.050$ & $\pm 0.208$ & $\pm 5.160$ & $\pm 0.144$& $\pm 0.323$ & $\pm 0.163$ & $\pm 0.389$ & $\pm 0.086$ \\ 
\hline
$12$ & $256$ & $491$ & $1.040$ & $0.139$ & $2.765$ & $0.613$ & $0.1584$  & $1.30826$ & $0.1600$ & $0.343$ & $1.357$ & $1.143$ & $-4.870$ & $1.806$& $1.486$ & $1.498$ & $0.885$ & $1.066$\\ 
 &  &  &  & & $\pm 0.025$ & $\pm 0.041$ &$\pm 0.0002$  & $\pm 0.00417$ & $\pm 0.0457$ &$\pm 0.069$ & $\pm 0.036$ & $\pm 0.105$ & $\pm 0.483$ & $\pm 0.087$& $\pm 0.081$ & $\pm 0.101$ & $\pm 0.211$ & $\pm 0.055$ \\ 
$16$ & $256$ & $440$ & $0.974$ & $0.453$ & $2.867$ & $1.413$ & $0.1592$  & $1.28935$ & $2.1767$ & $-35.927$ & $1.529$ & $0.119$ & $-86.394$ & $2.219$& $6.329$ & $1.906$ & $0.946$ & $1.322$\\ 
 &  &  &  & & $\pm 0.006$ & $\pm 0.080$ &$\pm 0.0001$  & $\pm 0.00206$ & $\pm 0.5515$ &$\pm 21.322$ & $\pm 0.006$ & $\pm 0.384$ & $\pm 27.812$ & $\pm 0.018$& $\pm 1.310$ & $\pm 0.023$ & $\pm 0.906$ & $\pm 0.032$ \\ 
\hline
$12$ & $192$ & $450$ & $0.863$ & $0.952$ & $2.905$ & $1.118$ & $0.1581$  & $1.30627$ & $1.1074$ & $-5.450$ & $1.540$ & $0.063$ & $-18.218$ & $2.271$& $2.295$ & $1.987$ & $0.220$ & $1.435$\\ 
 &  &  &  & & $\pm 0.008$ & $\pm 0.069$ &$\pm 0.0002$  & $\pm 0.00391$ & $\pm 0.1953$ &$\pm 2.435$ & $\pm 0.009$ & $\pm 0.227$ & $\pm 3.708$ & $\pm 0.025$& $\pm 0.338$ & $\pm 0.031$ & $\pm 0.361$ & $\pm 0.034$ \\ 
$16$ & $192$ & $399$ & $0.822$ & $0.984$ & $2.883$ & $1.307$ & $0.1587$  & $1.29607$ & $1.7245$ & $-19.092$ & $1.535$ & $-0.022$ & $-48.366$ & $2.237$& $4.543$ & $1.924$ & $1.295$ & $1.377$\\ 
 &  &  &  & & $\pm 0.009$ & $\pm 0.095$ &$\pm 0.0002$  & $\pm 0.00330$ & $\pm 0.4923$ &$\pm 13.253$ & $\pm 0.009$ & $\pm 0.422$ & $\pm 16.502$ & $\pm 0.026$& $\pm 1.051$ & $\pm 0.032$ & $\pm 0.768$ & $\pm 0.036$ \\ 
\hline\hline
\end{tabular}
\end{minipage}
              }
\end{table*} 

\begin{table*}
\rotatebox{90}{
\begin{minipage}[t]{1.0\textheight}
\caption{
Finite-size scaling analysis of $\Lambda^{\,}_{\perp}$ 
for the two-dimensional spin-directed $\mathbb{Z}^{\,}_{2}$ network model 
along the cut (\ref{eq: cut choice 2}) 
[B: $(0.5,5\pi/16,\delta^{2})$ for $\delta^{2}\in[0,0.5]$].
Only the values of $\Lambda^{\,}_{\parallel}$ satisfying 
$1.0<\Lambda^{\,}_{\parallel}<4.0$
enter the data set. Different data sets are chosen by varying the
minimum $M^{\,}_{\mathrm{min}}$ and maximum $M^{\,}_{\mathrm{max}}$
taken by the width $M$ 
of the two-dimensional spin-directed $\mathbb{Z}^{\,}_{2}$ network model. 
The quality of fit is measured by
$\chi^{2}_{\mathrm{min}}/N$ and $\mathcal{G}$.
The values $\nu$, $y$, $\delta^{2}_{\mathrm{c}}$, $\Lambda^{\mathrm{c}}_{\parallel}$,
$F^{p,0}_{\parallel}$, and $f^{p,q}_{\parallel}$ for the fitting parameters 
and their $\chi^{2}_{\mathrm{min}}/N$ are shown.
The numbers with $\pm$ are the statistical error bars (one sigma).
\label{tab:fitting parameters cut 2 parallel}
        }
\begin{tabular}{c c c|c c | r r|r| r r |r r | r r| r r| r r}
\hline\hline
$M^{\,}_{\mathrm{min}}$ 
&  
$M^{\,}_{\mathrm{max}}$ 
& 
$N$ 
& 
$\chi^{2}_{\mathrm{min}}/N$ 
& 
$\mathcal{G}$ 
&
$\nu$ 
& 
$|y|$ 
&
$\delta^{2}_{\mathrm{c}}$
& 
$\Lambda^{\mathrm{c}}_{\parallel}\equiv(1/F^{0,0}_{\parallel})$ 
& 
$f^{0,1}_{\parallel}$
& 
$F^{1,0}_{\parallel}$ 
& 
$f^{1,1}_{\parallel}$
& 
$F^{2,0}_{\parallel}$ 
& 
$f^{2,1}_{\parallel}$  
& 
$F^{3,0}_{\parallel}$ 
& 
$f^{3,1}_{\parallel}$
& 
$F^{4,0}_{\parallel}$ 
& 
$f^{4,1}_{\parallel}$\\ 
\hline
$16$ & $320$ & $450$ & $0.926$ & $0.752$ & $2.585$ & $1.355$  & $0.1595$ & $2.63191$ & $0.812$ & $0.622$ & $-3.609$ & $0.835$& $-2.811$ & $0.667$ & $-17.095$ & $0.266$ & $28.211$ \\ 
 & & & & & $\pm 0.008$ & $\pm 0.057$  & $\pm 0.0001$ & $\pm 0.00344$ & $\pm 0.105$ & $\pm 0.003$ & $\pm 0.498$ & $\pm 0.010$& $\pm 0.504$ & $\pm 0.017$ & $\pm 2.850$ & $\pm 0.028$ & $\pm 13.114$ \\ 
$24$ & $320$ & $392$ & $0.925$ & $0.717$ & $2.627$ & $1.102$  & $0.1591$ & $2.64418$ & $0.476$ & $0.640$ & $-1.779$ & $0.888$& $-2.039$ & $0.751$ & $-10.540$ & $0.312$ & $4.511$ \\ 
 & & & & & $\pm 0.018$ & $\pm 0.115$  & $\pm 0.0002$ & $\pm 0.00677$ & $\pm 0.129$ & $\pm 0.008$ & $\pm 0.516$ & $\pm 0.024$& $\pm 0.514$ & $\pm 0.041$ & $\pm 3.242$ & $\pm 0.039$ & $\pm 7.585$ \\ 
$32$ & $320$ & $341$ & $0.987$ & $0.358$ & $2.637$ & $1.089$  & $0.1590$ & $2.64916$ & $0.544$ & $0.643$ & $-1.715$ & $0.902$& $-2.504$ & $0.768$ & $-10.699$ & $0.326$ & $4.774$ \\ 
 & & & & & $\pm 0.028$ & $\pm 0.202$  & $\pm 0.0002$ & $\pm 0.01055$ & $\pm 0.282$ & $\pm 0.012$ & $\pm 0.916$ & $\pm 0.039$& $\pm 1.139$ & $\pm 0.064$ & $\pm 6.135$ & $\pm 0.048$ & $\pm 11.316$ \\ 
$48$ & $320$ & $295$ & $1.096$ & $0.045$ & $2.625$ & $1.387$  & $0.1591$ & $2.64608$ & $1.670$ & $0.638$ & $-5.847$ & $0.882$& $-5.912$ & $0.711$ & $-12.797$ & $0.359$ & $-50.833$ \\ 
 & & & & & $\pm 0.035$ & $\pm 0.478$  & $\pm 0.0003$ & $\pm 0.01358$ & $\pm 2.519$ & $\pm 0.016$ & $\pm 9.286$ & $\pm 0.048$& $\pm 7.724$ & $\pm 0.065$ & $\pm 19.002$ & $\pm 0.065$ & $\pm 102.505$ \\ 
\hline
$16$ & $256$ & $415$ & $0.904$ & $0.827$ & $2.588$ & $1.333$  & $0.1593$ & $2.63549$ & $0.801$ & $0.623$ & $-3.449$ & $0.837$& $-2.656$ & $0.666$ & $-15.829$ & $0.267$ & $25.730$ \\ 
 & & & & & $\pm 0.010$ & $\pm 0.062$  & $\pm 0.0001$ & $\pm 0.00428$ & $\pm 0.109$ & $\pm 0.004$ & $\pm 0.509$ & $\pm 0.012$& $\pm 0.489$ & $\pm 0.020$ & $\pm 2.812$ & $\pm 0.030$ & $\pm 12.860$ \\ 
$24$ & $256$ & $357$ & $0.911$ & $0.760$ & $2.663$ & $0.947$  & $0.1587$ & $2.66138$ & $0.376$ &$0.654$ & $-1.280$ &$0.934$& $-1.647$ & $0.822$ & $-6.982$ & $0.342$ & $2.410$ \\ 
 & & & & & $\pm 0.029$ & $\pm 0.127$  & $\pm 0.0003$ & $\pm 0.01149$ & $\pm 0.101$ & $\pm 0.013$ & $\pm 0.365$ & $\pm 0.041$& $\pm 0.375$ & $\pm 0.071$ & $\pm 2.208$ & $\pm 0.052$ & $\pm 5.035$ \\ 
$32$ & $256$ & $306$ & $0.970$ & $0.427$ & $2.721$ & $0.763$  & $0.1581$ & $2.68761$ & $0.307$ & $0.679$ & $-0.869$ & $1.021$& $-1.455$ & $0.952$ & $-4.433$ & $0.421$ & $0.027$ \\ 
 & & & & & $\pm 0.068$ & $\pm 0.217$  & $\pm 0.0005$ & $\pm 0.02842$ & $\pm 0.131$ & $\pm 0.033$ & $\pm 0.373$ & $\pm 0.112$& $\pm 0.502$ & $\pm 0.190$ & $\pm 2.298$ & $\pm 0.106$ & $\pm 3.875$ \\ 
\hline
$16$ & $192$ & $377$ & $0.886$ & $0.867$ & $2.561$ & $1.464$  & $0.1596$ & $2.62681$ & $1.038$ & $0.613$ & $-4.543$ & $0.812$& $-3.431$ & $0.628$ & $-22.355$ & $0.233$ & $51.727$ \\ 
 & & & & & $\pm 0.011$ & $\pm 0.074$  & $\pm 0.0002$ & $\pm 0.00483$ & $\pm 0.169$ & $\pm 0.004$ & $\pm 0.803$ & $\pm 0.013$& $\pm 0.748$ & $\pm 0.020$ & $\pm 4.719$ & $\pm 0.029$ & $\pm 23.579$ \\ 
$24$ & $192$ & $319$ & $0.920$ & $0.686$ & $2.638$ & $1.062$  & $0.1589$ & $2.65335$ & $0.491$ & $0.642$ & $-1.591$ & $0.907$& $-2.226$ & $0.777$ & $-9.744$ & $0.280$ & $10.521$ \\ 
 & & & & & $\pm 0.033$ & $\pm 0.161$  & $\pm 0.0004$ & $\pm 0.01315$ & $\pm 0.172$ & $\pm 0.014$ & $\pm 0.593$ & $\pm 0.044$& $\pm 0.687$ & $\pm 0.074$ & $\pm 4.032$ & $\pm 0.050$ & $\pm 10.589$ \\ 
$32$ & $192$ & $268$ & $0.982$ & $0.349$ & $2.697$ & $0.889$  & $0.1582$ & $2.68162$ & $0.442$ & $0.664$ & $-1.065$ & $0.991$& $-2.178$ & $0.903$ & $-6.721$ & $0.313$ & $8.302$ \\ 
 & & & & & $\pm 0.081$ & $\pm 0.301$  & $\pm 0.0007$ & $\pm 0.03407$ & $\pm 0.287$ & $\pm 0.035$ & $\pm 0.681$ & $\pm 0.123$& $\pm 1.278$ & $\pm 0.206$ & $\pm 5.229$ & $\pm 0.088$ & $\pm 12.495$ \\ 
\hline\hline
\end{tabular}
\end{minipage}
              }
\end{table*}

\begin{table*}
\rotatebox{90}{
\begin{minipage}[t]{1.0\textheight}
\caption{
Finite size scaling analysis of $\Lambda^{\,}_{\parallel}$ 
for the two-dimensional spin-directed $\mathbb{Z}^{\,}_{2}$ network model 
along the cut (\ref{eq: cut choice 3}) 
[C: $(0.6,\pi/4,\delta^{2})$ for $\delta^{2}\in[0,0.4]$].
Only the values of $\Lambda^{\,}_{\perp}$  satisfying 
$0.5<\Lambda^{\,}_{\perp}<2.0$ 
enter the data set. Different data sets are chosen by varying the
minimum $M^{\,}_{\mathrm{min}}$ and maximum $M^{\,}_{\mathrm{max}}$
taken by the width $M$ 
of the two-dimensional spin-directed $\mathbb{Z}^{\,}_{2}$ network model. 
The quality of fit is measured by
$\chi^{2}_{\mathrm{min}}/N$ and $\mathcal{G}$.
The values $\nu$, $y$, $\delta^{2}_{\mathrm{c}}$, $\Lambda^{\mathrm{c}}_{\perp}$,
$F^{p,0}_{\perp}$, and $f^{p,q}_{\perp}$ for the fitting parameters 
and their $\chi^{2}_{\mathrm{min}}/N$ are shown.
The numbers with $\pm$ are the statistical error bars (one sigma).
\label{tab:fitting parameters cut 3 perpendicular}
         }
\begin{tabular}{c c c|c c | r r|r| r r r  |r r r | r r| r r| r }
\hline\hline
$M^{\,}_{\mathrm{min}}$ 
&  
$M^{\,}_{\mathrm{max}}$ 
& 
$N$ 
& 
$\chi^{2}_{\mathrm{min}}/N$ 
& 
$\mathcal{G}$ 
&
$\nu$ 
& 
$|y|$ 
&
$\delta^{2}_{\mathrm{c}}$
& 
$\Lambda^{\mathrm{c}}_{\perp}\equiv(1/F^{0,0}_{\perp})$ 
& 
$f^{0,1}_{\perp}$ 
& 
$f^{0,2}_{\perp}$
& 
$F^{1,0}_{\perp}$ 
& 
$f^{1,1}_{\perp}$ 
& 
$f^{1,2}_{\perp}$
& 
$
F^{2,0}_{\perp}$ 
& 
$f^{2,1}_{\perp}$  
& 
$F^{3,0}_{\perp}$ & 
$f^{3,1}_{\perp}$
& 
$F^{4,0}_{\perp}$\\ 
\hline
$16$ & $256$ & $172$ & $0.570$ & $1.000$ & $2.837$ & $0.713$ & $0.2756$  & $1.24122$ & $0.1135$ & $-0.861$ & $1.420$ & $1.861$ & $-5.834$ & $1.852$& $1.835$ & $1.507$ & $3.884$ & $1.226$ \\ 
 &  &  &  & & $\pm 0.069$ & $\pm 0.138$ &$\pm 0.0004$  & $\pm 0.00760$ & $\pm 0.0579$ &$\pm 0.523$ & $\pm 0.097$ & $\pm 0.280$ & $\pm 3.660$ & $\pm 0.223$& $\pm 0.205$ & $\pm 0.302$ & $\pm 0.639$ & $\pm 0.205$ \\ 
$24$ & $256$ & $150$ & $0.536$ & $1.000$ & $2.810$ & $0.764$ & $0.2763$  & $1.22920$ & $-0.0080$ & $-0.374$ & $1.416$ & $2.138$ & $-8.279$ & $1.814$& $2.293$ & $1.436$ & $5.463$ & $1.098$ \\ 
 &  &  &  & & $\pm 0.107$ & $\pm 0.253$ &$\pm 0.0005$  & $\pm 0.00993$ & $\pm 0.1141$ &$\pm 0.916$ & $\pm 0.139$ & $\pm 0.949$ & $\pm 12.369$ & $\pm 0.337$& $\pm 0.555$ & $\pm 0.453$ & $\pm 1.618$ & $\pm 0.280$ \\ 
\hline
$16$ & $192$ & $162$ & $0.513$ & $1.000$ & $2.865$ & $0.819$ & $0.2764$  & $1.22782$ & $0.0032$ & $-0.729$ & $1.487$ & $1.993$ & $-8.833$ & $1.973$& $2.006$ & $1.674$ & $4.180$ & $1.323$ \\ 
 &  &  &  & & $\pm 0.068$ & $\pm 0.163$ &$\pm 0.0004$  & $\pm 0.00701$ & $\pm 0.0752$ &$\pm 0.638$ & $\pm 0.084$ & $\pm 0.453$ & $\pm 6.916$ & $\pm 0.206$& $\pm 0.308$ & $\pm 0.285$ & $\pm 0.899$ & $\pm 0.215$ \\ 
$24$ & $192$ & $140$ & $0.426$ & $1.000$ & $2.830$ & $0.938$ & $0.2776$  & $1.20775$ & $-0.4189$ & $3.267$ & $1.493$ & $2.702$ & $-19.406$ & $1.938$& $3.108$ & $1.599$ & $7.578$ & $1.167$ \\ 
 &  &  &  & & $\pm 0.094$ & $\pm 0.270$ &$\pm 0.0008$  & $\pm 0.01461$ & $\pm 0.2471$ &$\pm 4.782$ & $\pm 0.098$ & $\pm 1.621$ & $\pm 32.971$ & $\pm 0.264$& $\pm 1.190$ & $\pm 0.365$ & $\pm 3.573$ & $\pm 0.260$ \\ 
\hline\hline
\end{tabular}
\end{minipage}
              }
\end{table*}

\begin{table*}
\rotatebox{90}{
\begin{minipage}[t]{1.0\textheight}
\caption{
Finite size scaling analysis of $\Lambda^{\,}_{\parallel}$ 
for the two-dimensional spin-directed $\mathbb{Z}^{\,}_{2}$ network model 
along the cut (\ref{eq: cut choice 3}) 
[C: $(0.6,\pi/4,\delta^{2})$ for $\delta^{2}\in[0,0.4]$].
Only the values of $\Lambda^{\,}_{\perp}$  satisfying 
$1.5<\Lambda^{\,}_{\parallel}<4.5$
enter the data set. Different data sets are chosen by varying the
minimum $M^{\,}_{\mathrm{min}}$ and maximum $M^{\,}_{\mathrm{max}}$
taken by the width $M$ 
of the two-dimensional spin-directed $\mathbb{Z}^{\,}_{2}$ network model. 
The quality of fit is measured by
$\chi^{2}_{\mathrm{min}}/N$ and $\mathcal{G}$.
The values $\nu$, $y$, $\delta^{2}_{\mathrm{c}}$, $\Lambda^{\mathrm{c}}_{\parallel}$,
$F^{p,0}_{\parallel}$, and $f^{p,q}_{\parallel}$ for the fitting parameters 
and their $\chi^{2}_{\mathrm{min}}/N$ are shown.
The numbers with $\pm$ are the statistical error bars (one sigma).
\label{tab:fitting parameters cut 3 parallel}
        }
\begin{tabular}{c c c|c c | r r|r| r r r|r r r| r r | r r | r}
\hline\hline
$M^{\,}_{\mathrm{min}}$ 
&  
$M^{\,}_{\mathrm{max}}$ 
& 
$N$ 
& 
$\chi^{2}_{\mathrm{min}}/N$ 
& 
$\mathcal{G}$ 
&
$\nu$ 
& 
$|y|$ 
&
$\delta^{2}_{\mathrm{c}}$
& 
$\Lambda^{\mathrm{c}}_{\parallel}\equiv(1/F^{0,0}_{\parallel})$ 
& 
$f^{0,1}_{\parallel}$ 
& 
$f^{0,2}_{\parallel}$
& 
$F^{1,0}_{\parallel}$ 
& 
$f^{1,1}_{\parallel}$ 
& 
$f^{1,2}_{\parallel}$
& 
$F^{2,0}_{\parallel}$ 
& 
$f^{2,1}_{\parallel}$  
& 
$F^{3,0}_{\parallel}$ 
& 
$f^{3,1}_{\parallel}$
& 
$F^{4,0}_{\parallel}$ \\ 
\hline
$16$ & $256$ & $284$ & $0.993$ & $0.295$ & $2.513$ & $0.877$ & $0.2765$  & $2.74397$ & $-0.1021$ & $0.750$ & $0.495$ & $-1.312$ & $-0.003$ & $0.581$& $-1.405$ & $0.433$ & $-1.473$ & $0.190$ \\ 
 &  &  &  & & $\pm 0.068$ & $\pm 0.560$ &$\pm 0.0004$  & $\pm 0.01498$ & $\pm 0.1312$ &$\pm 2.186$ & $\pm 0.039$ & $\pm 1.985$ & $\pm 4.992$ & $\pm 0.084$& $\pm 1.386$ & $\pm 0.081$ & $\pm 1.542$ & $\pm 0.032$ \\ 
$24$ & $256$ & $238$ & $0.864$ & $0.804$ & $2.642$ & $0.768$ & $0.2760$  & $2.75834$ & $-0.0674$ & $0.770$ & $0.556$ & $-2.241$ & $8.971$ & $0.717$& $-1.997$ & $0.593$ & $-2.545$ & $0.256$ \\ 
 &  &  &  & & $\pm 0.080$ & $\pm 0.225$ &$\pm 0.0006$  & $\pm 0.02421$ & $\pm 0.1548$ &$\pm 1.371$ & $\pm 0.046$ & $\pm 1.089$ & $\pm 12.120$ & $\pm 0.105$& $\pm 0.772$ & $\pm 0.121$ & $\pm 1.123$ & $\pm 0.051$ \\ 
\hline
$16$ & $192$ & $277$ & $0.981$ & $0.341$ & $2.483$ & $0.991$ & $0.2767$  & $2.73993$ & $-0.1617$ & $1.569$ & $0.481$ & $-1.585$ & $-0.150$ & $0.553$& $-1.588$ & $0.405$ & $-1.700$ & $0.177$ \\ 
 &  &  &  & & $\pm 0.066$ & $\pm 0.732$ &$\pm 0.0005$  & $\pm 0.01841$ & $\pm 0.2463$ &$\pm 5.887$ & $\pm 0.038$ & $\pm 3.232$ & $\pm 9.851$ & $\pm 0.077$& $\pm 2.188$ & $\pm 0.073$ & $\pm 2.482$ & $\pm 0.030$ \\ 
$24$ & $192$ & $231$ & $0.859$ & $0.813$ & $2.644$ & $0.747$ & $0.2764$  & $2.74260$ & $-0.1485$ & $1.149$ & $0.563$ & $-2.229$ & $8.454$ & $0.727$& $-1.944$ & $0.605$ & $-2.513$ & $0.256$ \\ 
 &  &  &  & & $\pm 0.097$ & $\pm 0.260$ &$\pm 0.0008$  & $\pm 0.03223$ & $\pm 0.1824$ &$\pm 1.784$ & $\pm 0.060$ & $\pm 1.169$ & $\pm 12.712$ & $\pm 0.132$& $\pm 0.845$ & $\pm 0.151$ & $\pm 1.257$ & $\pm 0.059$ \\ 
\hline\hline
\end{tabular}
\end{minipage}
              }
\end{table*}

\begin{table*}
\begin{minipage}[t]{1.0\textwidth}
\caption{
Finite size scaling analysis of $\Lambda^{\,}_{\parallel}$ 
for the two-dimensional spin-directed $\mathbb{Z}^{\,}_{2}$ network model 
along the cut (\ref{eq: cut choice D}) [D: $(0.5,\theta,0.234)$].
Only the values of $\Lambda^{\,}_{\parallel}$  satisfying 
$2.0<\Lambda^{\,}_{\parallel}<5.0$
enter the data set. Different data sets are chosen by varying the
minimum $M^{\,}_{\mathrm{min}}$ and maximum $M^{\,}_{\mathrm{max}}$
taken by the width $M$ 
of the two-dimensional spin-directed $\mathbb{Z}^{\,}_{2}$ network model. 
The quality of fit is measured by
$\chi^{2}_{\mathrm{min}}/N$ and $\mathcal{G}$.
According to the scaling analysis along the one-dimensional cut
 (\ref{eq: cut choice 1}),
we use the values $\theta^{\,}_{\mathrm{c}}/\pi=0.25$ and
 $\Lambda^{\mathrm{c}}_{\parallel}=0.3657$ 
shown in Table \ref{tab:summary_FSS} as the known numbers.
The values $\nu$, $y$,
$F^{p,0}_{\parallel}$, and $f^{p,q}_{\parallel}$ for the fitting parameters 
and their $\chi^{2}_{\mathrm{min}}/N$ are shown.
The numbers with $\pm$ are the statistical error bars (one sigma).
\label{tab:fitting parameters cut D parallel}
}
\begin{tabular}{c c c|c c |r r|c| c r r|r r| r| r}
\hline\hline
$M_\text{min}$ &  $M_\text{max}$ & $N$ 
& $\chi^{2}_\text{min}/N$ & ${\cal G}$ 
&$\nu$ & $|y|$ 
&${\theta^{\,}_{\mathrm{c}}/\pi}^{(*)}$
& ${\Lambda^c_{\parallel} \equiv 1/F_{\parallel}^{0,0}}^{(*)} $ 
& $f_{\parallel}^{1,1}$ & $f_{\parallel}^{1,2}$
& $F_{\parallel}^{2,0}$ & $f_{\parallel}^{2,1}$ 
& $F_{\parallel}^{3,0}$
& $F_{\parallel}^{4,0}$\\
\hline
$16$ & $256$ & $78$ & $0.794$ & $0.742$ & $2.872$ & $0.387$ & $0.2500$  & $3.65700$ & $-0.0374$ & $0.095$ & $0.356$ & $1.025$ & $0.375$ & $0.141$\\ 
 &  &  &  & & $\pm 0.045$ & $\pm 0.046$ &$-$  & $-$ & $\pm 0.0062$ &$\pm 0.019$ & $\pm 0.024$ & $\pm 0.116$ & $\pm 0.017$ & $\pm 0.029$ \\ 
$24$ & $256$ & $66$ & $0.905$ & $0.414$ & $2.866$ & $0.329$ & $0.2500$  & $3.65700$ & $-0.0328$ & $0.072$ & $0.339$ & $1.034$ & $0.373$ & $0.141$\\ 
 &  &  &  & & $\pm 0.054$ & $\pm 0.066$ &$-$  & $-$ & $\pm 0.0076$ &$\pm 0.025$ & $\pm 0.035$ & $\pm 0.169$ & $\pm 0.019$ & $\pm 0.031$ \\ 
$32$ & $256$ & $55$ & $0.943$ & $0.290$ & $2.912$ & $0.299$ & $0.2500$  & $3.65700$ & $-0.0188$ & $0.026$ & $0.346$ & $0.932$ & $0.390$ & $0.151$\\ 
 &  &  &  & & $\pm 0.067$ & $\pm 0.110$ &$-$  & $-$ & $\pm 0.0104$ &$\pm 0.027$ & $\pm 0.055$ & $\pm 0.223$ & $\pm 0.025$ & $\pm 0.035$ \\ 
$48$ & $256$ & $45$ & $0.826$ & $0.461$ & $2.882$ & $0.678$ & $0.2500$  & $3.65700$ & $-0.1157$ & $0.869$ & $0.391$ & $1.700$ & $0.379$ & $0.142$\\ 
 &  &  &  & & $\pm 0.080$ & $\pm 0.176$ &$-$  & $-$ & $\pm 0.0679$ &$\pm 0.971$ & $\pm 0.032$ & $\pm 0.517$ & $\pm 0.030$ & $\pm 0.036$ \\ 
\hline
$16$ & $192$ & $73$ & $0.778$ & $0.755$ & $2.862$ & $0.417$ & $0.2500$  & $3.65700$ & $-0.0394$ & $0.107$ & $0.362$ & $1.035$ & $0.373$ & $0.137$\\ 
 &  &  &  & & $\pm 0.047$ & $\pm 0.052$ &$-$  & $-$ & $\pm 0.0069$ &$\pm 0.024$ & $\pm 0.024$ & $\pm 0.114$ & $\pm 0.017$ & $\pm 0.030$ \\ 
$24$ & $192$ & $61$ & $0.909$ & $0.382$ & $2.855$ & $0.372$ & $0.2500$  & $3.65700$ & $-0.0365$ & $0.090$ & $0.350$ & $1.032$ & $0.370$ & $0.137$\\ 
 &  &  &  & & $\pm 0.057$ & $\pm 0.080$ &$-$  & $-$ & $\pm 0.0091$ &$\pm 0.035$ & $\pm 0.035$ & $\pm 0.152$ & $\pm 0.020$ & $\pm 0.031$ \\ 
$32$ & $192$ & $50$ & $0.967$ & $0.232$ & $2.908$ & $0.367$ & $0.2500$  & $3.65700$ & $-0.0220$ & $0.035$ & $0.365$ & $0.914$ & $0.388$ & $0.152$\\ 
 &  &  &  & & $\pm 0.072$ & $\pm 0.140$ &$-$  & $-$ & $\pm 0.0136$ &$\pm 0.042$ & $\pm 0.050$ & $\pm 0.170$ & $\pm 0.026$ & $\pm 0.036$ \\ 
\hline\hline
\end{tabular}
\end{minipage}
\end{table*}

\end{widetext}

\end{appendices}



\clearpage
\setcounter{equation}{0}
\setcounter{table}{0}
\setcounter{figure}{0}
\setcounter{figure}{0}
\setcounter{section}{0}
\renewcommand{\theequation}{S\arabic{equation}}
\renewcommand{\thefigure}{S\arabic{figure}}
\renewcommand{\thetable}{S-\Roman{table}}

\begin{widetext}
{\large {\bf Supplemental Materials for ``Spin-directed network model
 for the surface states of weak three-dimensional $\mathbb{Z}^{\,}_{2}$
 topological insulators''}}


\setcounter{page}{1}

\vspace{0.2cm}
In this supplemental material, we present in more detail
the derivations of the two-dimensional Dirac Hamiltonians 
presented in Appendix A and B of the main paper. 
\vspace{0.2cm}


\section{
Quasi-one-dimensional model for the surface states of a
weak three-dimensional $\mathbb{Z}^{\,}_{2}$ topological insulator
        }
\label{suppl: Wire construction for the surface states of a 3D Z2 WTI}

\subsection{
Definition
           }

Figure~\ref{fig:z2pancake} 
depicts a two-dimensional $\mathbb{Z}^{\,}_{2}$
topological \textit{band} insulator. 
The interior of the ellipse shown
in Fig.~\ref{fig:z2pancake}(a) 
is called the bulk.
The boundary of the ellipse shown
in Fig.~\ref{fig:z2pancake}(a) 
is the edge. 
Figure~\ref{fig:z2pancake}
represents a model of noninteracting electrons such that
(i) the single-particle eigenstates with support in the bulk 
display a spectral gap $\Delta$
as is indicated in  Fig.~\ref{fig:z2pancake}(b), 
while (ii) the single-particle eigenstates with support on the edge
realize a two-fold degenerate dispersion that crosses the spectral
gap of the bulk states as is indicated in Fig.~\ref{fig:z2pancake}(b), 
are extended along the edge, but are exponentially localized in the direction
perpendicular to the edge. These edge states represent a single pair of
Kramers' degenerate electrons propagating with opposite velocities.
These edge states are also called helical states as the expectation values
of the electronic spins are opposite for each electron forming the 
Kramers' degenerate pair and change with the momenta $k$
of the electrons parallel to the edge.
The low-energy and long-wave-length effective Hamiltonian of the single pair
of helical states depicted in Fig.~\ref{fig:z2pancake}(a) is
\begin{subequations}
\begin{equation}
\hat{H}^{\,}_{\mathrm{Helical}}:=
\int\limits_{\mathrm{edge}} \mathrm{d}x\,
\left(
\hat{\Psi}^{\dag}
(-{i})\,
v\,
\sigma^{\,}_{3}\,
\partial^{\,}_{x}\,
\hat{\Psi}
\right)(x).
\label{seq: def single z2pancake hamiltonian}
\end{equation}
Units are chosen so that $\hbar=1$. The speed $v$ is positive
by convention. The operators
$\hat{\Psi}^{\dag}_{\alpha}(x)$
and
$\hat{\Psi}^{\,}_{\alpha}(x)$
create and destroy at the position $x$ along the edge
an electron with the projection
$\alpha=\uparrow,\downarrow$
of its spin along the spin quantization axis, respectively.
They make up the doublet of operators
$\hat{\Psi}^{\dag}(x)$
and
$\hat{\Psi}(x)$,
respectively. The Pauli matrices 
$\sigma^{\,}_{1}$,
$\sigma^{\,}_{2}$,
and
$\sigma^{\,}_{3}$
act on the spin components of the electrons.
The unit $2\times2$ matrix in spin space is denoted $\sigma^{\,}_{0}$.
Hamiltonian~(\ref{seq: def single z2pancake hamiltonian})
is invariant under the operation of time reversal defined by
\begin{equation}
\hat{\Psi}^{\dag}(x)=
\hat{\Psi}^{\prime\dag}(x)\,
K\,
\sigma^{\,}_{2},
\qquad
\hat{\Psi}(x)=
\sigma^{\,}_{2}\,K\,
\hat{\Psi}^{\prime}(x),
\end{equation}
\end{subequations}
where $K$ denotes the operation of complex conjugation.

A layered microscopic model that captures the tunneling of
helical edge states between adjacent layers for energy scales below
the bulk gap $\Delta$ 
is defined by the Hamiltonian
\begin{subequations}
\label{seq: def layered microscopic model}
\begin{equation}
\begin{split}
\hat{H}^{\,}_{\mathrm{layered}}:=&\,
\int\limits_{\mathrm{edge}} \mathrm{{d}}x\,
\sum_{n=1}^{2N}
\left[
\left(
\hat{\Psi}^{\dag}_{n}
(-{i})\,
v^{\,}_{n}\,
\sigma^{\,}_{3}\,
\partial^{\,}_{x}\,
\hat{\Psi}^{\,}_{n}
\right)(x)
+
\hat{\Psi}^{\dag}_{n}(x)\,
\mu^{\,}_{n}(x)\,
\sigma^{\,}_{0}\,
\hat{\Psi}^{\,}_{n}(x)
\right]
\\
&\,
+
\int\limits_{\mathrm{edge}} \mathrm{{d}}x\,
\sum_{n=1}^{2N-1}
\left[
\hat{\Psi}^{\dag}_{n+1}(x)
\left(
\frac{\lambda^{\,}_{n,0}}{2}
\sigma^{\,}_{0}+
{i}
\sum_{j=1}^{3}
\frac{\lambda^{\,}_{n,j}}{2}
\sigma^{\,}_{j}
\right)
\hat{\Psi}^{\,}_{n}(x)
+
\mathrm{H.c.}
\right].
\end{split}
\label{seq: def layered z2pancake hamiltonian}
\end{equation}
There is an even number of layers $2N$.
Each layer $n$ with $n=1,\cdots,2N$
has its own Fermi velocity $v^{\,}_{n}>0$
and chemical potential $\mu^{\,}_{n}\in\mathbb{R}$.
Any two consecutive layers are coupled by hopping matrix
elements parametrized by the four independent real-valued
couplings $\lambda^{\,}_{n,\mu}$ with $\mu=0,1,2,3$ and 
$n=1,\cdots,2N-1$. Open boundary conditions are chosen
along the layering axis.
Hamiltonian~(\ref{seq: def layered z2pancake hamiltonian})
is invariant under the operation of time reversal defined by
\begin{equation}
\hat{\Psi}^{\dag}_{n}(x)=
\hat{\Psi}^{\prime\dag}_{n}(x)\,
K\,
\sigma^{\,}_{2},
\qquad
\hat{\Psi}^{\,}_{n}(x)=
\sigma^{\,}_{2}\,K\,
\hat{\Psi}^{\prime}_{n}(x),
\label{seq: def time reversal}
\end{equation}
\end{subequations}
where $K$ denotes the operation of complex conjugation
and $n=1,\cdots,2N$. 
Hamiltonian~(\ref{seq: def layered z2pancake hamiltonian})
is depicted in Fig.~\ref{fig:layeredz2pancake}.

We can turn the layered model%
~(\ref{seq: def layered microscopic model}) 
into a layered microscopic model of a weak three-dimensional 
$\mathbb{Z}^{\,}_{2}$ topological \textit{band} insulator by demanding that
\begin{equation}
v^{\,}_{n}=
v^{\,}_{\mathrm{u},x}
+
(-1)^{n}\,
v^{\,}_{\mathrm{s}},
\qquad
\mu^{\,}_{n}=
\mu^{\,}_{\mathrm{u}}
+
(-1)^{n}\,
\mu^{\,}_{\mathrm{s}},
\qquad
\lambda^{\,}_{n,\mu}=
\left(1-\delta^{\,}_{n,2N}\right)
\left[
\lambda^{\,}_{\mathrm{u},\mu}
+
(-1)^{n}\,
\lambda^{\,}_{\mathrm{s},\mu}
\right],
\qquad
\mu=0,1,2,3,
\end{equation}
for $n=1,\cdots, 2N$. As we shall demonstrate, the continuum limit
\begin{subequations}
\begin{equation}
\lim_{\substack{\lambda^{\,}_{\mathrm{u},0}\to\infty\\ \mathfrak{a}^{\,}_{y}\to0}}
\frac{\lambda^{\,}_{\mathrm{u},j}}{\lambda^{\,}_{\mathrm{u},0}}=0,
\qquad
j=1,2,3,
\qquad
\lim_{\substack{\lambda^{\,}_{\mathrm{u},0}\to\infty\\ \mathfrak{a}^{\,}_{y}\to0}}
\frac{\lambda^{\,}_{\mathrm{s},\mu}}{\lambda^{\,}_{\mathrm{u},0}}=0,
\qquad
\mu=0,1,2,3,
\end{equation}
where
\begin{equation}
\lim_{\substack{\lambda^{\,}_{\mathrm{u},0}\to\infty\\ \mathfrak{a}^{\,}_{y}\to0}}
\frac{\lambda^{\,}_{\mathrm{u},0}}{2}\times(2\mathfrak{a}^{\,}_{y})\equiv
v^{\,}_{\mathrm{u},y}>0
\label{eq: infinite band width limit}
\end{equation}
\end{subequations}
is a finite non-vanishing and positive number
($\mathfrak{a}^{\,}_{y}$ is the interlayer lattice spacing),
is related to Dirac fermions in a four-dimensional representation of
the Clifford algebra in $(2+1)$-dimensional space and time.

\subsection{
Generic stacked microscopic model 
and its continuum limit in the stacking direction
           }

\subsubsection{
Definition
              }

Motivated by Hamiltonian%
~(\ref{seq: def layered z2pancake hamiltonian}),
we define Hamiltonian
\begin{subequations}
\begin{equation}
\hat{H}:=
\int\limits_{\mathrm{edge}} \mathrm{{d}}x\,
\sum_{n=1}^{2N}
\left\{
\hat{\Psi}^{\dag}_{n}\,
\left[
\mathsf{V}^{\,}_{n}\,
(-{i})\partial^{\,}_{x}\,
+
\mathsf{U}^{\,}_{n}
\right]
\hat{\Psi}^{\,}_{n}
+
\left(
\hat{\Psi}^{\dag}_{n}\,
\mathsf{T}^{\,}_{n}\,
\hat{\Psi}^{\,}_{n+1}
+
\hat{\Psi}^{\dag}_{n+1}\,
\mathsf{T}^{\dag}_{n}\,
\hat{\Psi}^{\,}_{n}
\right)
\right\},
\label{eq: def H generic}
\end{equation}
whereby periodic boundary conditions have been assumed in the layering index
$n$, i.e., $n+2N\equiv n$, 
the $2\times2$ matrices 
$\mathsf{V}^{\,}_{n}$ 
and
$\mathsf{U}^{\,}_{n}$ 
are Hermitean,
and the $2\times2$ matrix $\mathsf{T}^{\,}_{n}$ 
is arbitrary. Reversal of time is implemented by the transformation
\begin{equation}
\hat{\Psi}^{\dag}_{n}(x)=
\hat{\Psi}^{\prime\dag}_{n}(x)\,
K\,
\sigma^{\,}_{2},
\qquad
\hat{\Psi}^{\,}_{n}(x)=
\sigma^{\,}_{2}\,K\,
\hat{\Psi}^{\prime}_{n}(x).
\label{seq: def time reversal bis}
\end{equation}
\end{subequations}
If we write
\begin{subequations}
\begin{equation}
\mathsf{V}^{\,}_{n}=
v^{\,}_{n,0}\,\sigma^{\,}_{0}
+
v^{\,}_{n,1}\,\sigma^{\,}_{1}
+
v^{\,}_{n,2}\,\sigma^{\,}_{2}
+
v^{\,}_{n,3}\,\sigma^{\,}_{3},
\qquad 
v^{\,}_{n,\mu}\in\mathbb{R},
\qquad
\mu=0,1,2,3,
\end{equation}
\begin{equation}
\mathsf{U}^{\,}_{n}=
U^{\,}_{n,0}\,\sigma^{\,}_{0}
+
U^{\,}_{n,1}\,\sigma^{\,}_{1}
+
U^{\,}_{n,2}\,\sigma^{\,}_{2}
+
U^{\,}_{n,3}\,\sigma^{\,}_{3},
\qquad 
U^{\,}_{n,\mu}\in\mathbb{R},
\qquad
\mu=0,1,2,3,
\end{equation}
and
\begin{equation}
\mathsf{T}^{\,}_{n}=
t^{\prime}_{n,0}\,\sigma^{\,}_{0}
+
t^{\prime}_{n,1}\,\sigma^{\,}_{1}
+
t^{\prime}_{n,2}\,\sigma^{\,}_{2}
+
t^{\prime}_{n,3}\,\sigma^{\,}_{3}
+
{i}
\left(
t^{\prime\prime}_{n,0}\,\sigma^{\,}_{0}
+
t^{\prime\prime}_{n,1}\,\sigma^{\,}_{1}
+
t^{\prime\prime}_{n,2}\,\sigma^{\,}_{2}
+
t^{\prime\prime}_{n,3}\,\sigma^{\,}_{3}
\right),
\quad 
t^{\prime}_{n,\mu},t^{\prime\prime}_{n,\mu}\in\mathbb{R},
\quad
\mu=0,1,2,3,
\end{equation}
\end{subequations}
we find that
\begin{subequations}
\label{appeq: hat H as hat H e + hat H o}
\begin{equation}
\hat{H}=
\hat{H}^{(\mathrm{e})}
+
\hat{H}^{(\mathrm{o})}.
\end{equation}
Here,
\begin{equation}
\begin{split}
&
\hat{H}^{(\mathrm{e})}=
\int\limits_{\mathrm{edge}} \mathrm{d}x\,
\sum_{n=1}^{2N}
\left\{
\hat{\Psi}^{\dag}_{n}\,
\left[
\mathsf{V}^{(\mathrm{o})}_{n}\,
(-{i})\partial^{\,}_{x}
+
\mathsf{U}^{(\mathrm{e})}_{n}
\right]
\hat{\Psi}^{\,}_{n}
+
\left(
\hat{\Psi}^{\dag}_{n}\,
\mathsf{T}^{(\mathrm{e})}_{n}\,
\hat{\Psi}^{\,}_{n+1}
+
\hat{\Psi}^{\dag}_{n+1}\,
\mathsf{T}^{(\mathrm{e})\dag}_{n}\,
\hat{\Psi}^{\,}_{n}
\right)
\right\},
\\
&
\mathsf{V}^{(\mathrm{o})}_{n}=
v^{\,}_{n,1}\,\sigma^{\,}_{1}
+
v^{\,}_{n,2}\,\sigma^{\,}_{2}
+
v^{\,}_{n,3}\,\sigma^{\,}_{3},
\\
&
\mathsf{U}^{(\mathrm{e})}_{n}=
U^{\,}_{n,0}\,\sigma^{\,}_{0},
\\
&
\mathsf{T}^{(\mathrm{e})}_{n}=
t^{\prime}_{n,0}\,\sigma^{\,}_{0}
+
{i}
\left(
t^{\prime\prime}_{n,1}\,\sigma^{\,}_{1}
+
t^{\prime\prime}_{n,2}\,\sigma^{\,}_{2}
+
t^{\prime\prime}_{n,3}\,\sigma^{\,}_{3}
\right),
\end{split}
\end{equation}
is even under the transformation~(\ref{seq: def time reversal bis}),
while
\begin{equation}
\begin{split}
&
\hat{H}^{(\mathrm{o})}=
\int\limits_{\mathrm{edge}}\mathrm{d}x\,
\sum_{n=1}^{2N}
\left\{
\hat{\Psi}^{\dag}_{n}
\left[
\mathsf{V}^{(\mathrm{e})}_{n}\,
(-{i})\partial^{\,}_{x}
+
\mathsf{U}^{(\mathrm{o})}_{n}
\right]
\hat{\Psi}^{\,}_{n}
+
\left(
\hat{\Psi}^{\dag}_{n}\,
\mathsf{T}^{(\mathrm{o})}_{n}\,
\hat{\Psi}^{\,}_{n+1}
+
\hat{\Psi}^{\dag}_{n+1}\,
\mathsf{T}^{(\mathrm{o})\dag}_{n}\,
\hat{\Psi}^{\,}_{n}
\right)
\right\},
\\
&
\mathsf{V}^{(\mathrm{e})}_{n}=
v^{\,}_{n,0}\,\sigma^{\,}_{0},
\\
&
\mathsf{U}^{(\mathrm{o})}_{n}=
U^{\,}_{n,1}\,\sigma^{\,}_{1}
+
U^{\,}_{n,2}\,\sigma^{\,}_{2}
+
U^{\,}_{n,3}\,\sigma^{\,}_{3},
\\
&
\mathsf{T}^{(\mathrm{o})}_{n}=
t^{\prime}_{n,1}\,\sigma^{\,}_{1}
+
t^{\prime}_{n,2}\,\sigma^{\,}_{2}
+
t^{\prime}_{n,3}\,\sigma^{\,}_{3}
+
{i}
t^{\prime\prime}_{n,0}\,\sigma^{\,}_{0},
\end{split}
\end{equation}
\end{subequations}
is odd under the transformation~(\ref{seq: def time reversal bis}).

\begin{table}[t]
\begin{center}
\caption{
The first three tables from left to right give the
transformation laws of $\sigma^{\ }_{\mu}\otimes\tau^{\ }_{\nu}$
with $\mu,\nu=0,\cdots,3$
under complex conjugation $*$,
multiplication from the left and from the right by 
$\sigma^{\ }_{2}\otimes\tau^{\ }_{0}$,
and by composition of the two operations, respectively.
        }
\label{tab : trsf laws X mu nu before gauge trsf}
\begin{tabular}{c c c c c}
$\hphantom{\sigma^{\,}_{0}}\ $\vline& $\tau^{\ }_{0}$ & $\tau^{\ }_{1}$ & $\tau^{\ }_{2}$ & $\tau^{\ }_{3}$ \\
\hline
$\sigma^{\ }_{0}\ $\vline           &       $+$       &       $+$       &       $-$       &       $+$       \\
$\sigma^{\ }_{1}\ $\vline           &       $+$       &       $+$       &       $-$       &       $+$       \\
$\sigma^{\ }_{2}\ $\vline           &       $-$       &       $-$       &       $+$       &       $-$       \\
$\sigma^{\ }_{3}\ $\vline           &       $+$       &       $+$       &       $-$       &       $+$       
\end{tabular}
\qquad
\begin{tabular}{c c c c c}
$\hphantom{\sigma^{\,}_{0}}\ $\vline& $\tau^{\ }_{0}$ & $\tau^{\ }_{1}$ & $\tau^{\ }_{2}$ & $\tau^{\ }_{3}$ \\
\hline
$\sigma^{\ }_{0}\ $\vline           &       $+$       &       $+$       &       $+$       &       $+$       \\
$\sigma^{\ }_{1}\ $\vline           &       $-$       &       $-$       &       $-$       &       $-$       \\
$\sigma^{\ }_{2}\ $\vline           &       $+$       &       $+$       &       $+$       &       $+$       \\
$\sigma^{\ }_{3}\ $\vline           &       $-$       &       $-$       &       $-$       &       $-$       
\end{tabular}
\qquad
\begin{tabular}{c c c c c}
$\hphantom{\sigma^{\,}_{0}}\ $\vline& $\tau^{\ }_{0}$ & $\tau^{\ }_{1}$ & $\tau^{\ }_{2}$ & $\tau^{\ }_{3}$ \\
\hline
$\sigma^{\ }_{0}\ $\vline           &       $+$       &       $+$       &       $-$       &       $+$       \\
$\sigma^{\ }_{1}\ $\vline           &       $-$       &       $-$       &       $+$       &       $-$       \\
$\sigma^{\ }_{2}\ $\vline           &       $-$       &       $-$       &       $+$       &       $-$       \\
$\sigma^{\ }_{3}\ $\vline           &       $-$       &       $-$       &       $+$       &       $-$       
\end{tabular}
\end{center}
\end{table}

\subsubsection{
Gauge transformation
              }

We do the gauge transformation
\begin{equation}
\hat{\Psi}^{\dag}_{n}\to
(+{i})^{n}\,
\hat{\Psi}^{\dag}_{n},
\qquad
\hat{\Psi}^{\,}_{n}\to
(-{i})^{n}\,
\hat{\Psi}^{\,}_{n}.
\end{equation}
Under this gauge transformation
\begin{subequations}
\label{eq: def H generic bis}
\begin{equation}
\hat{H}:=
\int\limits_{\mathrm{edge}} \mathrm{d}x\,
\sum_{n=1}^{2N}
\left\{
\hat{\Psi}^{\dag}_{n}
\left[
\mathsf{V}^{\,}_{n}\,
(-{i})\partial^{\,}_{x}\,
+
\mathsf{U}^{\,}_{n}
\right]
\hat{\Psi}^{\,}_{n}
+
(-{i})
\left(
\hat{\Psi}^{\dag}_{n}\,
\mathsf{T}^{\,}_{n}\,
\hat{\Psi}^{\,}_{n+1}
-
\hat{\Psi}^{\dag}_{n+1}\,
\mathsf{T}^{\dag}_{n}\,
\hat{\Psi}^{\,}_{n}
\right)
\right\},
\label{eq: def H generic bis a}
\end{equation}
while the operation~(\ref{seq: def time reversal bis})
of time reversal reads
\begin{equation}
\hat{\Psi}^{\dag}_{n}(x)=
\hat{\Psi}^{\prime\dag}_{n}(x)\,
K\,
\sigma^{\,}_{2}\,
(-{i})^{n},
\qquad
\hat{\Psi}^{\,}_{n}(x)=
(+{i})^{n}\,
\sigma^{\,}_{2}\,K\,
\hat{\Psi}^{\prime}_{n}(x).
\label{eq: def H generic bis b} 
\end{equation}
\end{subequations}

\subsubsection{
Preparation for the continuum limit in the stacking direction
              }

The continuum limit in the stacking direction
consists in doing the replacements
\begin{subequations}
\label{eq: expansion for continuum limit}
\begin{equation}
\begin{split}
&
\hat{\Psi}^{\,}_{2n}(x)\to 
(2\mathfrak{a}^{\,}_{y})^{1/2}\,
\hat{\eta}^{\,}_{1}(x,y),
\qquad
\hat{\Psi}^{\,}_{2n\pm2}(x)\to 
(2\mathfrak{a}^{\,}_{y})^{1/2}\,
\left[
\hat{\eta}^{\,}_{1}(x,y)
\pm
(2\mathfrak{a}^{\,}_{y})
\left(\frac{\partial \hat{\eta}^{\,}_{1}}{\partial y}\right)(x,y)
+
\cdots
\right],
\\
&
\hat{\Psi}^{\,}_{2n+1}(x)\to 
(2\mathfrak{a}^{\,}_{y})^{1/2}\,
\hat{\eta}^{\,}_{2}(x,y),
\qquad
\hat{\Psi}^{\,}_{2n+1\pm2}(x)\to 
(2\mathfrak{a}^{\,}_{y})^{1/2}\,
\left[
\hat{\eta}^{\,}_{2}(x,y)
\pm
(2\mathfrak{a}^{\,}_{y})
\left(\frac{\partial \hat{\eta}^{\,}_{2}}{\partial y}\right)(x,y)
+
\cdots
\right],
\end{split}
\label{eq: expansion for continuum limit a}
\end{equation}
for the operators,
\begin{equation}
\begin{split}
&
\left.
\begin{array}{l}
\mathsf{V}^{\,}_{2n}=
V^{\,}_{+}(x,y)
+
V^{\,}_{-}(x,y)
\\
\\
\mathsf{V}^{\,}_{2n+1}=
V^{\,}_{+}(x,y)
-
V^{\,}_{-}(x,y)
\end{array}
\right\}
\qquad
V^{\,}_{\pm}(x,y)=
V^{\dag}_{\pm}(x,y)
\end{split}
\label{eq: expansion for continuum limit b}
\end{equation}
for the velocity matrices,
\begin{equation}
\begin{split}
&
\left.
\begin{array}{l}
\mathsf{U}^{\,}_{2n}=
U^{\,}_{+}(x,y)
+
U^{\,}_{-}(x,y)
\\
\\
\mathsf{U}^{\,}_{2n+1}=
U^{\,}_{+}(x,y)
-
U^{\,}_{-}(x,y)
\end{array}
\right\}
\qquad
U^{\,}_{\pm}(x,y)=
U^{\dag}_{\pm}(x,y)
\end{split}
\label{eq: expansion for continuum limit c}
\end{equation}
for the on-site potential matrices,
\begin{equation}
\begin{split}
&
\mathsf{T}^{\,}_{2n}=
W^{\,}_{+}(x,y)
+
W^{\,}_{-}(x,y),
\qquad
\mathsf{T}^{\dag}_{2n}=
W^{\dag}_{+}(x,y)
+
W^{\dag}_{-}(x,y),
\\
&
\mathsf{T}^{\,}_{2n+1}=
W^{\,}_{+}(x,y)
-
W^{\,}_{-}(x,y)
+
(2\mathfrak{a}^{\,}_{y})\,
\left(
\partial^{\,}_{y} W^{\,}_{+}
\right)(x,y)
-
(2\mathfrak{a}^{\,}_{y})\,
\left(
\partial^{\,}_{y} W^{\,}_{-}
\right)(x,y),
\\ 
&
\mathsf{T}^{\dag}_{2n-1}=
W^{\dag}_{+}(x,y)
-
W^{\dag}_{-}(x,y)
-
(2\mathfrak{a}^{\,}_{y})\,
\left(
\partial^{\,}_{y} W^{\dag}_{+}
\right)(x,y)
+
(2\mathfrak{a}^{\,}_{y})\,
\left(
\partial^{\,}_{y} W^{\dag}_{-}
\right)(x,y),
\end{split}
\label{eq: expansion for continuum limit d}
\end{equation}
for the tunneling matrices, and
\begin{equation}
\sum_{n=1}^{N}\,(2\mathfrak{a}^{\,}_{y})\equiv
\int\limits_{0}^{N(2\mathfrak{a}^{\,}_{y})}\mathrm{{d}}y
\label{eq: expansion for continuum limit e}
\end{equation}
\end{subequations}
once the continuum limit has been taken.
In anticipation of the continuum limit, we shall
already use Eq.~(\ref{eq: expansion for continuum limit e}).

\begin{table}[t]
\begin{center}
\caption{
The three tables from left to right give the
transformation laws of $\sigma^{\ }_{\mu}\otimes\tau^{\ }_{\nu}$
with $\mu,\nu=0,\cdots,3$
under complex conjugation $*$,
multiplication from the left and from the right by 
$\sigma^{\ }_{2}\otimes\tau^{\ }_{3}$,
and by composition of the two operations, respectively.
        }
\label{tab : trsf laws X mu nu}
\hfill
\begin{tabular}{c c c c c}
$\hphantom{\sigma^{\,}_{0}}\ $\vline& $\tau^{\ }_{0}$ & $\tau^{\ }_{1}$ & $\tau^{\ }_{2}$ & $\tau^{\ }_{3}$ \\
\hline
$\sigma^{\ }_{0}\ $\vline           &       $+$       &       $+$       &       $-$       &       $+$       \\
$\sigma^{\ }_{1}\ $\vline           &       $+$       &       $+$       &       $-$       &       $+$       \\
$\sigma^{\ }_{2}\ $\vline           &       $-$       &       $-$       &       $+$       &       $-$       \\
$\sigma^{\ }_{3}\ $\vline           &       $+$       &       $+$       &       $-$       &       $+$       
\end{tabular}
\hfill
\begin{tabular}{c c c c c}
$\hphantom{\sigma^{\,}_{0}}\ $\vline& $\tau^{\ }_{0}$ & $\tau^{\ }_{1}$ & $\tau^{\ }_{2}$ & $\tau^{\ }_{3}$ \\
\hline
$\sigma^{\ }_{0}\ $\vline           &       $+$       &       $-$       &       $-$       &       $+$       \\
$\sigma^{\ }_{1}\ $\vline           &       $-$       &       $+$       &       $+$       &       $-$       \\
$\sigma^{\ }_{2}\ $\vline           &       $+$       &       $-$       &       $-$       &       $+$       \\
$\sigma^{\ }_{3}\ $\vline           &       $-$       &       $+$       &       $+$       &       $-$       
\end{tabular}
\hfill
\begin{tabular}{c c c c c}
$\hphantom{\sigma^{\,}_{0}}\ $\vline& $\tau^{\ }_{0}$ & $\tau^{\ }_{1}$ & $\tau^{\ }_{2}$ & $\tau^{\ }_{3}$ \\
\hline
$\sigma^{\ }_{0}\ $\vline           &       $+$       &       $-$       &       $+$       &       $+$       \\
$\sigma^{\ }_{1}\ $\vline           &       $-$       &       $+$       &       $-$       &       $-$       \\
$\sigma^{\ }_{2}\ $\vline           &       $-$       &       $+$       &       $-$       &       $-$       \\
$\sigma^{\ }_{3}\ $\vline           &       $-$       &       $+$       &       $-$       &       $-$       
\end{tabular}
\hfill$\hphantom{AAA}$
\end{center}
\end{table}

With the help of the compacter notation
\begin{subequations}
\begin{equation}
\hat{\chi}:=
\begin{pmatrix}
\hat{\eta}^{\,}_{1}
\\
\hat{\eta}^{\,}_{2}
\end{pmatrix}
\end{equation}
and the decomposition 
\begin{equation}
A=
\frac{A+A^{\dag}}{2}
+
{i}
\left(
\frac{A-A^{\dag}}{2{i}}
\right)\equiv
\mathrm{Re}\,A
+
{i}\,
\mathrm{Im}\,A
\end{equation}
of any matrix $A$ into its real and imaginary parts
$\mathrm{Re}\,A=(\mathrm{Re}\,A)^{\dag}$
and
$\mathrm{Im}\,A=(\mathrm{Im}\,A)^{\dag}$,
respectively, we find that,
with the help of Table~\ref{tab : trsf laws X mu nu},
we may decompose the expansion up to first order in powers of
$\mathfrak{a}^{\,}_{y}$ of the single-particle Hamiltonian%
~(\ref{eq: def H generic bis a})
into its even and odd contributions under reversal of time%
~(\ref{eq: def H generic bis b}) according to
\begin{equation}
\hat{H}=
\int\limits_{edge}\mathrm{{d}}x
\int\limits_{0}^{N(2\mathfrak{a}^{\,}_{y})}\mathrm{{d}}y\,
\mathcal{H}(x,y),
\qquad
\mathcal{H}(x,y)=
\mathcal{H}^{(\mathrm{e})}(x,y)
+
\mathcal{H}^{(\mathrm{o})}(x,y).
\end{equation}
Here, we need to introduce the unit $2\times2$ matrix $\tau^{\,}_{0}$
and the Pauli matrices $\tau^{\,}_{1}$, $\tau^{\,}_{2}$, and $\tau^{\,}_{3}$
with the help of which
\begin{equation}
\begin{split}
\mathcal{H}^{(\mathrm{e})}(x,y)=&\,
\left[
V^{(\mathrm{o},0)}_{+}(x,y)\,
\otimes\,
\tau^{\,}_{0}\,
+
V^{(\mathrm{o},3)}_{-}(x,y)\,
\otimes\,
\tau^{\,}_{3}
\right]
(-{i})\partial^{\,}_{x}
\\
&\,
+
U^{(\mathrm{e},0)}_{+}(x,y)\,
\otimes\,
\tau^{\,}_{0}\,
+
U^{(\mathrm{e},3)}_{-}(x,y)\,
\otimes\,
\tau^{\,}_{3}
\\
&\,
+
2\,
\left(
\mathrm{Im}\,
W^{\,}_{+}
\right)^{(\mathrm{e},1)}(x,y)\,
\otimes
\tau^{\,}_{1}\,
+
2\,
\left(
\mathrm{Re}\,
W^{\,}_{-}
\right)^{(\mathrm{e},2)}(x,y)\,
\otimes
\tau^{\,}_{2}\,
\\
&\,
+
(2\mathfrak{a}^{\,}_{y})
\left\{
(-{i})\partial^{\,}_{y}\,
\left[
\mathrm{Re}\,
\left(
W^{\dag}_{+}
-
W^{\dag}_{-}
\right)
\right]^{(\mathrm{o},1)}
(x,y)
\otimes
\tau^{\,}_{1}
-
(-{i})\partial^{\,}_{y}\,
\left[
\mathrm{Im}\,
\left(
W^{\dag}_{+}
-
W^{\dag}_{-}
\right)
\right]^{(\mathrm{o},2)}
(x,y)
\otimes
\tau^{\,}_{2}
\right\}
\\
&\,
+
(2\mathfrak{a}^{\,}_{y})
\left\{
\left[
\mathrm{Re}\,
\Big(
W^{\dag}_{+}
-
W^{\dag}_{-}
\Big)
\right]^{(\mathrm{o},1)}(x,y)
\otimes
\tau^{\,}_{1}\,
(-{i})\partial^{\,}_{y}
-
\left[
\mathrm{Im}\,
\Big(
W^{\dag}_{+}
-
W^{\dag}_{-}
\Big)
\right]^{(\mathrm{o},2)}(x,y)
\otimes
\tau^{\,}_{2}\,
(-{i})\partial^{\,}_{y}
\right\}
\label{eq: mathcal{H} even under TRS a}
\end{split}
\end{equation}
and
\begin{equation}
\begin{split}
\mathcal{H}^{(\mathrm{o})}(x,y)=&\,
\left[
V^{(\mathrm{e},0)}_{+}(x,y)\,
\otimes\,
\tau^{\,}_{0}\,
+
V^{(\mathrm{e},3)}_{-}(x,y)\,
\otimes\,
\tau^{\,}_{3}
\right]
(-{i})\partial^{\,}_{x}
\\
&\,
+
U^{(\mathrm{o},0)}_{+}(x,y)\,
\otimes\,
\tau^{\,}_{0}\,
+
U^{(\mathrm{o},3)}_{-}(x,y)\,
\otimes\,
\tau^{\,}_{3}
\\
&\,
+
2\,
\left(
\mathrm{Im}\,
W^{\,}_{+}
\right)^{(\mathrm{o},1)}(x,y)\,
\otimes
\tau^{\,}_{1}\,
+
2\,
\left(
\mathrm{Re}\,
W^{\,}_{-}
\right)^{(\mathrm{o},2)}(x,y)\,
\otimes
\tau^{\,}_{2}\,
\\
&\,
+
(2\mathfrak{a}^{\,}_{y})
\left\{
(-{i})\partial^{\,}_{y}\,
\left[
\mathrm{Re}\,
\left(
W^{\dag}_{+}
-
W^{\dag}_{-}
\right)
\right]^{(\mathrm{e},1)}
(x,y)
\otimes
\tau^{\,}_{1}
-
(-{i})\partial^{\,}_{y}\,
\left[
\mathrm{Im}\,
\left(
W^{\dag}_{+}
-
W^{\dag}_{-}
\right)
\right]^{(\mathrm{e},2)}
(x,y)
\otimes
\tau^{\,}_{2}
\right\}
\\
&\,
+
(2\mathfrak{a}^{\,}_{y})
\left\{
\left[
\mathrm{Re}\,
\Big(
W^{\dag}_{+}
-
W^{\dag}_{-}
\Big)
\right]^{(\mathrm{e},1)}(x,y)
\otimes
\tau^{\,}_{1}\,
(-{i})\partial^{\,}_{y}
-
\left[
\mathrm{Im}\,
\Big(
W^{\dag}_{+}
-
W^{\dag}_{-}
\Big)
\right]^{(\mathrm{e},2)}(x,y)
\otimes
\tau^{\,}_{2}\,
(-{i})\partial^{\,}_{y}
\right\}.
\end{split}
\end{equation}
For any $A:=a^{\,}_{\mu}\,\sigma^{\,}_{\mu}=A^{\dag}$ 
with $a^{\,}_{\mu}\in\mathbb{R}$ for $\mu=0,1,2,3$,
the notation
\begin{equation}
A\otimes\tau^{\,}_{\nu}=
A^{(\mathrm{e},\nu)}\otimes\tau^{\,}_{\nu}
+
A^{(\mathrm{o},\nu)}\otimes\tau^{\,}_{\nu}
\end{equation}
is defined by demanding that
\begin{equation}
\begin{split}
&
\hbox{
for
$A^{(\mathrm{e},\nu)}\otimes\tau^{\,}_{\nu}$,
     }\qquad
\sigma^{\,}_{2}\otimes\tau^{\,}_{3}\,
\left(
A^{(\mathrm{e},\nu)}\otimes\tau^{\,}_{\nu}
\right)^{*}
\sigma^{\,}_{2}\otimes\tau^{\,}_{3}=
+
A^{(\mathrm{e},\nu)}\otimes\tau^{\,}_{\nu},
\\
&
\hbox{
for
$A^{(\mathrm{o},\nu)}\otimes\tau^{\,}_{\nu}$,
     }\qquad
\sigma^{\,}_{2}\otimes\tau^{\,}_{3}\,
\left(
A^{(\mathrm{o},\nu)}\otimes\tau^{\,}_{\nu}
\right)^{*}
\sigma^{\,}_{2}\otimes\tau^{\,}_{3}=
-
A^{(\mathrm{o},\nu)}\otimes\tau^{\,}_{\nu},
\end{split}
\end{equation}
\end{subequations}
hold for any $\nu=0,1,2,3$.

\subsubsection{
Continuum limit that delivers the Dirac Hamiltonian when
time-reversal symmetry holds
              }

We specialize to the case when time-reversal symmetry holds.
In other words, we seek a well-defined continuum limit for
the single-particle Hamiltonian 
Eq.\ (\ref{eq: mathcal{H} even under TRS a})
where the symmetry under conjugation by
$\sigma^{\,}_{2}\otimes\tau^{\,}_{3}\,K$
($K$ complex conjugation)
implies that (see Table~\ref{tab : trsf laws X mu nu})
\begin{equation}
\label{eq: mathcal{H} even under TRS} 
\begin{split}
&
V^{(\mathrm{o},0)}_{+}=
v^{\,}_{+,j}\,\sigma^{\,}_{j},
\qquad
V^{(\mathrm{o},3)}_{-}=
v^{\,}_{-,j}\,\sigma^{\,}_{j},
\qquad
v^{\,}_{\pm,j}\in\mathbb{R},
\qquad j=1,2,3,
\\
&
U^{(\mathrm{e},0)}_{+}=
u^{\,}_{+,0}\,\sigma^{\,}_{0},
\qquad
U^{(\mathrm{e},3)}_{-}=
u^{\,}_{-,0}\,\sigma^{\,}_{0},
\qquad
u^{\,}_{\pm,0}\in\mathbb{R},
\\
&
\left(
\mathrm{Im}\,
W^{\,}_{+}
\right)^{(\mathrm{e},1)}=
w^{\prime\prime}_{+,j}\,\sigma^{\,}_{j},
\qquad
\left(
\mathrm{Re}\,
W^{\,}_{-}
\right)^{(\mathrm{e},2)}=
w^{\prime}_{-,0}\,\sigma^{\,}_{0},
\\
&
\left[
\mathrm{Re}\,
\Big(
W^{\dag}_{+}
-
W^{\dag}_{-}
\Big)
\right]^{(\mathrm{o},1)}=
+
\left(
w^{\prime}_{+,0}
-
w^{\prime}_{-,0}
\right)\,
\sigma^{\,}_{0},
\qquad
\left[
\mathrm{Im}\,
\Big(
W^{\dag}_{+}
-
W^{\dag}_{-}
\Big)
\right]^{(\mathrm{o},2)}=
-
\left(
w^{\prime\prime}_{+,j}
-
w^{\prime\prime}_{-,j}
\right)\,
\sigma^{\,}_{j}.
\end{split}
\end{equation}
Recall here that, for any $\nu=0,1,2,3$, 
evenness or oddness under time-reversal is
not imposed on the $4\times4$ complex matrices 
\begin{equation}
\begin{split}
&
W^{\,}_{\pm}\otimes\tau^{\nu}\equiv
\left(
w^{\prime}_{\pm,\mu}\,\sigma^{\,}_{\mu}
+
{i}
w^{\prime\prime}_{\pm,\mu}\,\sigma^{\,}_{\mu}
\right)
\otimes\tau^{\nu},
\qquad
w^{\prime}_{\pm,\mu},
w^{\prime\prime}_{\pm,\mu}
\in\mathbb{R},
\qquad
\mu=0,1,2,3,
\end{split}
\label{eq: mathcal{H} even under TRS b}
\end{equation}
but on their real and imaginary parts
\begin{equation}
\left(\mathrm{Re}\,W^{\,}_{\pm}\right)\otimes\tau^{\nu}=
w^{\prime}_{\pm,\mu}\,\sigma^{\,}_{\mu}\otimes\tau^{\nu},
\qquad
\left(\mathrm{Im}\,W^{\,}_{\pm}\right)\otimes\tau^{\nu}=
w^{\prime\prime}_{\pm,\mu}\,\sigma^{\,}_{\mu}\otimes\tau^{\nu},
\end{equation}
respectively.

The continuum limit involves the limit
$\mathfrak{a}^{\,}_{y}\to0$. To make sense of this limit
on the two last lines of the right-hand side of Eq.%
~(\ref{eq: mathcal{H} even under TRS a}),
we first assume the additive decompositions
into a constant (denoted by a straight overline)
and position dependent (denoted by a wiggly overline),
\begin{subequations}
\begin{equation}
v^{\,}_{\pm,\mu}(x,y)=
\bar{v}^{\,}_{\pm,\mu}
+
\tilde{v}^{\,}_{\pm,\mu}(x,y)
\end{equation}
for the ``Fermi velocities''
\begin{equation}
u^{\,}_{\pm,\mu}(x,y)=
\bar{u}^{\,}_{\pm,\mu}
+
\tilde{u}^{\,}_{\pm,\mu}(x,y)
\end{equation}
for the on-site potentials, and
\begin{equation}
w^{\prime}_{\pm,\mu}(x,y)=
\bar{w}^{\prime}_{\pm,\mu}
+
\tilde{w}^{\prime}_{\pm,\mu}(x,y),
\qquad
w^{\prime\prime}_{\pm,0}(x,y)=
\bar{w}^{\prime\prime}_{\pm,\mu}
+
\tilde{w}^{\prime\prime}_{\pm,\mu}(x,y),
\end{equation}
\end{subequations}
for the tunneling amplitudes. All the wiggly fields are smooth functions,
i.e., their $y$ derivatives are assumed to be of order 
$(\mathfrak{a}^{\,}_{y})^{0}=1$.
Second, we take the limit of infinite band width along the $y$ direction
by which
\begin{subequations}
\begin{equation}
\bar{w}^{\prime}_{+,0}\times(2\mathfrak{a}^{\,}_{y})\equiv v^{\,}_{\mathrm{u},y}
\label{eq:definition v_{u,y}}
\end{equation}
is held fixed to a non-vanishing and positive number
as $\bar{w}^{\prime}_{+,0}\to\infty$ and $\mathfrak{a}^{\,}_{y}\to0$
[recall Eq.~(\ref{eq: infinite band width limit})],
whereas
\begin{equation}
\frac{v^{\,}_{\pm,j}(x,y)}{\bar{w}^{\prime}_{+,0}},
\frac{u^{\,}_{\pm,0}(x,y)}{\bar{w}^{\prime}_{+,0}},
\frac{w^{\prime}_{-,0}(x,y)}{\bar{w}^{\prime}_{+,0}},
\frac{w^{\prime\prime}_{\pm,j}(x,y)}{\bar{w}^{\prime}_{+,0}}\to0,
\qquad
j=1,2,3,
\label{eq:w vanish}
\end{equation}
\end{subequations}
in the same limit for any $(x,y)\in\mathbb{R}^{2}$. 
Consequently, the penultimate line and the last line, except for the term
$
v^{\,}_{u,y}\sigma^{\,}_{0}\otimes\tau^{\,}_{1}(-{i})\partial^{\,}_{y}
$, 
on the right-hand side of Eq.%
~(\ref{eq: mathcal{H} even under TRS a})
vanish in this limit.
Third, we assume for simplicity but without loss of generality that
\begin{subequations}
\begin{equation}
\bar{v}^{\,}_{-,3}\equiv v^{\,}_{\mathrm{s},x}
\end{equation}
is a finite and non-vanishing positive number, while
\begin{equation}
\bar{v}^{\,}_{+,j}=
\bar{v}^{\,}_{-,1}=
\bar{v}^{\,}_{-,2}=0,
\qquad
j=1,2,3.
\end{equation}
\end{subequations}
Fourth, we consider the limit
\begin{equation}
\frac{\tilde{v}^{\,}_{\pm,\mu}(x,y)}{v^{\,}_{\mathrm{s},x}}\to0,
\qquad
\mu=0,1,2,3,
\end{equation}
for any $(x,y)\in\mathbb{R}^{2}$. 
This limit justifies treating a Dirac point as a fixed point in
the sense of the renormalization group.
Fifth, we assume a finite and non-vanishing dimerization ***
\begin{equation}
\bar{w}^{\prime}_{-,0}\neq0,\infty.
\end{equation}
The Dirac point is then captured by the single-particle Hamiltonian
\begin{align}
\mathcal{H}^{\mathrm{AII}}_{\mathrm{{d}}}\equiv&\,
\bar{\mathcal{H}}^{(\mathrm{e})}
\nonumber\\
=&\,
v^{\,}_{\mathrm{s},x}\,
\sigma^{\,}_{3}\otimes\tau^{\,}_{3}\,
(-{i})\partial^{\,}_{x}
+
v^{\,}_{\mathrm{u},y}\,
\sigma^{\,}_{0}\otimes\tau^{\,}_{1}\,
(-{i})\partial^{\,}_{y}
\nonumber\\
&\,
+
\bar{u}^{\,}_{+,0}\,
\sigma^{\,}_{0}\otimes\tau^{\,}_{0}
+
\bar{u}^{\,}_{-,0}\,
\sigma^{\,}_{0}\otimes\tau^{\,}_{3}
+
{\sum_{j=1}^{3}}
2\,
\bar{w}^{\prime\prime}_{+,j}\,
\sigma^{\,}_{j}\otimes\tau^{\,}_{1}
+
2\,
\bar{w}^{\prime}_{-,0}\,
\sigma^{\,}_{0}\otimes\tau^{\,}_{2}.
\end{align}
The parameters $v^{\,}_{\mathrm{s},x}$ and $v^{\,}_{\mathrm{u},y}$
enter as anisotropic Dirac velocities.
The parameter $\bar{u}^{\,}_{+,0}$ enters as a chemical potential.
The parameter $2\,\bar{w}^{\prime}_{-,0}$ enters as a mass.
The mass term anticommutes with all terms except for the chemical potential.
The remaining 4 terms with the parameters $\bar{u}^{\,}_{-,0}$
and $\bar{w}^{\prime\prime}_{+,j}$ with $j=1,2,3$
do not anticommute with all the gamma matrices multiplying the
first derivatives in position space. 

\begin{figure}[t]
\centering
\includegraphics[width=0.3\textwidth]{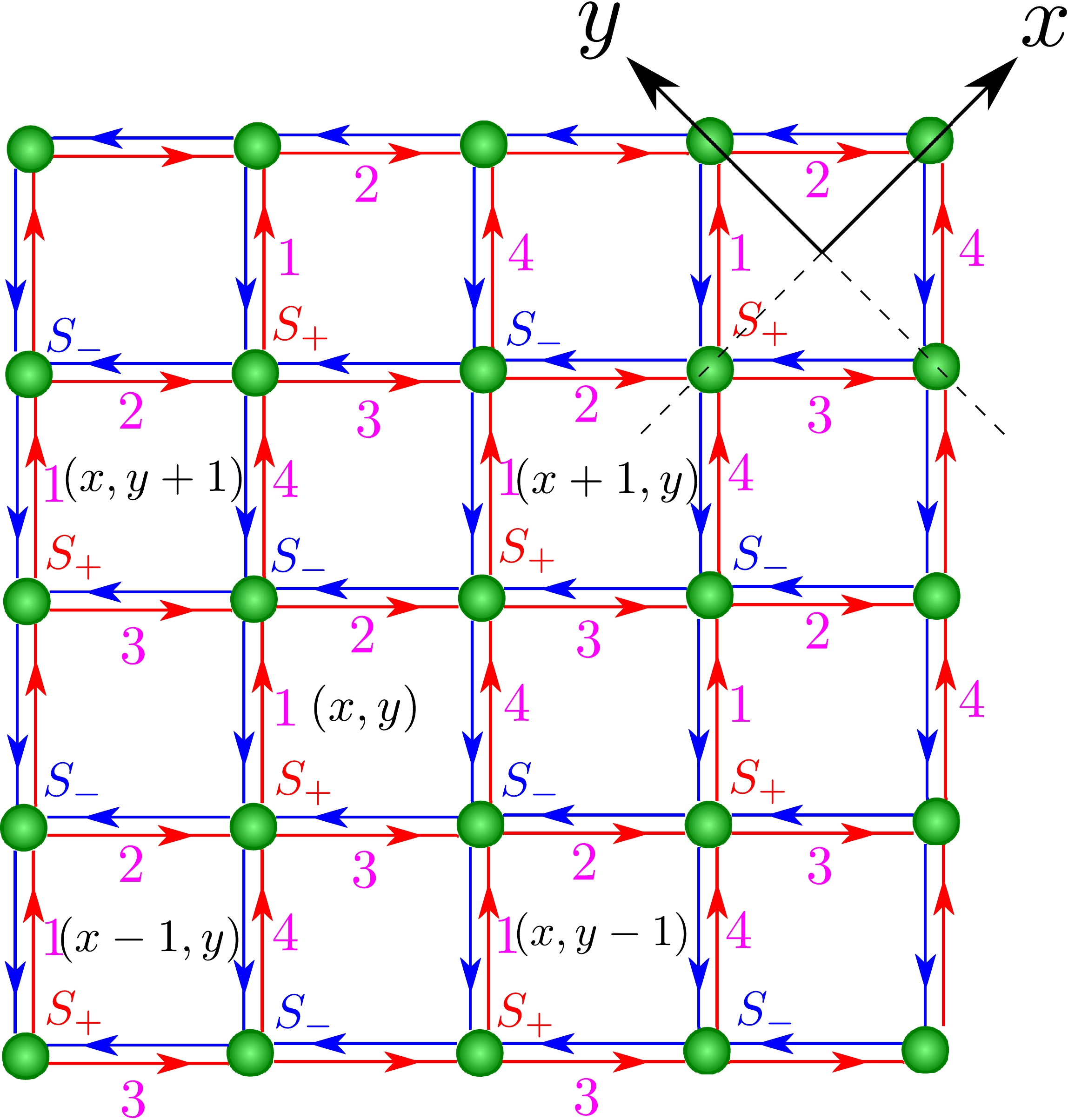}
\caption{
(Color online:)
The vertex representation in Fig.~\ref{fig:network directed Z2}(a)
of the two-dimensional spin-directed $\mathbb{Z}^{\,}_{2}$ network model
is displayed after rotating counterclockwise
Fig.~\ref{fig:network directed Z2}(a)
by $\pi/4$. The vertices at which the elementary scattering 
shown in Fig.~\ref{fig:network-dimer} takes place
form a bipartite square lattice denoted $\Lambda$.
Each of the two interpenetrating sublattices of $\Lambda$
is a square lattice. They are denoted $\Lambda^{\,}_{+}$ and $\Lambda^{\,}_{-}$.
The vertices of sublattices  $\Lambda^{\,}_{+}$ represent the
scattering matrix $S^{\,}_{+}$ obtained from
Eq.~(\ref{eq:S-matrix-dimer})
with the substitution $t\to t^{\,}_{+}$ and $r\to r^{\,}_{+}$.
The vertices of sublattices $\Lambda^{\,}_{-}$ represent the
scattering matrix $S^{\,}_{-}$ obtained from
Eq.~(\ref{eq:S-matrix-dimer}) 
with the substitution $t\to t^{\,}_{-}$ and $r\to r^{\,}_{-}$.
The lattice dual to $\Lambda$,
denoted $\Lambda^{\star}$,
is made of the sites at the center of the plaquettes of
$\Lambda$. 
The four edges of a plaquette of $\Lambda$ are either numbered
3, 4, 2, 1 or 2, 1, 3, 4 in a counterclockwise fashion.
The sites of $\Lambda^{\star}$ denoted $(x,y)$ 
are the centers of the plaquettes from $\Lambda$ numbered
3, 4, 2, 1 counterclockwise. This leaves all the sites of $\Lambda^{\star}$
unnumbered if they are centers of plaquettes in $\Lambda$ numbered 
2, 1, 3, 4 counterclockwise. 
Pairs of helical plane waves that form a Kramers' degenerate
doublet are represented by arrows on the nearest-neighbor links
of $\Lambda$. The color code is red for spin up and blue for spin down.
The eight plane waves along the edges of the plaquette
centered on the site $(x,y)$
from the dual lattice $\Lambda^{\star}$
are assigned the coordinate $(x,y)$.
        }
\label{sfig:network-lattice}
\end{figure}

\section{
Dirac Hamiltonian from the two-dimensional 
spin-directed $\mathbb{Z}^{\,}_{2}$ network model
        }
\label{suppl: Dirac Hamiltonian from the directed Z2 network model}

Starting from the two-dimensional CC network model,
Ho and Chalker derived in Ref.~\onlinecite{Ho96S}
the random Dirac Hamiltonian studied in Ref.~\onlinecite{Ludwig96S}
on its own merits. The two-dimensional $\mathbb{Z}^{\,}_{2}$
network model for a strong two-dimensional $\mathbb{Z}^{\,}_{2}$
topological insulator was related to a random Dirac Hamiltonian
in Ref.~\onlinecite{Ryu10S}. In this Appendix, we derive from
the two-dimensional spin-directed $\mathbb{Z}^{\,}_{2}$ network model
a random Dirac Hamiltonian.

\subsection{Block-off-diagonal representation of the scattering matrix}
 
To this end and along the same lines as in Ref.~\onlinecite{Ryu10S},
we are going to reformulate
the two-dimensional spin-directed $\mathbb{Z}^{\,}_{2}$ network model
in a form more suitable for taking a continuum limit that will yield
a Dirac Hamiltonian.

With the notations and conventions explained
in Fig.~\ref{sfig:network-lattice},
we can drop the incoming and outgoing labels on the plane waves
in an unambiguous way and write, for example,
\begin{subequations}
\begin{equation}
\psi^{\,}_{1,\uparrow}(x,y)=
e^{+{i}\phi^{\,}_{0}}
\left[
r^{\,}_{+}\,
e^{+{i}\phi^{\,}_{3}}\,
\psi^{\,}_{2,\uparrow}(x-1,y)
+
t^{\,}_{+}\,
e^{+{i}\phi^{\,}_{1}}\,
\cos\theta\,
\psi^{\,}_{4,\uparrow}(x-1,y)
+
t^{\,}_{+}\,
e^{+{i}\phi^{\,}_{2}}\,
\sin\theta\,
\psi^{\,}_{3,\downarrow}(x,y)
\right]
\end{equation}
and (we need to make the identification $S^{\dag}\to S^{\,}_{-}$)
\begin{equation}
\psi^{\,}_{1,\downarrow}(x,y)=
e^{-{i}\phi^{\,}_{0}}
\left[
r^{\,}_{-}\,
e^{-{i}\phi^{\,}_{3}}\,
\psi^{\,}_{2,\downarrow}(x,y)
-
t^{\,}_{-}\,
e^{-{i}\phi^{\,}_{2}}\,
\sin\theta\,
\psi^{\,}_{3,\uparrow}(x,y+1)
+
t^{\,}_{-}\,
e^{+{i}\phi^{\,}_{1}}\,
\cos\theta\,
\psi^{\,}_{4,\downarrow}(x,y+1)
\right].
\end{equation}
\end{subequations}
Carrying the same manipulations for edges
2, 3, and 4 of the plaquette centered at $(x,y)$
in the dual lattice of Fig.~\ref{sfig:network-lattice},
suggests the introduction of the eight-component spinor
\begin{subequations}
\begin{equation}
\Psi(x,y):=
\begin{pmatrix}
\Psi^{\,}_{+}(x,y)
\\
\Psi^{\,}_{-}(x,y)
\end{pmatrix},
\qquad
\Psi^{\,}_{+}(x,y):=
\begin{pmatrix}
\psi^{\,}_{1,\uparrow}(x,y) 
\\
\psi^{\,}_{4,\downarrow}(x,y) 
\\
\psi^{\,}_{3,\uparrow}(x,y) 
\\
\psi^{\,}_{2,\downarrow}(x,y) 
\end{pmatrix},
\qquad
\Psi^{\,}_{-}(x,y):=
\begin{pmatrix}
\psi^{\,}_{1,\downarrow}(x,y) 
\\
\psi^{\,}_{4,\uparrow}(x,y) 
\\
\psi^{\,}_{3,\downarrow}(x,y) 
\\
\psi^{\,}_{2,\uparrow}(x,y)
\end{pmatrix},
\end{equation}
the $8\times8$ operator-valued matrix
\begin{equation}
\hat{\mathcal{S}}:=
\begin{pmatrix}
0
&
\hat{S}^{\,}_{+-}
\\
\hat{S}^{\,}_{-+}
&
0
\end{pmatrix},
\end{equation}
where the $4\times4$ operator-valued matrices are given by
\begin{equation}
\hat{S}^{\,}_{+-}:=
e^{+{i}\phi^{\,}_{0}}
\begin{pmatrix}
0 
& 
t^{\,}_{+}\, 
e^{+{i}\phi^{\,}_{1}}\,
\cos\theta\,
\hat{s}^{x}_{-} 
& 
t^{\,}_{+}\, 
e^{+{i}\phi^{\,}_{2}}\,
\sin\theta
& 
r_+\,
e^{+{i}\phi^{\,}_{3}}\, 
\hat{s}^{x}_{-} 
\\
t_+ 
e^{+{i}\phi^{\,}_{1}}\,
\cos\theta\, 
\hat{s}^{x}_{+} 
& 
0 
& 
r^{\,}_{+}\, 
e^{-{i}\phi^{\,}_{3}}\, 
\hat{s}^{x}_{+}
& 
-
t^{\,}_{+}\,
e^{-{i}\phi^{\,}_{2}}\,
\sin\theta  
\\
-t^{\,}_{+}\,
e^{+{i}\phi^{\,}_{2}}\,
\sin\theta 
& 
r^{\,}_{+}\, 
e^{-{i}\phi^{\,}_{3}}\,
\hat{s}^{x}_{-} 
& 
0 
& 
-
t^{\,}_{+}\,
e^{-{i}\phi^{\,}_{1}}\, 
\cos\theta\,
\hat{s}^{x}_{-} 
\\
r^{\,}_{+}\, 
e^{+{i}\phi^{\,}_{3}}\, 
\hat{s}^{x}_{+} 
& 
t^{\,}_{+}\, 
e^{-{i}\phi^{\,}_{2}}\,
\sin\theta 
& 
-
t^{\,}_{+}\, 
e^{-{i}\phi^{\,}_{1}}\,
\cos\theta\,
\hat{s}^{x}_{+} 
& 
0
\end{pmatrix}
\end{equation}
and
\begin{equation}
\hat{S}^{\,}_{-+}:=
e^{+{i}\phi^{\prime}_{0}}
\begin{pmatrix}
0 
& 
+t^{\,}_{-}\, 
e^{+{i}\phi^{\prime}_{1}}\, 
\cos\theta\, 
\hat{s}^{y}_{+} 
& 
-t^{\,}_{-}\, 
e^{-{i}\phi^{\prime}_{2}}\, 
\sin\theta\,
\hat{s}^{y}_{+}
& 
r^{\,}_{-} 
e^{-{i}\phi^{\prime}_{3}}
\\
t^{\,}_{-} 
e^{+{i}\phi^{\prime}_{1}}\,
\cos\theta\,
\hat{s}^{y}_{-}\, 
& 
0 
& 
r^{\,}_{-}\, 
e^{+{i}\phi^{\prime}_{3}} 
& 
t^{\,}_{-} 
e^{+{i}\phi^{\prime}_{2}}\,
\sin\theta 
\hat{s}^{y}_{-}
\\
t^{\,}_{-} 
e^{-{i}\phi^{\prime}_{2}} 
\sin\theta\,
\hat{s}^{y}_{-} 
& 
r^{\,}_{-}\, 
e^{+{i}\phi^{\prime}_{3}} 
& 
0 
& 
-
t^{\,}_{-}\,
e^{-{i}\phi^{\prime}_{1}}\, 
\cos\theta\,
\hat{s}^{y}_{-}
\\
r^{\,}_{-}\, 
e^{-{i}\phi^{\prime}_{3}} 
& 
-
t^{\,}_{-}\, 
e^{+{i}\phi^{\prime}_{2}}\,
\sin\theta\, 
\hat{s}^{y}_{+} 
& 
-
t^{\,}_{-}\, 
e^{-{i}\phi^{\prime}_{1}}\, 
\cos\theta\,
\hat{s}^{y}_{+} 
& 
0
\end{pmatrix},
\end{equation}
while we have introduced the translation operators
\begin{equation}
\left(\hat{s}^{x}_{\pm}f\right)(x,y)=f(x\pm1,y),
\qquad
\left(\hat{s}^{y}_{\pm}f\right)(x,y)=f(x,y\pm1),
\end{equation} 
for we may then interpret the linear equation
\begin{equation}
\Psi(x,y;t+\delta t):=
\hat{\mathcal{U}}\,
\Psi(x,y;t),
\qquad
\hat{\mathcal{U}}:=
\hat{\mathcal{S}}^{2}=
\begin{pmatrix}
\hat{S}^{\,}_{+-}\,
\hat{S}^{\,}_{-+}
&
0
\\
0
&
\hat{S}^{\,}_{-+}\,
\hat{S}^{\,}_{+-}
\end{pmatrix}=
\hat{\mathcal{U}}^{\dag},
\end{equation}
\end{subequations}
as a unitary time-evolution by the discrete time step $\delta t\equiv1$ 
for the two-dimensional spin-directed $\mathbb{Z}^{\,}_{2}$ network model.
Following Ho and Chalker,%
~\cite{Ho96S}
we may define the Hamiltonian
\begin{equation}
\hat{\mathcal{H}}:=
+{i}\left(\hat{\mathcal{U}}-1\right)\equiv
\begin{pmatrix}
\hat{H}^{\,}_{+-}
&
0
\\
0
&
\hat{H}^{\,}_{-+}
\end{pmatrix}
\label{appeq: full Hamiltonian for directed Z2 network} 
\end{equation}
for the two-dimensional spin-directed $\mathbb{Z}^{\,}_{2}$ network model.
Without loss of generality, 
we are after the upper $4\times4$ Hermitean block
\begin{equation}
\hat{H}^{\,}_{+-}\equiv
\begin{pmatrix}
\hat{H}^{\,}_{A}
&
\hat{H}^{\,}_{AB}
\\
\hat{H}^{\dag}_{AB}
&
\hat{H}^{\,}_{B}
\end{pmatrix},
\qquad
\hat{H}^{\,}_{A}=\hat{H}^{\dag}_{A},
\qquad
\hat{H}^{\,}_{B}=\hat{H}^{\dag}_{B},
\label{appeq: upper block Hamiltonian for directed Z2 network} 
\end{equation}
for the two-dimensional spin-directed $\mathbb{Z}^{\,}_{2}$ network model.
In particular, we seek a continuum limit of $\hat{H}^{\,}_{+-}$.

\subsection{
Dirac Hamiltonian from 
the two-dimensional spin-directed $\mathbb{Z}^{\,}_{2}$ network model
close to $\theta=\pi/2$
           }

We are going to deduce Hamiltonian%
~(\ref{appeq: upper block Hamiltonian for directed Z2 network})
from expanding 
Eq.~(\ref{appeq: full Hamiltonian for directed Z2 network})
when the parameters of 
the two-dimensional spin-directed $\mathbb{Z}^{\,}_{2}$ network model
are in the close vicinity to the line that is parametrized by $t^{2}$
in the three-dimensional parameter space%
~(\ref{eq: def parameter space is 3D}) and given by
\begin{equation}
(t^{2},\theta,\delta^{2})=(t^{2},\pi/2,0).
\label{sappeq: critical line delta-pi/2=delta2=0}
\end{equation}
We have seen in Sec.%
~\ref{subsubsec: Limit theta=pi/2 without dimerization}
that the two-dimensional spin-directed $\mathbb{Z}^{\,}_{2}$ network model
decouples along this line into two critical CC network models. 
In turn, the critical CC network model is related to a Dirac Hamiltonian.%
~\cite{Ho96S} For this reason we seek the continuum limit of
Hamiltonian%
~(\ref{appeq: upper block Hamiltonian for directed Z2 network})
close to the critical line%
~(\ref{sappeq: critical line delta-pi/2=delta2=0}).

The expansions
\begin{subequations}
\begin{equation}
\begin{split}
&
e^{+{i}\phi^{\,}_{0}}=
1+{i}\phi^{\,}_{0}+\cdots,
\qquad
e^{+{i}\phi^{\,}_{1}}=
1+{i}\phi^{\,}_{1}+\cdots,
\qquad
e^{+{i}\phi^{\,}_{2}}=
1+{i}\phi^{\,}_{2}+\cdots,
\qquad
e^{+{i}\phi^{\,}_{3}}=
1+{i}\phi^{\,}_{3}+\cdots,
\\
&
e^{+{i}\phi^{\prime}_{0}}=
1+{i}\phi^{\prime}_{0}+\cdots,
\qquad
e^{+{i}(\pi+\phi^{\prime}_{1})}=
-(1+{i}\phi^{\prime}_{1}+\cdots),
\qquad
e^{+{i}\phi^{\prime}_{2}}=
1+{i}\phi^{\prime}_{2}+\cdots,
\qquad
e^{+{i}\phi^{\prime}_{3}}=
1+{i}\phi^{\prime}_{3}+\cdots,
\end{split}
\end{equation}
for the random phases, the expansions
\begin{equation}
t^{\,}_{+}\, 
t^{\,}_{-}= 
t^{2}
+
\cdots, 
\quad
r^{\,}_{+}\, 
r^{\,}_{-} =
r^{2}
+
\cdots,
\quad
t^{\,}_{\pm}\,
r^{\,}_{\mp}=
t\,r\, 
\pm 
\frac{\delta^{2}}{2t\,r}
+
\cdots,
\quad
\cos\left(\frac{\pi}{2}-\theta'\right)=\theta'+\cdots,
\quad
\sin\left(\frac{\pi}{2}-\theta'\right)=1+\cdots,
\end{equation}
for the products of the transmission and reflection amplitudes
and for the deviations of $\theta$ about $\pi/2$
($\theta=\frac{\pi}{2}-\theta'$),
and the gradient expansions
\begin{equation}
s^{x}_{\pm}= 
1
\pm 
\partial^{\,}_{x}
+
\cdots,
\qquad
s^{y}_{\pm}= 
1
\pm 
\partial^{\,}_{y}
+
\cdots,
\end{equation}
\end{subequations}
for the shift operators, deliver the following continuum limit
for the three $2\times2$ blocks entering the right-hand side of Eq.%
~(\ref{appeq: upper block Hamiltonian for directed Z2 network}).
With the notations
\begin{subequations}
\begin{equation}
p^{\,}_{x,y}\equiv 
-{i}\partial^{\,}_{x,y},
\qquad
m\equiv 
\frac{\delta^{2}}{t\,r},
\qquad
A^{\,}_{x}\equiv 
\phi^{\,}_{3}-\phi^{\prime}_{3},  
\qquad
A^{\,}_{y}\equiv 
\phi^{\,}_{2}-\phi^{\prime}_{2},
\qquad
\lambda^{\,}_{\phi}\equiv 
\phi^{\,}_{1}
-
\phi^{\prime}_{1},
\qquad
V^{\,}_{0}\equiv
-
\phi^{\,}_{0}
-
\phi^{\prime}_{0},
\end{equation}
one verifies that
\begin{equation}
\begin{split}
&
\hat{H}^{\,}_{A}=
-
t\,
r\,
\left[
\left(
p^{\,}_{x}
-
A^{\,}_{x}
\right)
- 
\left(
p^{\,}_{y}
-
A^{\,}_{y}
\right)
\right]
\sigma^{\,}_{1} 
-
m\,
\sigma^{\,}_{2}
+
\left[
r^{2}\,
(p^{\,}_{x}-A^{\,}_{x}) 
+ 
t^{2} 
(p^{\,}_{y}-A^{\,}_{y})
\right]
\sigma^{\,}_{3}, 
\\
&
\hat{H}^{\,}_{B}=
+
t\, 
r\,
\left[ 
\left(
p^{\,}_{x}
+
A^{\,}_{x}
\right) 
+
\left(
p^{\,}_{y}
+
A^{\,}_{y}
\right)
\right]
\sigma^{\,}_{1} 
+
m\,
\sigma^{\,}_{2} 
+
\left[
r^{2}\,
\left(
p^{\,}_{x}
+
A^{\,}_{x}
\right) 
- 
t^{2} 
\left(
p^{\,}_{y}
+
A^{\,}_{y}
\right)
\right]
\sigma^{\,}_{3},
\\
&
\hat{H}^{\,}_{AB}=
+
\theta^{\prime}\,
t\,
\left(
2
+
{i}\lambda^{\,}_\phi
\right)
\left(
{i}
r\,\sigma^{\,}_{0}
-
t\,\sigma^{\,}_{2}
\right).
\end{split}
\end{equation}
\end{subequations}
Since we assume that $\theta^{\prime}$ is small,
we neglect the other small variables, 
except for $\phi^{\,}_{1}$ and
$\phi^{\prime}_{1}$, 
to derive $\hat{H}^{\,}_{AB}$.
With the help of a second set of Pauli matrices
$\tau^{\,}_{1}$, $\tau^{\,}_{2}$, and $\tau^{\,}_{3}$, 
together with the unit
$2\times2$ matrix $\tau^{\,}_{0}$
and the unitary transformation defined by
\begin{subequations}
\label{appeq: H+- if theta=pi/2}
\begin{equation}
R(t):= 
\left(
\sigma^{\,}_{0}
\otimes
\frac{\tau^{\,}_{0}+\tau^{\,}_{3}}{2} 
+
e^{-{i}\alpha(t)\,\sigma^{\,}_{2}}
\otimes
\frac{\tau^{\,}_{0}-\tau^{\,}_{3}}{2} 
\right)\,
\left(
e^{-{i}\frac{\pi}{4}\sigma^{\,}_{1}}
e^{-{i}\frac{\pi}{2}\sigma^{\,}_{2}}
\right)
\otimes
\tau^{\,}_{0},
\qquad
\alpha(t):= 
\arcsin\,t,
\label{appeq: H+- if theta=pi/2 a}
\end{equation}
we find
\begin{equation}
\begin{split}
{\cal H}_{\theta=\pi/2} \equiv
\hat{H}^{\,}_{+-}=&\,
\left[
t\,
r
\left(
p^{\,}_{x}
-
p^{\,}_{y}
\right) 
\sigma^{\,}_{1}
+ 
\left( 
r^{2}\,p^{\,}_{x} 
+ 
t^{2}\, p^{\,}_{y}
\right) 
\sigma^{\,}_{2}
\right] 
\otimes 
\tau^{\,}_{0} 
-
\left[
t\,
r 
\left(
A^{\,}_{x}
-
A^{\,}_{y}
\right) 
\sigma^{\,}_{1}
+ 
\left(
r^2\, 
A^{\,}_{x} 
+ 
t^{2}\, 
A^{\,}_{y}
\right) 
\sigma^{\,}_{2} 
\right] 
\otimes 
\tau^{\,}_{3} 
\\
&\,
- 
m\,\sigma^{\,}_{3}\otimes\tau^{\,}_{3} 
- 
\theta^{\prime}\,
t\,
\left(
\lambda^{\,}_{\phi}\,
\sigma^{\,}_0 \otimes \tau^{\,}_{1}
+ 
2\,
\sigma^{\,}_{0} 
\otimes
\tau^{\,}_{2}
\right)
+ 
V^{\,}_{0} \,
\sigma^{\,}_{0} 
\otimes 
\tau^{\,}_{0}.
\end{split}
\label{sappeq: H+- if theta=pi/2 b}
\end{equation}
\end{subequations}
This Hamiltonian is invariant under reversal of time, i.e.,
\begin{equation}
\mathcal{T}\,
{\cal H}_{\theta=\pi/2}\,
\mathcal{T}^{-1}= 
{\cal H}_{\theta=\pi/2},
\qquad 
\mathcal{T}:= 
{i}\sigma^{\,}_{2}\otimes\tau^{\,}_{1}\, 
K,
\end{equation}
where $K$ denotes the operation of complex conjugation.
The mass $m$ that encodes the dimerization multiplies the matrix
$\sigma^{\,}_{3}\otimes\tau^{\,}_{3}$
that anticommutes with all other contributions to the 
continuum limit~(\ref{sappeq: H+- if theta=pi/2 b})
with $V^{\,}_{0}=0$. Hence, dimerization opens a spectral gap in the
spectrum of Hamiltonian~(\ref{sappeq: H+- if theta=pi/2 b}).
At the isotropic point defined by 
$t^{2}=r^{2}=1/2$,
Hamiltonian~(\ref{sappeq: H+- if theta=pi/2 b}) 
becomes the Dirac Hamiltonian
\begin{subequations}
\begin{equation}
{\cal H}_{\theta=\pi/2}=
\frac{1}{\sqrt{2}}
\left(
p^{\prime}_{x}\,\sigma^{\,}_{1} 
+ 
p^{\prime}_{y}\,\sigma^{\,}_{2}
\right)
\otimes 
\tau^{\,}_{0} 
-
\frac{1}{\sqrt{2}}
\left(
A^{\prime}_{x}\,\sigma^{\,}_{1} 
+ 
A^{\prime}_{y}\,\sigma^{\,}_{2}
\right) 
\otimes 
\tau^{\,}_{3}   
- 
m\, 
\sigma^{\,}_{3} \otimes \tau^{\,}_{3}
- 
\frac{\theta^{\prime}}{\sqrt{2}}
\left(
\lambda^{\,}_{\phi}\,
\sigma^{\,}_0 \otimes \tau^{\,}_{1}
+ 
2\,
\sigma^{\,}_{0}\otimes\tau^{\,}_{2} 
\right)
+
V^{\,}_{0} 
\sigma^{\,}_{0} 
\otimes 
\tau^{\,}_{0},
\end{equation}
where 
\begin{equation}
p^{\prime}_{x}\equiv 
\frac{
p^{\,}_{x}-p^{\,}_{y}
     }
     {
\sqrt{2}
     },
\qquad
p^{\prime}_{y}\equiv 
\frac{
p^{\,}_{x}+p^{\,}_{y}
     }
     {
\sqrt{2}
     }, 
\qquad
A^{\prime}_{x}\equiv 
\frac{
A^{\,}_{x}-A^{\,}_{y}
     }
     {
\sqrt{2}
     }, 
\qquad
A^{\prime}_{y}\equiv 
\frac{
A^{\,}_{x}+A^{\,}_{y}
     }
     {
\sqrt{2}
     }.
\end{equation}
\end{subequations}

\subsection{
Dirac Hamiltonian from
the two-dimensional spin-directed $\mathbb{Z}^{\,}_{2}$ network model
close to $\theta=0$}

We are going to deduce Hamiltonian%
~(\ref{appeq: upper block Hamiltonian for directed Z2 network})
from expanding 
Eq.~(\ref{appeq: full Hamiltonian for directed Z2 network})
when the parameters of 
the two-dimensional spin-directed $\mathbb{Z}^{\,}_{2}$ network model
are in the close vicinity to the line that is parametrized by $t^{2}$
in the three-dimensional parameter space%
~(\ref{eq: def parameter space is 3D}) and given by
\begin{equation}
(t^{2},\theta,\delta^{2})=(t^{2},0,0).
\label{sappeq: line at theta=delta2=0}
\end{equation}
We have seen in Sec.%
~\ref{subsubsec: Limit theta=0 without dimerization}
that the two-dimensional spin-directed $\mathbb{Z}^{\,}_{2}$ network model
decouples along this line into two two-dimensional directed CC network model.
In turn, the two-dimensional directed CC network model is critical.
For this reason we seek the continuum limit of
Hamiltonian%
~(\ref{appeq: upper block Hamiltonian for directed Z2 network})
close to the line~(\ref{sappeq: line at theta=delta2=0}).

The spin up and down quantum numbers are separately conserved 
along the line~(\ref{sappeq: line at theta=delta2=0}).
This suggests the use of the basis for the scattering states defined by
\begin{equation}
\begin{pmatrix}
\psi^{\,}_{1,\uparrow}(x,y) 
\\
\psi^{\,}_{3,\uparrow}(x,y) 
\\
\psi^{\,}_{2,\downarrow}(x,y) 
\\
\psi^{\,}_{4,\downarrow}(x,y)
\end{pmatrix} 
=
\hat{S}^{\,}_{+-}
\begin{pmatrix}
\psi^{\,}_{1,\downarrow}(x,y) 
\\
\psi^{\,}_{3,\downarrow}(x,y) 
\\
\psi^{\,}_{2,\uparrow}(x,y) 
\\
\psi^{\,}_{4,\uparrow}(x,y)
\end{pmatrix},
\qquad
\begin{pmatrix}
\psi^{\,}_{1,\downarrow}(x,y) 
\\
\psi^{\,}_{3,\downarrow}(x,y) 
\\
\psi^{\,}_{2,\uparrow}(x,y) 
\\
\psi^{\,}_{4,\uparrow}(x,y)
\end{pmatrix} 
=
\hat{S}^{\,}_{-+}
\begin{pmatrix}
\psi^{\,}_{1,\uparrow}(x,y) 
\\
\psi^{\,}_{3,\uparrow}(x,y) 
\\
\psi^{\,}_{2,\downarrow}(x,y) 
\\
\psi^{\,}_{4,\downarrow}(x,y)
\end{pmatrix}.
\label{appeq: def basis for theta=0 case}
\end{equation}

The expansions
\begin{subequations}
\begin{equation}
\begin{split}
&
e^{+{i}\phi^{\,}_{0}}=
1+{i}\phi^{\,}_{0}+\cdots,
\qquad
e^{+{i}\phi^{\,}_{1}}=
1+{i}\phi^{\,}_{1}+\cdots,
\qquad
e^{+{i}\phi^{\,}_{2}}=
1+{i}\phi^{\,}_{2}+\cdots,
\qquad
e^{+{i}\phi^{\,}_{3}}=
1+{i}\phi^{\,}_{3}+\cdots,
\\
&
e^{+{i}\phi^{\prime}_{0}}=
1+{i}\phi^{\prime}_{0}+\cdots,
\qquad
e^{+{i}\phi^{\prime}_{1}}=
1+{i}\phi^{\prime}_{1}+\cdots,
\qquad
e^{+{i}(\pi+\phi^{\prime}_{2})}=
-
\left(
1+{i}\phi^{\prime}_{2}+\cdots
\right),
\qquad
e^{+{i}\phi^{\prime}_{3}}=
1+{i}\phi^{\prime}_{3}+\cdots,
\end{split}
\end{equation}
for the random phases, the expansions
\begin{equation}
t^{\,}_{+}\, 
t^{\,}_{-}= 
t^{2}
+
\cdots, 
\quad
r^{\,}_{+}\, 
r^{\,}_{-} =
r^{2}
+
\cdots,
\quad
t^{\,}_{\pm}\,
r^{\,}_{\mp}=
t\,r\, 
\pm 
\frac{\delta^{2}}{2t\,r}
+
\cdots,
\quad
\cos\theta=1+\cdots,
\quad
\sin\theta=\theta+\cdots,
\end{equation}
for the products of the transmission and reflection amplitudes
and for the deviations of $\theta$ about $0$,
and the gradient expansions
\begin{equation}
s^{x}_{\pm}= 
1
\pm 
\partial^{\,}_{x}
+
\cdots,
\qquad
s^{y}_{\pm}= 
1
\pm 
\partial^{\,}_{y}
+
\cdots,
\end{equation}
\end{subequations}
for the shift operators, deliver the following continuum limit
for the three $2\times2$ blocks entering the right-hand side of Eq.%
~(\ref{appeq: upper block Hamiltonian for directed Z2 network})
represented in the basis~(\ref{appeq: def basis for theta=0 case}).
With the notations
\begin{subequations}
\begin{equation}
p^{\,}_{x,y}\equiv 
-{i}\partial^{\,}_{x,y},
\qquad
m\equiv 
\frac{\delta^{2}}{t\,r},
\qquad
A^{\,}_{x}\equiv 
\phi^{\,}_{3}-\phi^{\prime}_{3},  
\qquad
A^{\,}_{y}\equiv 
\phi^{\,}_{1}-\phi^{\prime}_{1},
\qquad
\lambda^{\,}_{\phi}\equiv 
\phi^{\,}_{2} 
+ 
\phi^{\prime}_{2},
\qquad
V^{\,}_{0}\equiv
-
\phi^{\,}_{0}
-
\phi^{\prime}_{0},
\end{equation}
one verifies that
\begin{equation}
\begin{split}
&
\hat{H}^{\,}_{A}=
p^{\,}_{x}\,
\sigma^{\,}_{0}
+
t\,
r\,
\left[
A^{\,}_{x}
+ 
(p^{\,}_{y}-A^{\,}_{y})
\right]
\sigma^{\,}_{1} 
-
m\,
\sigma^{\,}_{2}
-
\left[
r^{2}\,
A^{\,}_{x}
-
t^{2}\,
\left(
p^{\,}_{y}
-
A^{\,}_{y}
\right) 
\right]
\sigma^{\,}_{3}, 
\\
&
\hat{H}^{\,}_{B}=
-
p^{\,}_{x}\,
\sigma^{\,}_{0}
-
t\, 
r\,
\left[ 
A^{\,}_{x}
+
\left(
p^{\,}_{y}
+
A^{\,}_{y}
\right)
\right]
\sigma^{\,}_{1} 
+
m\,
\sigma^{\,}_{2} 
-
\left[
r^{2}\,
A^{\,}_{x}
-
t^{2} 
\left(
p^{\,}_{y}
+
A^{\,}_{y}
\right)
\right]
\sigma^{\,}_{3},
\\
&
\hat{H}^{\,}_{AB}=
-
\theta\,
t\,
\left(
+
2
+
{i}
\lambda^{\,}_{\phi}\,
\right)
\left(
{i}t\,
\sigma^{\,}_{0}
+
r\,
\sigma^{\,}_{2}
\right).
\end{split}
\end{equation}
\end{subequations}
Since we assume that $\theta$ is small,
we neglect the other small variables, except for $\phi_2$ and
$\phi_2^{\prime}$, 
to derive $\hat{H}_{AB}$.
With the help of a second set of Pauli matrices
$\tau^{\,}_{1}$, $\tau^{\,}_{2}$, and $\tau^{\,}_{3}$, together with the unit
$2\times2$ matrix $\tau^{\,}_{0}$
and the unitary transformation defined by
\begin{subequations}
\label{appeq: H+- if theta=0}
\begin{equation}
R(t):= 
\left(
\sigma^{\,}_{0}
\otimes
\frac{\tau^{\,}_{0}-\tau^{\,}_{3}}{2}
+
e^{+{i}\beta(t)\,\sigma^{\,}_{2}}
\otimes
\frac{\tau^{\,}_{0}-\tau^{\,}_{3}}{2}
\right)
e^{-{i}\frac{\pi}{4}\sigma^{\,}_{1}}\,
\otimes
\tau^{\,}_{0},
\qquad
\beta(t):= 
\arccos\,t,
\label{appeq: H+- if theta=0 a}
\end{equation}
we find
\begin{equation}
\begin{split}
{\cal H}_{\theta=0} \equiv
\hat{H}^{\,}_{+-}=&\,
p^{\,}_{x}\,
\sigma^{\,}_{0}
\otimes
\tau^{\,}_{3}
+
\left(
t\,
r\,
p^{\,}_{y}\,
\sigma^{\,}_{1}
+
t^{2}\,
p^{\,}_{y}\,
\sigma^{\,}_{2}
\right)
\otimes
\tau^{\,}_{0}
+
\left[
t\,
r
\left(
A^{\,}_{x}
-
A^{\,}_{y}
\right) 
\sigma^{\,}_{1}
- 
\left( 
r^{2}\,
A^{\,}_{x} 
+ 
t^{2}\,
A^{\,}_{y}
\right) 
\sigma^{\,}_{2}
\right] 
\otimes 
\tau^{\,}_{3} 
\\
&\,
+ 
m\,
\sigma^{\,}_{3}
\otimes
\tau^{\,}_{3} 
+
\theta\,
t\,
\left(
\lambda^{\,}_{\phi}\,
\sigma^{\,}_{0} \otimes \tau^{\,}_{1}
+ 
2\,
\sigma^{\,}_{0} 
\otimes
\tau^{\,}_{2}
\right)
+ 
V^{\,}_{0}\,
\sigma^{\,}_{0} 
\otimes 
\tau^{\,}_{0}.
\end{split}
\label{sappeq: H+- if theta=0 b}
\end{equation}
\end{subequations}
This Hamiltonian is invariant under reversal of time, i.e.,
\begin{equation}
\mathcal{T}\,
{\cal H}_{\theta=0}\,
\mathcal{T}^{-1}= 
{\cal H}_{\theta=0},
\qquad 
\mathcal{T}:= 
{i}\sigma^{\,}_{2}\otimes\tau^{\,}_{1}\, 
K,
\end{equation}
where $K$ denotes the operation of complex conjugation.

As was the case for
the continuum limit~(\ref{sappeq: H+- if theta=pi/2 b}),
the term
$V^{\,}_{0}\,\sigma^{\,}_{0}\otimes\tau^{\,}_{0}$
acts as a chemical potential, 
for it commutes with all contributions
to the continuum limit~(\ref{sappeq: H+- if theta=0 b}).
We shall set $V^{\,}_{0}=0$ when deciding if a gap at energy 0 
is opened by dimerization.
In comparison to the continuum limit~(\ref{sappeq: H+- if theta=pi/2 b}),
the term $p^{\,}_{x}\,\sigma_{0}\otimes\tau^{\,}_{3}$
has appeared that commutes with all contributions
to the continuum limit~(\ref{sappeq: H+- if theta=0 b})
except for the term
$
\theta\,
t\,
\left(
\lambda^{\,}_{\phi}\,
\sigma^{\,}_{0} \otimes \tau^{\,}_{1}
+ 
2\,
\sigma^{\,}_{0} 
\otimes
\tau^{\,}_{2}
\right)$.
If we set  $t=A^{\,}_{x}=A^{\,}_{y}=V^{\,}_{0}=0$,
we find the two (two-fold degenerate) gapless dispersions $|p^{\,}_{x}\pm m|$.
More generally, a branch of excitation is expected to cross
the energy 0 at some $m$-dependent value of the momentum 
when $\theta=V^{\,}_{0}=0$.
As the coupling $m$ is caused by dimerization,
dimerization thus fails to open a gap if we set $\theta=V^{\,}_{0}=0$.
On the other hand, because the term
$\theta\,
t\,
\left(
\lambda^{\,}_{\phi}\,
\sigma^{\,}_{0} \otimes \tau^{\,}_{1}
+ 
2\,
\sigma^{\,}_{0} 
\otimes
\tau^{\,}_{2}
\right)$
anticommutes with both 
$m\,\sigma^{\,}_{3}\otimes\tau^{\,}_{3}$ 
and
$p^{\,}_{x}\,\sigma_{0}\otimes\tau^{\,}_{3}$,
we expect that a sufficiently large $\theta$ opens a gap
for a given $m$.

\medskip
\end{widetext}


\begin{thebibliography}{99}
\bibitem{footnote: order limits matter here}
The order of limits matters here.
We assume that we can describe the disordered surface states of
a strong $\mathbb{Z}^{\,}_{2}$ topological insulator
by a Dirac Hamiltonian in a two-dimensional representation
of the Clifford algebra with all random potentials
compatible with the time-reversal symmetry
white-noise correlated. The disorder strength is 
measured by the variances $g^{2}$ of the random potentials.
The strong disorder limit by which the variances 
$g^{2}$ are made arbitrarily large within the Dirac approximation
presumes that $g\ll\Delta$ still holds whereby $\Delta$
is the bulk band gap. The opposite limit for which the disorder
strength is larger than the bulk band gap destroys all
topological attributes.
\bibitem{footnote critical directed CC}
Critical transport in the direction orthogonal to
the direction along which transport is perfect (directed)
in the directed CC network model for a
weak three-dimensional Chern insulator
was established
by Chalker and Dohmen in Ref.%
~\onlinecite{Chalker95}. 
They quantified how the transport along the direction that
corresponds to the stacking direction 
of a weak three-dimensional Chern insulator
depends on the ratio of the height and the perimeter
in the cylindrical geometry of 
Fig.~\ref{fig:layeredz2pancake}.
\end{thebibliography}
\end{document}